\documentclass{aa}  

\usepackage{lscape}             
\usepackage{placeins}           
\usepackage{pdflscape}
\usepackage{graphicx}
\usepackage{txfonts}
\usepackage{array}
\usepackage{blindtext}
\usepackage{hyperref}
\usepackage{rotating}
\usepackage{graphicx,capt-of}
\usepackage{longtable}
\usepackage{comment}
\usepackage{subcaption}
\usepackage{afterpage}
\usepackage{float}
\usepackage{hyperref}
\usepackage{bm}
\usepackage{multirow}
\usepackage{tcolorbox}
\usepackage{makecell}
\usepackage{tabularx}

\usepackage{natbib}
\bibpunct{(}{)}{;}{a}{}{,} 

\hypersetup{%
        colorlinks=true, linkcolor=blue, 
        breaklinks,
        citecolor = blue,
        unicode=true,
              }

\usepackage{caption}

\begin{document}

   \title{Multi-chromatic observations of classical Cepheids using the CHARA Array interferometer}

   \subtitle{Surface brightness-colour relation, projection factor, and limb-darkening}

%
%
%

   \author{M.C. Bailleul\inst{1}\corrauth{manon.bailleul@oca.eu}        
        \and N.~Nardetto\inst{1} \email{nicolas.nardetto@oca.eu}
        \and D.~Mourard\inst{1} \email{denis.mourard@oca.eu}
        \and A.~Mérand\inst{2} \email{Antoine.Merand@eso.org}
        \and P.~Kervella\inst{3,4} \email{pierre.kervella@obspm.fr}
        \and A.~Gallenne\inst{5} \email{alexandre.gallenne@gmail.com}
        \and G.~Bras\inst{3} \email{garance.bras@obspm.fr}
        \and J.~Monnier\inst{6} \email{monnier@umich.edu}
        \and K.~Sivkova\inst{3,4} \email{Ekaterina.Sivkova@obspm.fr}
        \and B.~Apostolova\inst{1} \email{brankica.apostolova@oca.eu}
        \and L.~Breuval\inst{7} \email{lbreuval@stsci.edu}
        \and W.~Kiviaho\inst{3,4} \email{wilma.kiviaho@obspm.fr}
        \and W.~Gieren\inst{8} \email{wgieren@astro-udec.cl}
        \and V.~Hocdé\inst{1} \email{vincent.hocde@oca.eu}
        \and R.~S. Rathour\inst{1} \email{rajeev.rathour@oca.eu}
        \and G.~Pietrzyński\inst{9} \email{pietrzyn@camk.edu.pl}
        \and P.~Bério\inst{1} \email{philippe.berio@oca.eu}
        \and F.~Morand\inst{1} \email{frederic.morand@oca.eu}
        \and D.~Salabert\inst{1} \email{david.salabert@oca.eu}
        \and J.~Jones\inst{10} \email{jjones176@gsu.edu}
        \and C.~Lanthermann\inst{10} \email{clanthermann@gsu.edu}
        \and N.~Ebrahimkutty\inst{1} \email{nayeem.ebrahimkutty@oca.eu}
        \and J.~Jon\'ak\inst{1} \email{juraj.jonak@oca.eu}
        \and H.~Nowacki\inst{1} \email{hugo.nowacki@oca.eu}
        \and R.~V.~Iba\~nez Bustos\inst{1} \email{romina.ibanez@oca.eu}
        \and M.~Vrard\inst{1} \email{mathieu.vrard@oca.eu}
        \and S.~Kraus\inst{11} \email{S.Kraus@exeter.ac.uk}
        \and N.~Anugu\inst{10} \email{nanugu@gsu.edu}
        \and M.~Gutierrez\inst{6} \email{mgutie60@ucsc.edu}
        \and N.~Ibrahim\inst{6} \email{inoura@umich.edu}
        \and T.~Gardner\inst{12} \email{Tyler.Gardner@colorado.edu}
        }
   \institute{Université Côte d’Azur, Observatoire de la Côte d’Azur, CNRS, Laboratoire Lagrange, Nice, France
   \and European Southern Observatory, 85748 Garching, Munich, Germany
   \and LIRA, Observatoire de Paris, Universit\'e PSL, Sorbonne Universit\'e, Universit\'e Paris Cit\'e, CY Cergy Paris Universit\'e, CNRS, 5 place Jules Janssen, 92195 Meudon, France
   \and French-Chilean Laboratory for Astronomy, IRL 3386, CNRS, Casilla 36-D, Santiago, Chile
   \and Instituto de Alta Investigaci\'on, Universidad de Tarapac\'a, Casilla 7D, Arica, Chile
   \and Astronomy Department, University of Michigan, Ann Arbor, MI 48109, USA
   \and European Space Agency (ESA), ESA Office, Space Telescope Science Institute, 3700 San Martin Drive, Baltimore, MD 21218, USA
   \and Universidad de Concepción, Departamento Astronomía, Casilla 160-C, Concepción, Chile
   \and Nicolaus Copernicus Astronomical Center, Polish Academy of Sciences, ul. Bartycka 18, 00-716 Warszawa, Poland
   \and The CHARA Array of Georgia State University, Mount Wilson Observatory, Mount Wilson, CA 91203, USA
   \and Astrophysics Group, Department of Physics \& Astronomy, University of Exeter, Stocker Road, Exeter, EX4 4QL, UK
   \and Cooperative Institute for Research in Environmental Sciences, 216 UCB, University of Colorado Boulder,  Boulder, CO, 80309, USA
   }

   \date{...}

 
  \abstract
   {The Baade-Wesselink (BW) method, also called the parallax of pulsation method, compares the linear and angular variations of Cepheids to derive their distance. This method is limited, however, by the projection factor, which relates the observed radial velocity to the true pulsation velocity of the star. In this context, interferometry is crucial. Firstly, to apply the BW method to bright Cepheids, and secondly, to calibrate the surface brightness-colour relations (SBCRs). These relations can then be used to derive the angular diameter curve as a function of the pulsation phase of faint Cepheids.
   }
   {Using simultaneous observations from the CHARA Array interferometer in the K, H, and R bands with the MYSTIC, MIRC-X, and SPICA combiners, respectively, we aim to understand the physics of Cepheid atmospheres better.}
   {Applying a specific method to multi-chromatic simultaneous interferometric observations of Cepheids, we derived robust limb-darkened angular diameters that were then used to calibrate the SBCR and study the projection factor. We also developed a strategy to measure the limb-darkening of Cepheids in R, H, and K bands. These measurements were then used to constrain the geometrical component of the projection factor.}
   {From the limb-darkened angular diameter curves of three Cepheids, we decreased the scatter of the SBCR in $V-K$ colour to 0.0011 magnitude and to 0.0040 for the SBCR in $G_{BP}-G_{RP}$. These SBCRs are particularly robust because they are based on multi-chromatic diameters and are homogeneous, which previous calibrations for Cepheids were not.
   For the very first time, we derived the limb-darkening of Cepheids in R, H, and K bands. For $\delta$ Cep, we derived an R-based projection factor of $1.275 \pm 0.051$, and the geometrical part obtained from the measured limb-darkening coefficient in R band was estimated to $1.420 \pm 0.016$ (on average), as expected from stellar static and hydrodynamical atmosphere models. The limb-darkening coefficients obtained in H and K band are consistent with models.}
   {These results demonstrate that multi-chromatic interferometry can improve the accuracy of the BW method. It is therefore essential to continue the CHARA survey of Cepheids in the coming years.}


   \maketitle
   \nolinenumbers

\section{Introduction}
\label{sec:Introduction}
In 1868, Fizeau was the first to propose the idea of using interferometry to measure the angular diameters of stars \citep{fizeau_prix_1868}. However, it was not until 1891, when Michelson determined the angular diameter of the Jupiter moons \citep{michelson_measurement_1891}, that the effectiveness of this technique was proven, and it was not until 1921 that the angular size of a star, the red supergiant Betelgeuse, was measured for the first time \citep{michelson_measurement_1921}. We had to wait until 1997 to finally resolve the angular variation in the prototype of classical Cepheids variables, $\delta$~Cep \citep{mourard_mean_1997}. Almost 30 years later, the angular size variation has now been resolved at better than 3$\sigma$ significance for only seven classical Cepheids. 

Interferometric observations are of great importance to calibrate the surface brightness-colour relation (SBCR), which allows us to derive the angular size of a star using photometry alone. SBCRs were used, for instance, to derive the distance of the Magellanic Clouds from eclipsing binaries \citep{pietrzynski_eclipsing-binary_2013,pietrzynski_distance_2019,graczyk_distance_2020}, which is particularly important for the determination of the Hubble–Lemaître constant \citep{riess_comprehensive_2022, riess_jwst_2024,h0dn_collaboration_local_2026,di_valentino_cosmoverse_2025}. SBCRs are also used in exoplanet studies, to derive the size of exoplanet host stars \citep{gent_sapp_2022,mauro_characterization_2022}, for asteroseismic targets \citep{valle_testing_2024,campante_tess_2019}, and more particularly, in the context of the Baade-Wesselink (BW) method of Cepheid distance determinations \citep{wesselink_surface_1969}. The BW method is simple: the comparison of the angular and linear variations in a Cepheid allows us to directly obtain its distance. This method relies on three distinct approaches depending on how the angular variation is determined: direct interferometric observations \citep{lane_longbaseline_2002, kervella_cepheid_2004}, SBCRs \citep{storm_calibrating_2011,storm_calibrating_2011-1}, and a more recent technique that combines multiple photometric bands, velocimetry, and interferometry \citep[Spectro-Photo-Interferometry for pulsating stars, SPIPS][]{merand_cepheid_2015}. The linear radius variation in the Cepheid is obtained using radial velocity measurements, but it is based on a numerical factor to convert disk-integrated radial velocities into pulsation velocities, the projection-factor \citep{nardetto_self_2004}. 
The $p$ factor is composed of two main components \cite{nardetto_high-resolution_2007}: $p=p_{0}\mathrm{f_{dyn}}$, where $p_0$ corresponds to the geometrical part linked to the limb-darkening of the star, whereas $\mathrm{f_{dyn}}$ is a dynamical part that includes corrections related with the differential velocity between the optical layer associated with the photosphere and the gas velocity layer associated with the line-forming region in the atmosphere of the star.
The determination of this $p$ factor is the main weakness of the BW technique. Several studies have shown that the projection factors of Cepheids are highly dispersed (to approximately 10\%) for all pulsation periods \citep{gallenne_observational_2017, trahin_inspecting_2021}, which limits the precision of the BW method \citep{nardetto_harps-n_2023}. Furthermore, the geometrical projection factor depends on the limb-darkening, and thus, on the wavelength \citep{nardetto_high-resolution_2009}. The BW method is indeed commonly used by combining radial velocity measurements secured in the visible domain with a specific projection factor adapted to visible, in order to retrieve the proper achromatic photospheric velocity. It is thus particularly important to derive the limb-darkening in the visible domain to infer its effect on the geometric projection factor. 

Sect.~\ref{sec:Simultaneous observation using the MIRC-X, MYSTIC, and SPICA/CHARA instruments} presents the interferometric observations of three Cepheids. In Sect.~\ref{sec:Measurement of the limb-darkened diameter of Cepheids} we derive the limb-darkened angular diameters from polychromatic observations. Based on these measurements, SBCRs are calibrated in the $(V-K)$ and ($G_{BP}-G_{RP}$) colour systems in Sect.~\ref{sec:Surface brightness–colour relation}. For the first time, we measure limb-darkening coefficients of Cepheids in R, H, and K bands in Sect.~\ref{sec:Measuring the limb-darkening of classical cepheids}. In Sect.~\ref{sec:New calculation of the projection factor of delta Cep} we convert the limb-darkening coefficients into geometric projection factors ($p_0$) for the R, H, and K bands. On the other hand, a spectroscopic V-band $p$ factor, which includes geometric and dynamical components for $\delta$~Cep, is determined using the inverse Baade–Wesselink approach. Finally, we summarise our results  in Sect.~\ref{sec:Conclusion}.

\section{Simultaneous observations using the MIRC-X, MYSTIC, and SPICA/CHARA instruments}
\label{sec:Simultaneous observation using the MIRC-X, MYSTIC, and SPICA/CHARA instruments}
The Center for High Angular Resolution Astronomy (CHARA) Array, located at Mount Wilson, in Southern California, is a project led by the Georgia State University in the United States. This array of telescopes enables long-baseline interferometric observations (up to 331 meters). It consists of six telescopes, each equipped with a mirror with a diameter of one meter \citep{ten_brummelaar_first_2005}. Our observations were made simultaneously using the CHARA latest-generation instruments, the Michigan InfraRed Combiner-eXeter (MIRC-X) \citep{anugu_mirc-x_2020}, Michigan Young STar Imager at CHARA (MYSTIC) \citep{monnier_mystic_2018,setterholm_mystic_2023}, and Stellar Parameters and Images with a Cophased Array (SPICA) combiners \citep{mourard_arxiv_2026}, which operate in the H, K, and R bands, respectively. Our sample of Cepheids consists of the brightest and best-resolved Cepheids with the CHARA Array: $\delta$~Cep (P=5.4 days, Vmag=3.75), $\eta$ Aql (P=7.2 days, Vmag=3.80) and $\zeta$ Gem (P=10.1 days, Vmag=3.79). The configuration of the telescopes we used during the observations depended on the observability of the target at each date and is reported in Table~\ref{tab:obs_log}, together with the calibrator stars. The angular diameters of the calibrators in H band are indicated in Table~\ref{tab:calibrators}. The observations were collected between CHARA periods 2024B and 2025B with the aim of covering the entire pulsation cycle for each of the Cepheids. 

All the observations were performed with the same gain in the camera of MIRC-X and MYSTIC between science and calibrators, with the exception of the nights of 29 and 30 September 2024 for $\delta$~Cep and $\zeta$ Gem. The results obtained for these two nights should be interpreted with caution.
The MIRC-X and MYSTIC observations were reduced using the remote data reduction machine in Atlanta, and the calibration was performed using the IDL code part of the data reduction software. We also applied the standard MIRC calibration errors to the data\footnote{\url{https://chara.gsu.edu/tutorials/mirc-data-reduction}}. As indicated in the manual\footnote{MIRC-X MYSTIC Pipeline User Manual v0.9.6, by JB. Le Bouquin, C. L. Davies}, we applied a wavelength correction by dividing the measured wavelength by a factor that depended on the year of observation. For MIRC-X, the factor was $1.0054 \pm 0.0006$ for all years except 2025, and it was $0.999 \pm 0.001$ for 2025 observations. For MYSTIC, the factor was $1.0067 \pm 0.0007$ (for all years).
The night of 15 March 2025 was not used because of reduction and calibration issues.
The SPICA data were reduced and calibrated using the SPICA data reduction pipeline \citep{mourard_arxiv_2026}. We checked all the calibrators to ensure that their closure phase (CP or T3PHI) was equal to zero and that there was no sign of binarity. Furthermore, all calibrators are unresolved. The final calibrated science files can be found in the Optical interferometry DataBase (OiDB)\footnote{\url{https://oidb.jmmc.fr}}.

\section{Limb-darkened angular diameter based on multi-chromatic observations}
\label{sec:Measurement of the limb-darkened diameter of Cepheids}

Limb-darkening (LD) is an optical phenomenon observed in stars, where the brightness gradually decreases from the centre of the stellar disk toward its edge. This effect arises from the temperature gradient in the stellar atmosphere.
Observations at the disk centre reach deeper and hotter, and thus, brighter layers, and those near the limb sample cooler layers, resulting in a lower observed intensity. The intensity profile of a star can then be approximated using various limb-darkening laws. These laws are functions of $\mu=\cos{\beta}$, where $\beta$ is defined as the angle between the line of sight and the surface normal. At the centre of the star, $\mu=1$, where the intensity is highest, and at the limb $\mu=0$. We define:
\begin{itemize}
    \item $\theta_{\mathrm{UD}}$: A uniform disk (UD) angular diameter that corresponds to an effective angular diameter that does not take the limb-darkening of the star into account ($I(\mu)=1$). Its value depends on the photometric band.
    \item $\theta_{\mathrm{LD}}^{\mathrm{root}}$: An LD angular diameter associated with a square-root law description of the intensity profile: $I(\mu)=1-c_{\lambda}(1-\mu)-d_{\lambda}(1-\sqrt{\mu})$, with fixed LD coefficients. This diameter was estimated for each photometric bands separately.
    \item $\theta_{\mathrm{LD}}^{\mathrm{poly}}$: A multi-chromatic LD diameter derived using a square-root law and by fixing the LD coefficients in R, H, and K. This diameter was derived by combining data from all photometric bands,
\end{itemize}
where $c_{\lambda}$ and $d_{\lambda}$ are the limb-darkening coefficients associated with the square-root law and were taken from the tables of \cite{claret_gravity_2011} based on Atlas atmosphere models \citep{kurucz_model_1979}. The input parameters of these tables are the effective temperature, the surface gravity, the metallicity, and the micro-turbulence velocity. We considered solar metallicity for the three Cepheids in our sample and a micro-turbulence velocity of 1 km/s, and the $T_{\text{eff}}$ and $\log g$ quantities were derived at the specific pulsating phase of the observation from the SPIPS models \citep{merand_cepheid_2015} presented in \cite{trahin_inspecting_2021}.
The square-root law was chosen as it is known to describe the intensity profile correctly \citep{kervella_radii_2017,ebrahimkutty_optimised_2024}.
We considered the squared visibilities ($V^2$) and the CP for each measurements. All fitting procedures were carried out using the Python code \texttt{pmoired}\footnote{\url{https://github.com/amerand/PMOIRED}} \citep{merand_flexible_2022}, and a bootstrapping procedure was used for a better estimate of the uncertainties. To be conservative and provide more realistic errors on the angular diameters presented in this section, we considered uncertainties derived from the fitting of $V^2$ values alone, that is, we did not consider CP, and when necessary, we set the minimum uncertainty to 0.001 mas for the angular diameters. 
For the $V^2$, we set a minimum relative uncertainty of 0.05 for the three instruments to avoid systematics. For the CP, we set the absolute minimum and maximum uncertainty to 0.5° and 30°, respectively. Setting a minimum overrides the data, while a maximum error ignores data with errors larger than the limit.
A background was also considered in H and K bands following the standard method when we analysed the data of the MIRC-X and MYSTIC instruments \citep{anugu_mirc-x_2020}. This background was taken into account to reduce residual bias between the two instruments from the calibration.\\

\begin{figure}[h!]
\centering
    \begin{subfigure}[t]{0.43\textwidth}
        \centering
        \includegraphics[width=\linewidth]{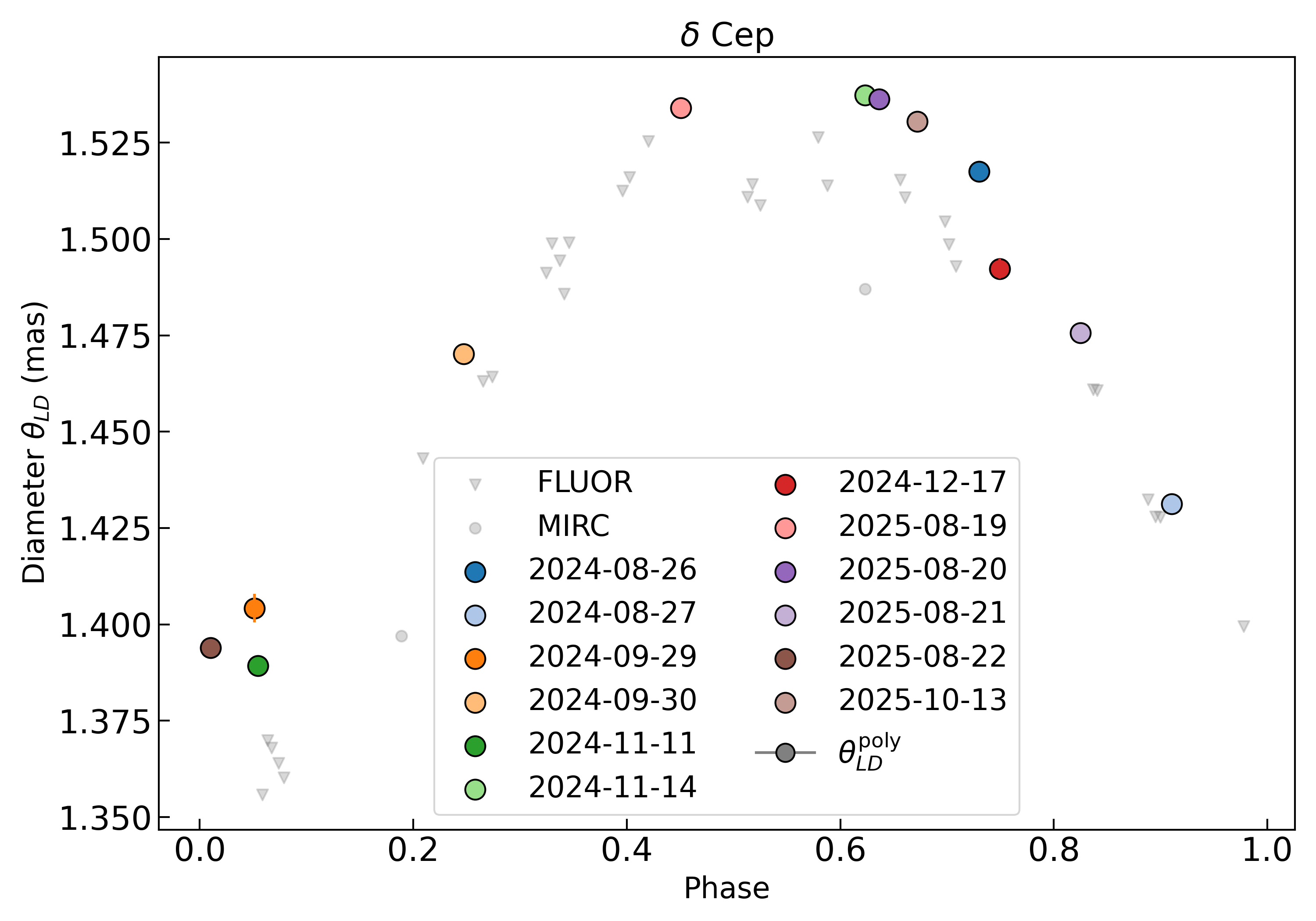}
    \end{subfigure}
    \begin{subfigure}[t]{0.43\textwidth}
        \centering
        \includegraphics[width=\linewidth]{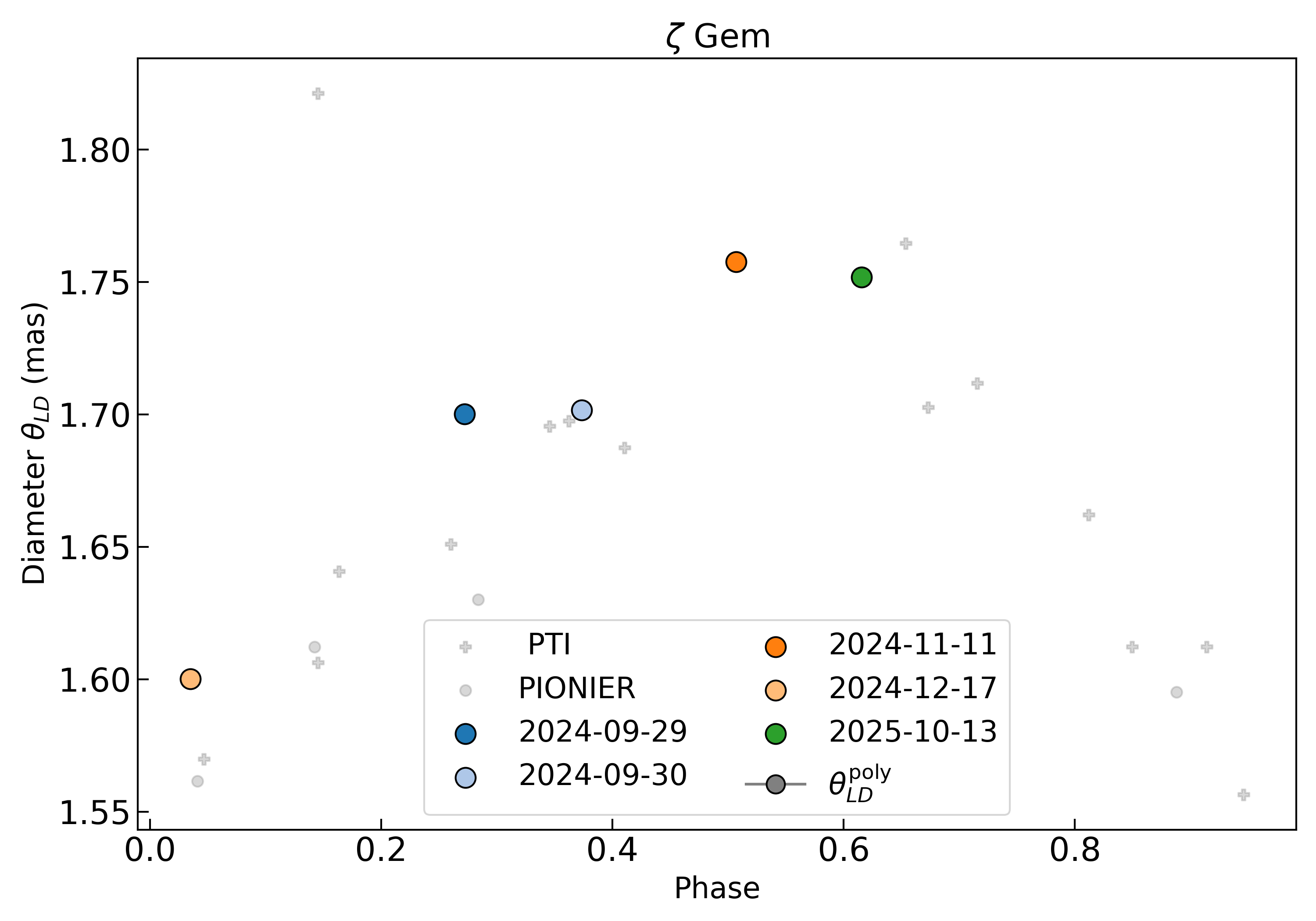}
    \end{subfigure}
   \begin{subfigure}[t]{0.43\textwidth}
        \centering
        \includegraphics[width=\linewidth]{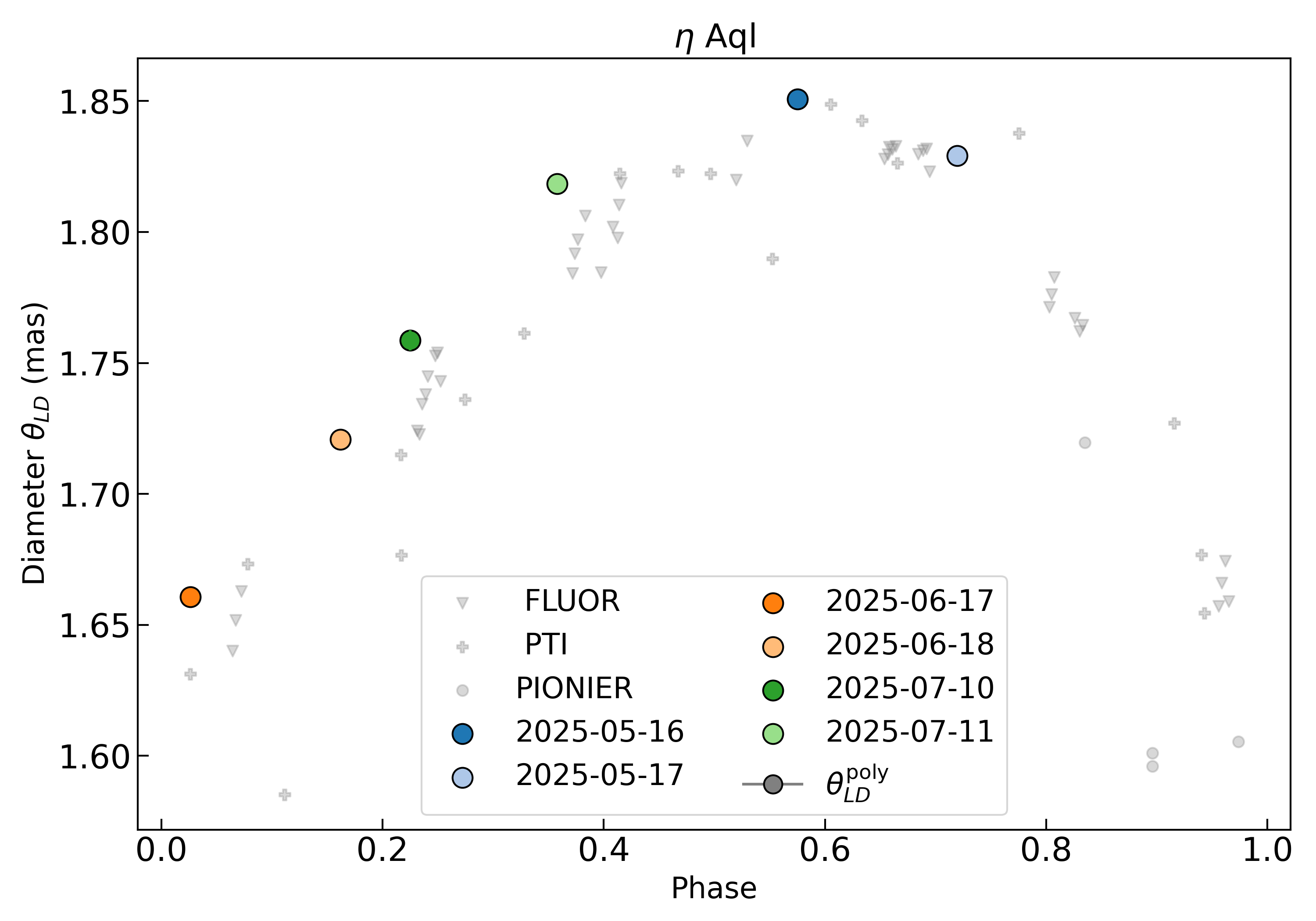}
    \end{subfigure}
    \caption{$\theta_{\mathrm{LD}}^{\mathrm{poly}}$ as a function of the pulsation phase for the three Cepheids in our sample. The uncertainties are smaller than the size of the markers (see the legend and Sect.\ref{sec:Measurement of the limb-darkened diameter of Cepheids} for more details).}
    \label{fig:diameter_curves_all}
\end{figure}

The pulsation phase associated with the observing dates was calculated using O-C diagrams from \cite{csornyei_study_2022} to correct for the period change induced by the evolution of the Cepheid in the HR diagram. The phases were calculated using $\phi_{i} = \frac{\mathrm{JD}_{i} - \mathrm{T}_{0} - \mathrm{OC}_{i}}{\mathrm{P}_{\mathrm{ref}}} \pmod 1$, where $\mathrm{T}_{0}$ is the reference epoch, and $\mathrm{P}_{\mathrm{ref}}$ is the pulsation period. The term $\mathrm{OC}_{i}$ represents the phase correction derived from a parabolic fit of the O-C diagram.
The $\theta_{\mathrm{LD}}^{\mathrm{poly}}$ angular diameter is plotted as a function of the pulsation phase in Fig.\ref{fig:diameter_curves_all} and is compared to previous angular diameters in the literature (in grey). The results are also indicated in tables ~\ref{tab:results_delta_cep} to \ref{tab:results_eta_aql}. After verification, the data for $\zeta$ Gem on the night of 29 September 2024 appear to be incorrectly calibrated, and were not taken into account in the rest of the paper (see figures \ref{fig:diameter_curves_all}, and \ref{fig:all_V2_per_nights} for this night).

\section{Surface brightness–colour relation}
\label{sec:Surface brightness–colour relation}
The flux density received from a star per unit of solid angle, also called surface brightness ($F_{\lambda}$), can be related to its LD diameter and its apparent magnitude corrected for extinction $(m_{\lambda,0})$. This was first highlighted by \cite{wesselink_surface_1969}, then by \cite{barnes_stellar_1976}, and later by \cite{fouque_improved_1997}, who derived this equation:  
\begin{equation}
    F_{\lambda} = 4.2196 - 0.1 m_{\lambda_{0}} - 0.5\log(\mathrm{\theta_{LD}}),
\end{equation}
where 4.2196 is a constant that depends on solar parameters \citep{mamajek_iau_2015}.
They also showed that surface brightness and stellar colour indices expressed in magnitude are linked. 

Using the $\theta_{\mathrm{LD}}^{\mathrm{poly}}$ calculated in Sect.~\ref{sec:Measurement of the limb-darkened diameter of Cepheids} and the same method and photometric datasets presented in \cite{bailleul_surface_2025}, we derived the following SBCR in the $(V-K)$ and ($G_{BP}-G_{RP}$) colour systems: 
\begin{align}
    &F_{V} = -0.1296_{\pm0.0011}(V-K)_{0} + 3.9480_{\pm0.0017} \\
    &F_{G_{BP}}= -0.3476_{\pm0.0128}(G_{BP}-G_{RP})_{0} + 4.0354_{\pm0.0109},
\end{align}
with an RMS of 0.0011 and 0.0040 mag. The mean colour (i.e. the pivot) is 1.4893 and 0.8490 for the SBCR ($V$, $V-K$) and the SBCR ($G_{BP}$, $G_{BP}-G_{RP}$), respectively. The colour ranges are [1.0790, 1.8459] for $(V-K)_{0}$ and [0.7289, 0.9887] for $(G_{BP}-G_{RP})_{0}$.
This new calibration of the SBCR ($V$, $V-K$) based on robust multi-chromatic angular diameters reduces the scatter of the relation by a factor of about 3 compared to \cite{bailleul_surface_2025}. \\

\begin{figure}
    \centering
    \includegraphics[width=1\linewidth]{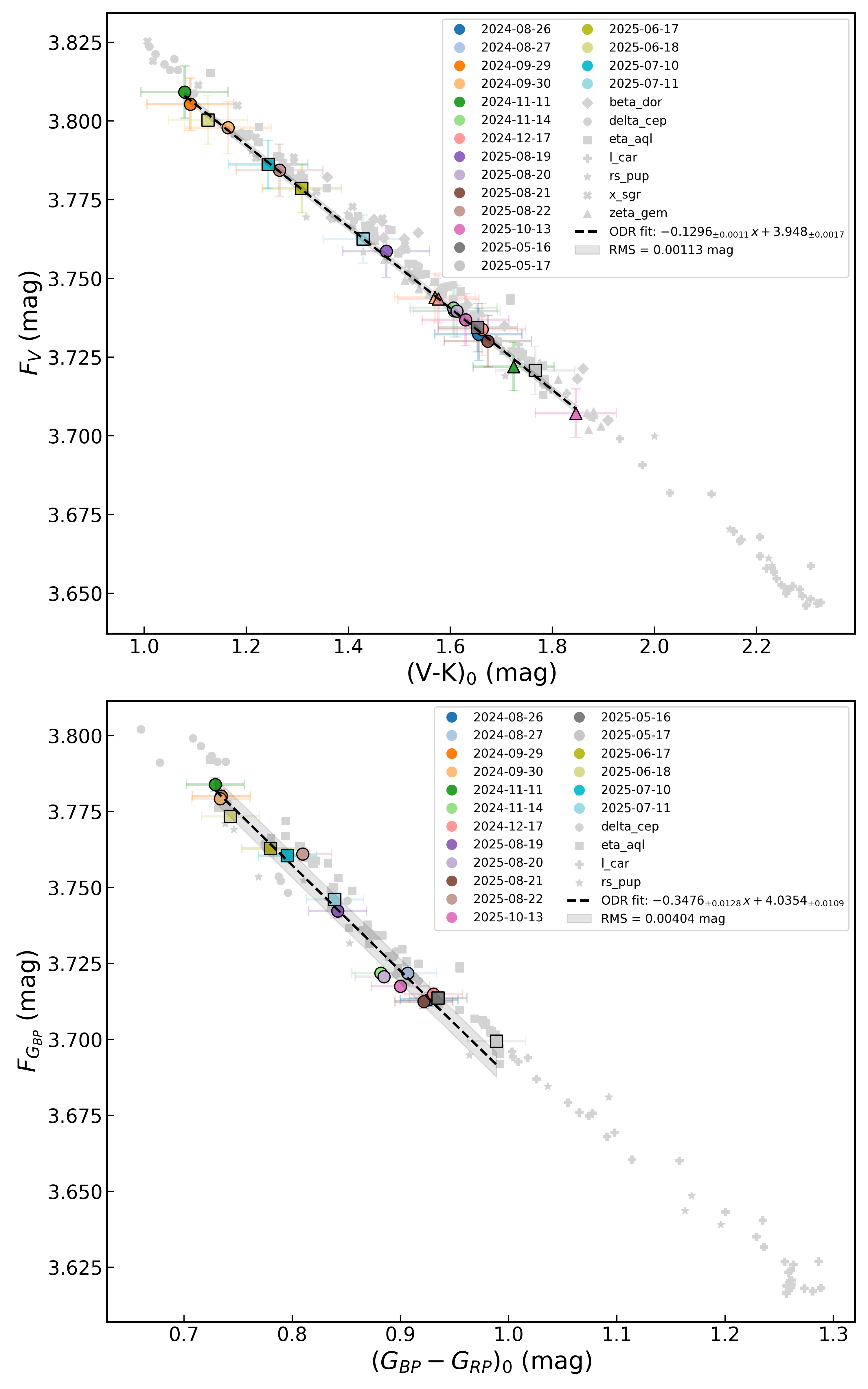}
    \caption{Surface brightness-colour relation for $\delta$~Cep, $\zeta$ Gem and $\eta$ Aql in the $(V-K)$ and ($G_{BP}-G_{RP}$) colour system using $\theta_{\mathrm{LD}}^{\mathrm{poly}}$ (coloured), compared to the results from \cite{bailleul_surface_2025} (grey).}
    \label{fig:sbcr}
\end{figure}

In Fig.~\ref{fig:sbcr} we over-plot the surface brightnesses values from \cite{bailleul_surface_2025} as a function of the $(V-K)$ and ($G_{BP}-G_{RP}$) colours (grey dots). The only difference is that an error was found in \cite{bailleul_surface_2025}, as the Fiber Linked Unit for Optical Recombination (FLUOR) angular diameters of $\delta$~Cep were converted from $\mathrm{\theta_{UD}}$ into $\mathrm{\theta_{LD}}$, whereas the diameters available in \cite{merand_projection_2005} (their Table 3) were already limb-darkened diameters. This difference does not change the results presented in \cite{bailleul_surface_2025}, although the RMS of the relations is slightly improved. This mistake is corrected in Fig.~\ref{fig:sbcr}. The SBCRs in this work based on $\theta_{\mathrm{LD}}^{\mathrm{poly}}$ are found to be shifted towards lower values compared to the previous calibration. This might be connected to the fact that the $\theta_{\mathrm{LD}}^{\mathrm{poly}}$ angular diameters found in this work are systematically larger than the values found in the literature (see also in Fig.~\ref{fig:diameter_curves_all}). An explanation could be that in order to use an homogeneous approach, \cite{bailleul_surface_2025} used the relation of \cite{hanbury_brown_effects_1974}, which is a mathematical approximation, to convert UD diameters (from the literature) into LD diameters. \cite{nardetto_calibrating_2020} indeed highlighted the  bias between the analytic and direct fit diameter (with linear law), using the tool \texttt{LITpro}. They found a difference of 0.4 to 0.8\% between the two estimates. Another aspect to consider is that in the case of $\delta$~Cep, $\mathrm{\theta_{LD}}$ from \cite{merand_projection_2005} were derived using LD coefficients taken from different tables, namely \citet{claret_non-linear_2000}, assuming a constant effective temperature and surface gravity for this star, and therefore, a constant LD coefficient, which might also affect the derived diameters. \\

An SBCR dedicated to Cepheids is particularly important in the context of the BW method. A recent study based on stellar atmosphere models showed for standard non-pulsating stars that SBCRs depend not only on temperature, but also on their luminosity class \citep{salsi_theoretical_2022}. In addition, \cite{bailleul_surface_2025} showed in their Figure 7 that they obtained a significative difference (within their common $V-K$ colour range) between their SBCR, dedicated to Cepheids, and the one derived for giants reported by \cite{salsi_progress_2021}. \cite{kiman_accurate_2024} established a recent calibration of the SBCR ($G_{BP}$, $G_{BP}-G_{RP}$), but for main-sequence stars alone, which makes a direct comparison with Cepheids not really relevant. To our knowledge, no SBCR is currently derived in the \textit{Gaia} bands for giant stars.

\section{Measuring the limb-darkening of classical cepheids}
\label{sec:Measuring the limb-darkening of classical cepheids}

In this section, we derive the limb-darkening coefficients of Cepheids in R, H, and K bands using a multi-chromatic approach. We used the $V^2$ and CP, in H, K, and R bands (when available) to derive the LD coefficients and the LD diameter. Following previous methods, we also considered a background for the MIRC-X and MYSTIC measurements. The fitted LD angular diameter is common between the different instruments, unlike the LD coefficients. We fitted six parameters: $\theta_{\mathrm{LD}}^{\mathrm{pow}}$, $\alpha_R$, $\alpha_H$, $\alpha_K$, $\mathrm{bkg_H}$, and $\mathrm{bkg_K}$, with $\alpha_R$ only when SPICA data were available.
Including the CP in the fit imposed a valuable additional constraint on the diameter because the closures jump from 0° to 180° when the visibilities reach the first zero. This requires good-quality CPs at this transition point, but when this is the case, they are better defined than the first zero in the visibility data, and the fit converges better. We used the bootstrapping method of \texttt{pmoired} to derive the parameters and associated uncertainties. The bootstrapping method is a statistical method for resampling the data with replacement. As explained in \texttt{pmoired} {tutorials\footnote{\url{https://github.com/amerand/PMOIRED_examples}}, the method is based on one spectral vector of one observable for one telescope configuration. Spectral vectors of $V^{2}$ or T3PHI (in our case) are different baselines (pairs), and are indivisible. For each iteration (each fit), spectral vectors can appear several times or not at all. This allowed us to take spectral correlation into account when we estimated uncertainties. The final results are the average standard deviation and correlation between all the fits performed (we chose 1000 iterations).\\

We chose to use a single power law to derive the diameter ($\theta_{\mathrm{LD}}^{\mathrm{pow}}$) and the LD coefficients ($\alpha_{\lambda}$) to reduce the number of parameters in the fitting procedure. As demonstrated by \cite{kervella_radii_2017}, the power law is the best compromise: it is a single-parameter law, but it describes visibility as well as the square-root law. The law is of the form
\begin{equation}
    I(\mu)=\mu^{\alpha_{\lambda}}.
\end{equation}

Firstly, the method works better when the star is resolved in the different bands. Secondly, as already mentioned, the method needs good-quality CP, with a well-defined jump. To illustrate our method, we present two examples. Firstly, we show in the appendix an example of the fit (30 September 2024, $\zeta$ Gem) with (figure ~\ref{fig:zeta_gem_20241111}) and without (figure ~\ref{fig:zeta_gem_20241111_wCP}) taking the CP into account for an observation with very high-quality CPs and for a resolved star in H and K band. Taking as reference the $\theta_{\mathrm{LD}}^{\mathrm{poly}}$ derived in Sect.~\ref{sec:Measurement of the limb-darkened diameter of Cepheids}, with fixed LD coefficients, we note that the diameter derived when the fit uses the CP corresponds better to this reference. This means that the diameter is better constrained, and the fit can estimate the LD coefficients better. Moreover, and importantly, the correlation from the bootstrap is reduced.
The second example in Fig.~\ref{fig:eta_aql_20250617} shows a night with poor-quality CP (17 June 2025, $\eta$~Aql). The fit is not able to reach a coefficient in the K-band. Finally, we compare in Fig.~\ref{fig:diam_comparison} the newly derived diameters with those obtained in Sect.~\ref{sec:Measurement of the limb-darkened diameter of Cepheids}. \\

In order to compare our results (LD coefficients) to expectations from models, we adopted two different approaches.
\begin{itemize}
    \item Firstly, we used the same SATLAS spherical models as described in \cite{neilson_spherically-symmetric_2013,neilson_spherically_2013} at the specific phase of observation in terms of effective temperature and surface gravity. According to the period–radius–mass relation given in \cite{bono_improving_2001}, the masses of the three stars in our sample range from 4.5 to 6 M$_{\odot}$. We then considered a 5 solar mass model for each star. Next, we fitted a power law directly on the intensity distribution profile of the SATLAS models we obtained. This provided an estimate of the LD coefficients in each band.
    \item Secondly, in the case of $\delta$~Cep alone, we considered hydrodynamical simulations. \cite{nardetto_self_2004} developed a hydrodynamical model of $\delta$~Cep and \cite{nardetto_probing_2006} calculated the associated intensity profile in the visible domain as a function of the pulsation phase. From this intensity profile variation, we extracted the power-law limb-darkening coefficient in the visible.
\end{itemize}  
\begin{figure*}[h!]
    \centering
    \includegraphics[width=1\linewidth]{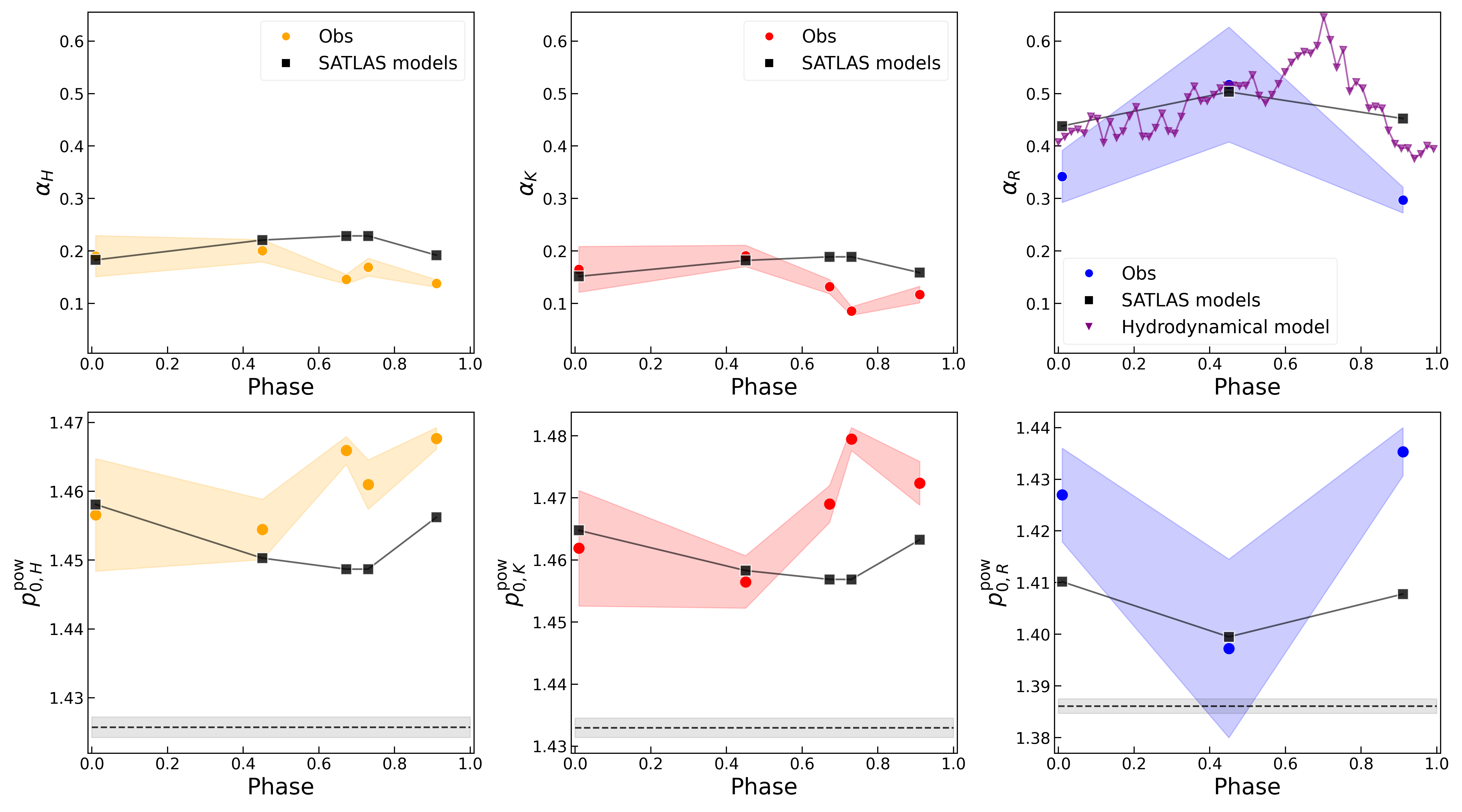}
    \caption{Bootstrapping results for the nights with good-quality CP for $\delta$~Cep. We indicate the LD coefficients at the top as a function of the pulsation phase derived from the CHARA observations and from the SATLAS models, respectively. The bottom parts show the corresponding geometrical $p$ factor, and the dashed line correspond to the $p_{0}$-period relation of \cite{neilson_cepheid_2012}. From left to right, we plot the H-, K-, and R-band results.}
    \label{fig:LDs_p0_results_deltaCep}
\end{figure*}

Fig.~\ref{fig:LDs_p0_results_deltaCep}, \ref{fig:LDs_p0_results_zetaGem}, and \ref{fig:LDs_p0_results_etaAql} show the results for $\delta$~Cep, $\zeta$~Gem, and $\eta$~Aql. The results are also indicated in Table \ref{tab:LD_results_powerLaw}.
In the top panel we show the fitted LD coefficients ($\alpha_H$, $\alpha_K$, $\alpha_R$) as a function of the pulsation phase (dots). The coefficients derived from SATLAS models are also plotted (squares). For $\delta$~Cep, we also show the result from the hydrodynamical model (triangles). 
The lower panel is the equivalent geometric projection factor (as derived in Sect.~\ref{subsec:The geometrical p-factor}). \\
\begin{table}[ht]
\centering
\caption{Fitted $\theta_{\mathrm{LD}}^{\mathrm{pow}}$ and $\alpha_{\lambda}$ coefficients for the three stars.}
\setlength{\tabcolsep}{2pt}
\renewcommand{\arraystretch}{1.25}
\begin{tabular}{lccccc}
\hline
Phase & $\theta_{LD}^{\text{pow}}$ (mas) & $\alpha_R$ & $\alpha_H$ & $\alpha_K$ & $\chi^2$ \\
\hline \hline
\multicolumn{6}{c}{$\delta$~Cep} \\
\hline
0.911 & $1.436_{\pm 0.002}$ & $0.297_{\pm 0.025}$ & $0.138_{\pm 0.007}$ & $0.117_{\pm 0.016}$ & $1.22$ \\
0.730 & $1.516_{\pm 0.003}$ &  & $0.169_{\pm 0.017}$ & $0.086_{\pm 0.008}$ & $1.14$ \\
0.672 & $1.531_{\pm 0.003}$ &  & $0.146_{\pm 0.009}$ & $0.132_{\pm 0.014}$ & $1.55$ \\
0.450 & $1.550_{\pm 0.005}$ & $0.517_{\pm 0.109}$ & $0.200_{\pm 0.021}$ & $0.191_{\pm 0.020}$ & $2.05$ \\
0.010 & $1.409_{\pm 0.008}$ & $0.342_{\pm 0.050}$ & $0.190_{\pm 0.039}$ & $0.165_{\pm 0.044}$ & $1.59$ \\
\hline
\multicolumn{6}{c}{$\zeta$ Gem} \\
\hline
0.373 & $1.704_{\pm 0.001}$ & & $0.201_{\pm 0.005}$ & $0.171_{\pm 0.004}$ & $1.76$ \\
0.616 & $1.747_{\pm 0.009}$ & & $0.150_{\pm 0.030}$ & $0.090_{\pm 0.036}$ & $1.31$ \\
0.507 & $1.759_{\pm 0.002}$ & & $0.207_{\pm 0.010}$ & $0.131_{\pm 0.011}$ & $3.29$ \\
\hline
\multicolumn{6}{c}{$\eta$ Aql} \\
\hline
0.162 & $1.720_{\pm 0.003}$ & & $0.120_{\pm 0.012}$ & $0.096_{\pm 0.015}$ & $2.12$ \\
0.575 & $1.854_{\pm 0.002}$ & & $0.143_{\pm 0.008}$ & $0.120_{\pm 0.009}$ & $1.93$ \\
0.719 & $1.825_{\pm 0.005}$ & & $0.126_{\pm 0.013}$ & $0.111_{\pm 0.019}$ & $0.92$ \\
0.358 & $1.822_{\pm 0.001}$ & & $0.140_{\pm 0.009}$ & $0.132_{\pm 0.002}$ & $2.60$ \\
0.225 & $1.759_{\pm 0.015}$ & & $0.133_{\pm 0.039}$ & $0.107_{\pm 0.057}$ & $10.86$ \\
\hline
\end{tabular}
\label{tab:LD_results_powerLaw}
\end{table}

We draw several conclusions.
Firstly, we derived limb-darkening coefficients of Cepheids in the R, H, and K bands. The results, at first order, agree well with the SATLAS models, although we generally obtained lower values than predicted by the model.
\begin{table}
    \centering
    \caption{Average ratio of the LD coefficients from our observations, compared to SATLAS (in italic). The reported uncertainties correspond to the standard deviation.}
    \begin{tabular}{cccc}
        Ratio & $\delta$ Cep & $\zeta$ Gem & $\eta$ Aql \\
        \hline
        $\alpha_{H}/\alpha_{K}$ & \makecell{1.29 $\pm$ 0.34 \\ \textit{1.21}} & \makecell{1.48 $\pm$ 0.21 \\ \textit{1.22}} & \makecell{1.18 $\pm$ 0.07 \\ \textit{1.21}}\\
        \hline
        $\alpha_{H}/\alpha_{R}$ & \makecell{0.47 $\pm$ 0.07\\ \textit{0.43}} & - & - \\
        \hline
    \end{tabular}
    \label{tab:ratio_LD_coeffs}
\end{table}
We found ratios of the $\alpha$ coefficients in H and K bands and H and R bands that agree well on average with the  models (see Table~\ref{tab:ratio_LD_coeffs}) for the three stars. 
Secondly, this method can only be used when the CP are taken into account. Without the CP, the fit does not converge at all to a solution for the LD coefficients in some cases. It also allowed us to reduce the correlation between the different fit parameters.
Thirdly, in Fig.~\ref{fig:LDs_p0_results_deltaCep}, we compare the observations (in R band) to the hydrodynamical models (in V band). The agreement is excellent, even though we miss some precise measurements (in particular for the phase between 0.1 and 0.4 and around 0.7) to firmly conclude a general consistency as a function of the pulsation phase. Fourthly, the derived LD coefficients in H and K are better constrained for $\eta$~Aql and $\zeta$ Gem than for $\delta$~Cep, probably because the stars are more highly resolved. For $\zeta$~Gem, the results agree better with the model, but we lack some measurements to cover the cycle properly. \\

More complex features might affect the visibilities and/or the closure phase measurements, such as circumstellar environment (CSE), binarity, spots, or any structure that might affect the intensity distribution of the star. We do not observe (or do not have the sufficient precision to detect it) a decrease in $V^2$ at low spatial frequencies that might indicate the presence of a CSE around the stars. We detected no modulation in the CP either that would indicate a companion (see e.g. \cite{nardetto_orbital_2024} for $\delta$~Cep). However, it cannot be excluded that spots on the stellar surface affect the position of the first zero and the visibility of the second lobe, which might slightly bias the measurements of the diameter and the limb-darkening.

\section{The projection factor}
\label{sec:New calculation of the projection factor of delta Cep}

\subsection{The Baade-Wesselink method}
As explained in Sect.~\ref{sec:Introduction}, a projection factor ($p$ or $p$ factor) is needed in the BW method to convert the observed radial velocity $V_{\text{rad}}$ into the true pulsation velocity $V_{\text{puls}}$, with $p=V_{\text{puls}}/V_{\text{rad}}$.
The angular diameter of the Cepheid is then provided by 
\begin{equation}
    \theta_{LD}(\phi_{i}) = \overline{\theta} + 9.3009 \frac{p}{d} \int_{0}^{P} V_{\text{rad}}(\phi_{i})d\phi_{i} \text{ [mas]},
\end{equation}
where the mean limb-darkened angular diameter ($\overline{\theta}$, in mas) and the $p$ factor (averaged over the pulsation) are fitted parameters, and $d$ is the distance to the star in parsecs.

\subsection{Applying the inverse Baade–Wesselink method to $\delta$ Cep}
\label{subsection:Applying the Baade–Wesselink method to delta Cep}

Using $\theta_{\mathrm{LD}}^{\mathrm{poly}}$ derived in Sect.~\ref{sec:Measurement of the limb-darkened diameter of Cepheids}, and radial velocities (RV), we applied the inverse BW method to $\delta$~Cep. We adjusted the average angular diameter and the $p$ factor, but the distance was fixed. We also took a phase shift in the fit between angular diameters and radial velocities into account in order to obtain the best possible fit between the two curves \citep{merand_projection_2005}. We decided to compare two distance estimates: d$_{1} = 272 \pm3\text{ (stat.)} \pm 5 \text{ (syst.)}$ pc from \cite{majaess_new_2012}, and d$_{2}=279.5 \pm 11.6$ pc from the very precise parallax of the $\delta$~Cep companion \citep{kervella_multiplicity_2019, trahin_inspecting_2021}.
We applied the method for two different RV sets: \cite{anderson_revealing_2015} (A15) and \cite{nardetto_harps-n_2017} (N17). The radial velocities from A15 were obtained using the High-Efficiency and high-Resolution Mercator Echelle Spectrograph (HERMES), and those from N17 were derived from observations by the High Accuracy Radial velocity Planet Searcher for the Northern Hemisphere (HARPS-N). They were observed in the visible domain. Our final estimates are indicated in Table~\ref{tab:p-factor}, and we used a bootstrapping uncertainty estimation. We found a mean angular diameter very similar for both, $\overline{\theta}$(A15)=$1.482 \pm 0.002$ mas and $\overline{\theta}$(N17)=$1.482 \pm 0.003$ mas (the choice of the distance does not affect this value). Figure~\ref{fig:BW_method_vrad_diam} shows the result of the fitting procedure for N17 RV and the d$_{2}$ distance. \\

\begin{figure}[h!]
    \centering
    \includegraphics[width=1\linewidth]{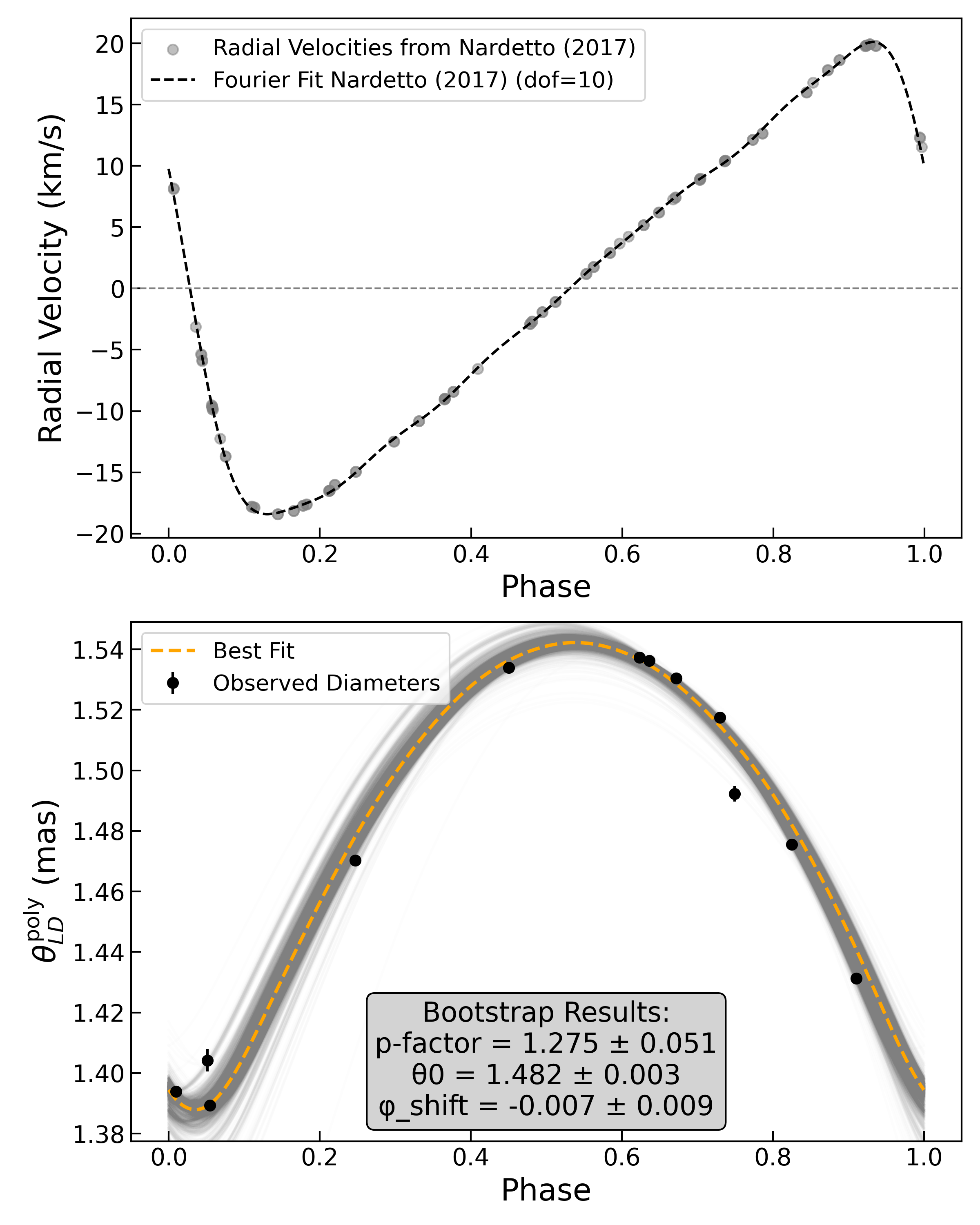}
    \caption{$p$ factor and mean angular diameter determination of $\delta$~Cep using the inverse BW method, the RV from \cite{nardetto_harps-n_2017}, and the distance from \cite{kervella_multiplicity_2019}. The dotted line corresponds to the fit, and the solid grey lines show all the bootstrapping iterations.}
    \label{fig:BW_method_vrad_diam}
\end{figure}

The projection factor of $\delta$~Cep was first derived observationally by \cite{merand_projection_2005} using an inverse BW method based on FLUOR/CHARA interferometric measurements and radial velocities from \cite{bersier_fundamental_1994} and \cite{barnes_radial_2005}. They found a $p$ factor of 1.270 $\pm$ 0.060. This result was at that time consistent with the previous theoretical prediction of \cite{nardetto_self_2004}. Later, \cite{nardetto_harps-n_2017} revisited the determination of the $p$ factor of $\delta$~Cep using different individual spectral lines and also the so-called cc-g radial velocity, corresponding to a Gaussian fit of the cross-correlated line profile. With the interferometric data from \cite{merand_projection_2005} and the previously mentioned HARPS-N spectra, they obtained $p_{\mathrm{cc-g}}=1.239 \pm 0.034\text{ (stat.)} \pm0.023\text{ (syst.)}$. We summarise the $p_{\mathrm{cc-g}}$-factors of $\delta$~Cep in Table~\ref{tab:p-factor}. We also include the estimate from \cite{merand_cepheid_2015} based on the SPIPS approach and the value found by \cite{groenewegen_projection_2007}, who also used an inverse BW approach. Our determination is consistent with previous results.

\begin{table}
    \centering
    \caption{$p$ factor determinations for $\delta$~Cep from a Gaussian fit of the cross-correlated line profile (cc-g).}
    \begin{tabular}{l|l}
        $p$ factor & Reference\\
        \hline \hline 
         1.270 $\pm$ 0.060 & \cite{merand_projection_2005}\\
         1.245 $\pm$ 0.050 & \cite{groenewegen_projection_2007} \\
         1.290 $\pm$ 0.050 & \cite{merand_cepheid_2015}\\
         1.239 $\pm$ 0.034 & \cite{nardetto_harps-n_2017}\\
         1.251  $\pm$ 0.047 & This work, A15 + d$_{1}$ \\
         1.285  $\pm$ 0.046 & This work, A15 + d$_{2}$ \\
         1.241 $\pm$ 0.047 & This work, N17 + d$_{1}$ \\
         1.275 $\pm$ 0.051 & This work, N17 + d$_{2}$ \\
        \hline
    \end{tabular}
    \label{tab:p-factor}
\end{table}

\subsection{The geometrical $p$ factor}
\label{subsec:The geometrical p-factor}

As introduced in Sect.~\ref{sec:Introduction}, the $p$ factor has two main components: $p=p_{0}\mathrm{f_{dyn}}$. The geometric component $p_{0}$ corresponds to an integration of the pulsation velocity projected along the line of sight and weighted by the intensity distribution of the stellar surface, that is, by the limb-darkening. To simplify the calculation, when we assume $V_{\text{puls}}=1$, then the geometric projection factor is given by

\begin{equation}
    p_{0}=\frac{1}{V_{\text{rad}}} = \left( \frac{\int^{1}_{0}I(\mu)\mu^{2}d\mu}{\int^{1}_{0}I(\mu)\mu d\mu} \right)^{-1}
    \label{eq:vrad_pfactor}.
\end{equation}

When the intensity profile is defined by a linear law, it leads to $p_{0}=\frac{1}{V_{\text{rad}}}=\frac{6-2u_{\lambda}}{4-u_{\lambda}}$ \citep{getting_characteristic_1934}. No equivalent analytic formula in the literature is associated with a power-law description of the limb-darkening. By integrating Eq.~\ref{eq:vrad_pfactor} with the power law, we derived
\begin{align}
    p_{0,\lambda}^{\text{pow}} & =\frac{\alpha_{\lambda} + 3}{\alpha_{\lambda} + 2} \label{eq:p0_alpha} \\
    \sigma(p_{0,\lambda}^{\text{pow}}) & =\frac{\sigma(\alpha_{\lambda})}{(\alpha_{\lambda}+2)^{2}}
    \label{p0_uncer},
\end{align}
with $\alpha_{\lambda}$, and $\sigma(\alpha_{\lambda})$ the limb-darkening coefficient in the $\lambda$-band, associated with the power law, and its associated uncertainty. From the LD coefficients calculated in Sect.~\ref{sec:Measuring the limb-darkening of classical cepheids} (observations) and SATLAS models, we calculated the corresponding $p_0$ with Eq.~\ref{eq:p0_alpha}.
In the bottom part of Fig.~\ref{fig:LDs_p0_results_deltaCep}, \ref{fig:LDs_p0_results_zetaGem}, and \ref{fig:LDs_p0_results_etaAql} we indicate the corresponding $p_0$ with the same labels as for the LD coefficients. We also compared our results to a time-independent $p_{0}$ obtained from the $p_{0}$-period relation of \cite{neilson_cepheid_2012} (dashed black line). This relation was derived by direct integration of the SATLAS intensity profile and is therefore independent of LD laws. The different values are generally consistent, but all the values of $p_{0}$ using the power-law description appear to be slightly higher than the value calculated directly by integration of the intensity profiles \citep{neilson_cepheid_2012}. The power law was used in order to analyse our interferometric data, which probably led to an intrinsic approximation that can only be overcome by reconstructing the image of the object, which is beyond the scope of this study.

\section{Conclusion}
\label{sec:Conclusion}
We simultaneously monitored three classical Cepheids ($\delta$~Cep, $\eta$ Aql, and $\zeta$ Gem) with the MYSTIC, MIRC-X, and SPICA instruments on the CHARA Array interferometer. They operate in the K, H, and R bands, respectively. We summarise our results below.

\begin{itemize}
\item We derived limb-darkened angular diameters based on multi-chromatic observations that are particularly robust compared to previous measurements in the literature.
\item We calibrated the SBCR ($V$, $V-K$) with an RMS of 0.0011 mag, which is three times better than the RMS obtained for the same relation by \cite{bailleul_surface_2025}. We also improved the SBCR in the \textit{Gaia} bands. These SBCRs are particularly precise and robust due to their homogeneity compare to previous calibrations \citep{kervella_cepheid_2004,bailleul_surface_2025} and because a multi-chromatic approach was used to derive the angular diameters. However, these SBCRs still need to be supplemented with further observations to cover a wider range of Cepheids (and thus, of pulsation periods).
\item Importantly, the SBCR ($V$, $V-K$) we derived is shifted to lower magnitude values than the SBCR presented by \cite{bailleul_surface_2025} because previous measurements were based on the approximate law of \cite{hanbury_brown_effects_1974} to convert the UD into LD angular diameters. 
\item For the first time, we measured limb-darkening coefficients of Cepheids in R, H, and K bands applying a multi-chromatic method, which required the inclusion of closure phases. The inclusion of the closure phase jump from 0 to 180 degrees at the first zero of visibilities added some constraints on the LD angular diameter, which then reduced the correlation with the fitted LD coefficients in R, H, and K bands, respectively, even though the second lobe is not robustly measured by SPICA (or MYSTIC). 
\item In R band, the measured limb-darkening coefficients as a function of the pulsation phase (three measurements) are consistent with the quasi static SATLAS atmosphere models of the Cepheids and are also consistent with the hydrodynamical model. In H and K bands, the results are also consistent with the SATLAS atmosphere model.
\item We derived the geometric projection factors associated with limb-darkening coefficients. In the R band, which is particularly important for the BW method, we found $p_0[R]=1.420 \pm 0.016$ (on average) for $\delta$~Cep, which is again consistent with models (SATLAS and hydrodynamic).
\item Applying the inverse BW approach, we found a projection factor of $1.275 \pm 0.051$ (N17 + d$_{2}$) for $\delta$~Cep, which is consistent with previous estimates in the literature. For the first time, we directly estimated the dynamical part of the projection factor from observations. Averaged over the pulsation, it is $\mathrm{f_{dyn}}=0.898 \pm 0.037$.
\item These preliminary results indicate that the limb-darkening coefficients of Cepheids in R, H, and K bands are rather consistent at first order with expectations from stellar models. However, these results are demanding in terms of data: they require simultaneous R, H, K observations, high S/N at large spatial frequencies, good closure phases, and monitoring to cover the entire pulsating cycle. It is therefore essential to continue the CHARA survey of Cepheids in the coming years.
\end{itemize}

\begin{acknowledgements}
The authors acknowledge the support of the French Agence Nationale de la Recherche (ANR), under grant ANR-23-CE31-0009-01 (Unlock-pfactor) and the financial support from ``Programme National de Physique Stellaire'' (PNPS) of CNRS/INSU, France. This research has made use of the SIMBAD and VIZIER (available at \href{http://cdsweb.u-strasbg.fr/}{http://cdsweb.u-strasbg.fr/}) databases at CDS, Strasbourg (France), and of the electronic bibliography maintained by the NASA/ADS system. This work has made use of data from the European Space Agency (ESA) mission {\it Gaia} (\url{https://www.cosmos.esa.int/gaia}), processed by the {\it Gaia} Data Processing and Analysis Consortium (DPAC, \url{https://www.cosmos.esa.int/web/gaia/dpac/consortium}). Funding for the DPAC has been provided by national institutions, in particular the institutions participating in the {\it Gaia} Multilateral Agreement. 
SPICA has been funded by CNRS, Observatoire de la Côte d’Azur, Université Côte d’Azur, Région Sud, and the University of Aarhus.  This project has been partly funded from the European Research Council (ERC)503 under the European Union’s Horizon 2020 research and innovation programme (Grant agreement No. 101019653).
This work is based upon observations obtained with the Georgia State University Center for High Angular Resolution Astronomy Array at Mount Wilson Observatory. The CHARA Array is supported by the National Science Foundation under Grant No. AST-2034336 and AST-2407956. Institutional support has been provided from the GSU College of Arts and Sciences, Office of the Provost, and Office of the Vice President for Research and Economic Development. Time at the CHARA Array was granted through the NOIRLab community access program (NOIRLab 2024B-600652, 2025A-417676, 2025B-177453, 2026B-561504: ; PI: M.Bailleul, N.Nardetto). SK acknowledges funding for MIRC-X received funding from the European Research Council (ERC) under the European Union's Horizon 2020 research and innovation programme (Starting Grant No. 639889 and Consolidated Grant No. 101003096). JDM acknowledges funding for the development of MIRC-X (NASA-XRP NNX16AD43G, NSF-AST 2009489) and MYSTIC (NSF-ATI 1506540, NSF-AST 1909165). This research has made use of the Jean-Marie Mariotti Center at https://jmmc.fr/, Optical Interferometry Database, and texttt{SearchCal} service, which involves the JSDC and JMDC catalogues.
AG acknowledges the support of the Agencia Nacional de Investigación Científica y Desarrollo (ANID) through the FONDECYT Regular grant 1241073.
\end{acknowledgements}


%
\bibliographystyle{aa} 
\bibliography{P3_biblio_2} 

\begin{appendix}

\section{Observing log}
\begin{table}[h]
    \centering
    \caption{Observing log listing the dates of observations, the configuration of the array, and the calibrators used (see also Table~\ref{tab:calibrators}).}
    \label{tab:obs_log}
    \begin{tabular}{lll}
        \hline
        Date & Configuration & CALs\\
        \hline
        \multicolumn{3}{c}{$\delta$ Cep} \\
        \hline
        2024-08-26 & S1-S2-E1-E2-W1-W2 & C1 \\
        2024-08-27 & S1-S2-E1-E2-W1-W2 & C1,2 \\
        2024-09-29 & S1-S2-E1-E2-W2 & C1,2\\
        2024-09-30 & S1-S2-E1-E2 & C1,2,4 \\
        2024-11-11 & S2-E1-E2-W1-W2 & C1,5\\
        2024-11-14 & S2-E1-E2-W1-W2 & C1 \\
        2024-12-17 & S2-E1-E2-W1-W2 & C1 \\
        2025-08-19 & S2-E1-E2-W1-W2 & C1,2 \\
        2025-08-20 & S2-E1-E2-W1-W2 & C1,2 \\
        2025-08-21 & S1-S2-E1-E2-W1-W2 & C1,2 \\
        2025-08-22 & S1-S2-E1-E2-W1-W2 & C1,2 \\
        2025-10-13 & S2-E1-E2-W1-W2 & C1,2,5\\
        \hline
        \multicolumn{3}{c}{$\zeta$ Gem} \\
        \hline
        2024-09-29 & S1-S2-E1-E2-W2 & C3\\
        2024-09-30 & S1-S2-E1-E2-W2 & C1,4\\
        2024-11-11 & S1-S2-E1-E2-W2 & C4\\
        2024-12-17 & S1-S2-E1-E2-W2 & C4,6\\
        2025-10-13 & S2-E1-E2-W1-W2 & C3,4\\
        \hline
        \multicolumn{3}{c}{$\eta$ Aql} \\
        \hline
        2025-05-15 & S1-S2-E1-E2-W1-W2 & C7 \\
        2025-05-16 & S1-S2-E1-E2-W1-W2 & C7 \\
        2025-05-17 & S1-S2-E1-E2-W1-W2 & C7 \\
        2025-06-17 & S1-S2-E1-E2-W1-W2 & C7 \\
        2025-06-18 & S1-S2-E1-E2-W1-W2 & C7 \\
        2025-07-10 & S1-S2-E1-E2-W1-W2 & C7 \\
        2025-07-11 & S1-S2-E1-W1-W2 & C7,8 \\
        \hline
    \end{tabular}
    \tablefoot{Usually the same configurations and calibrators are used for the three instruments, but as the data quality can differ from one instrument to the others, some baselines may not appear in the visibility curves presented in the paper (see Sect.~\ref{sec:Measurement of the limb-darkened diameter of Cepheids}). }
\end{table}

\begin{table}[h]
    \centering
    \caption{Calibrators with their uniform disk angular diameter in H-band (in milliarcsecond) and spectral type.}
    \label{tab:calibrators}
    \setlength{\tabcolsep}{2.5pt}
    \begin{tabular}{lcccc}
        \hline 
        HD & Number & $\mathrm{\theta_{UD}}$ (H band) & Sp.type & H mag \\
        \hline
        HD 3360 & C1 & 0.295 $\pm$ 0.031 & B2IV & 4.25 \\
        HD 213558 & C2 & 0.543 $\pm$ 0.051 & A1V & 3.87 \\
        HD 38899 & C3 & 0.303 $\pm$ 0.030 & B9IV & 4.98 \\
        HD 46300 & C4 & 0.417 $\pm$ 0.035 & A0Ib & 4.41 \\
        HD 214454 & C5 & 0.682 $\pm$ 0.050 & A9V & 3.98 \\
        HD 67228 & C6 & 0.718 $\pm$ 0.067 & G2IV & 3.91 \\
        HD 195810 & C7 & 0.322 $\pm$ 0.030 & B6III & 4.55 \\
        HD 145570 & C8 & 0.355 $\pm$ 0.026 & A1V & 4.89 \\
        \hline
    \end{tabular}
    \tablefoot{The diameters and spectral types are taken from the JSDC catalogue \citep{bourges_jmmc_2014}, except HD 46300, for which it is extracted from \texttt{SearchCal} \citep{chelli_pseudomagnitudes_2016}. The angular diameter in R and K bands are not listed here but are easily found in JSDC catalogue\footnote{\url{https://vizier.cds.unistra.fr/viz-bin/VizieR-3?-source=II/346/jsdc_v2}}.}
\end{table}

\onecolumn
\section{Bootstrapping results nights per nights for each star}

\begin{center}
    \scriptsize
    
    \includegraphics[width=0.45\textwidth]{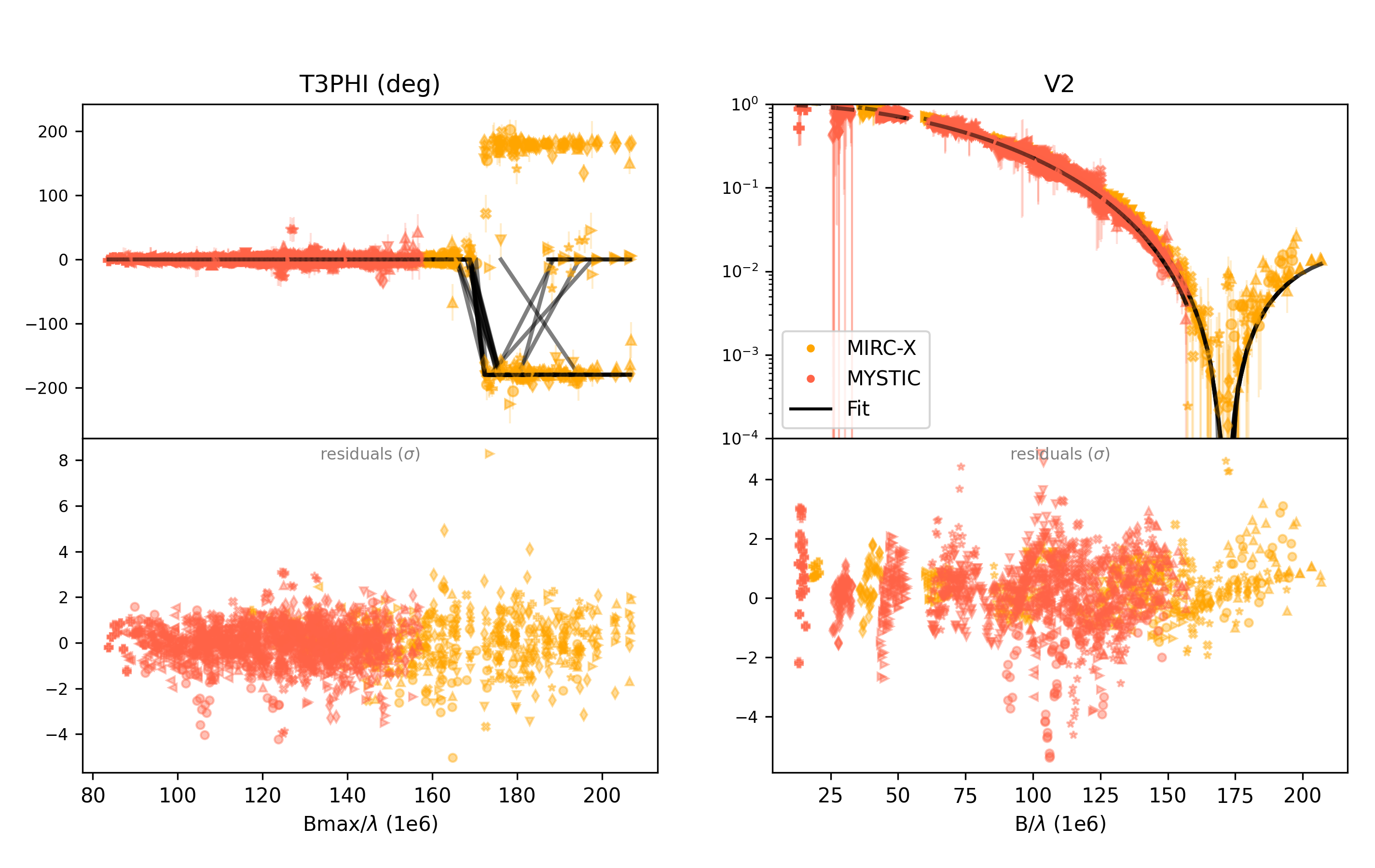}
    \hfill
    \includegraphics[width=0.45\textwidth]{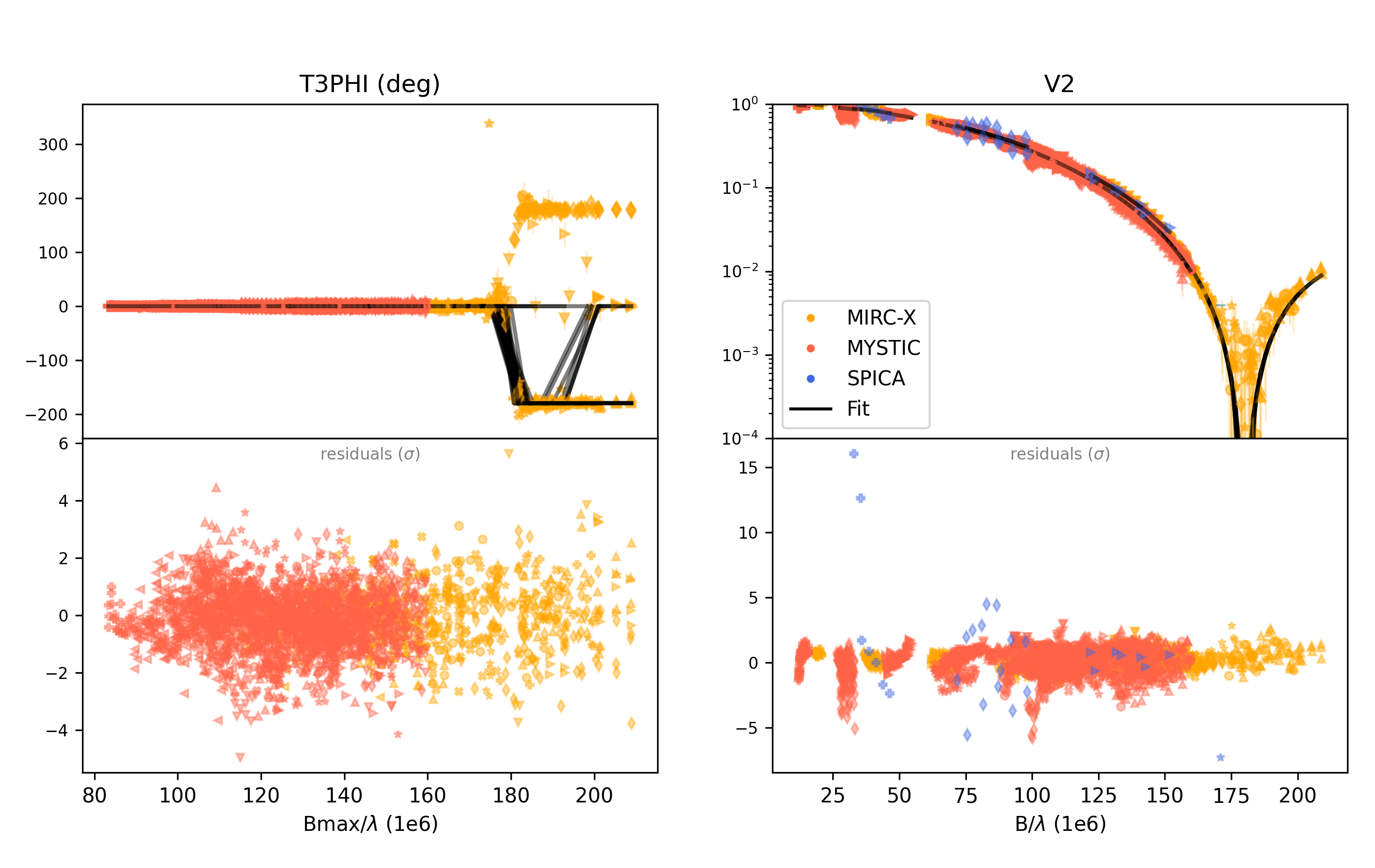}
    
    \begin{minipage}[t]{0.45\textwidth} \centering \textit{(a)} $^{*}\delta$ Cep: 2024-08-26 \end{minipage}
    \hfill
    \begin{minipage}[t]{0.45\textwidth} \centering \textit{(b)} $^{*}\delta$ Cep: 2024-08-27 \end{minipage}

    \includegraphics[width=0.45\textwidth]{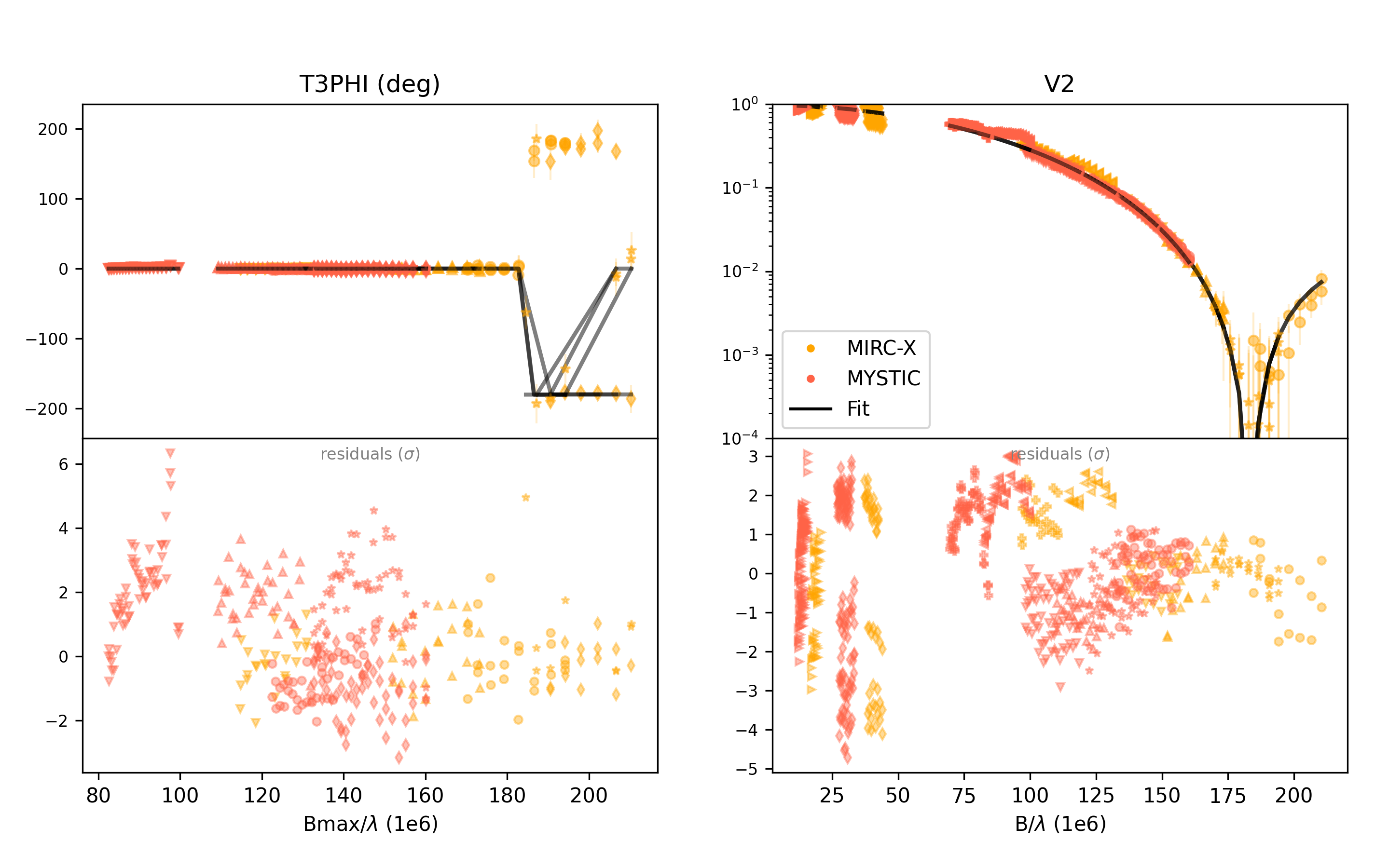}
    \hfill
    \includegraphics[width=0.45\textwidth]{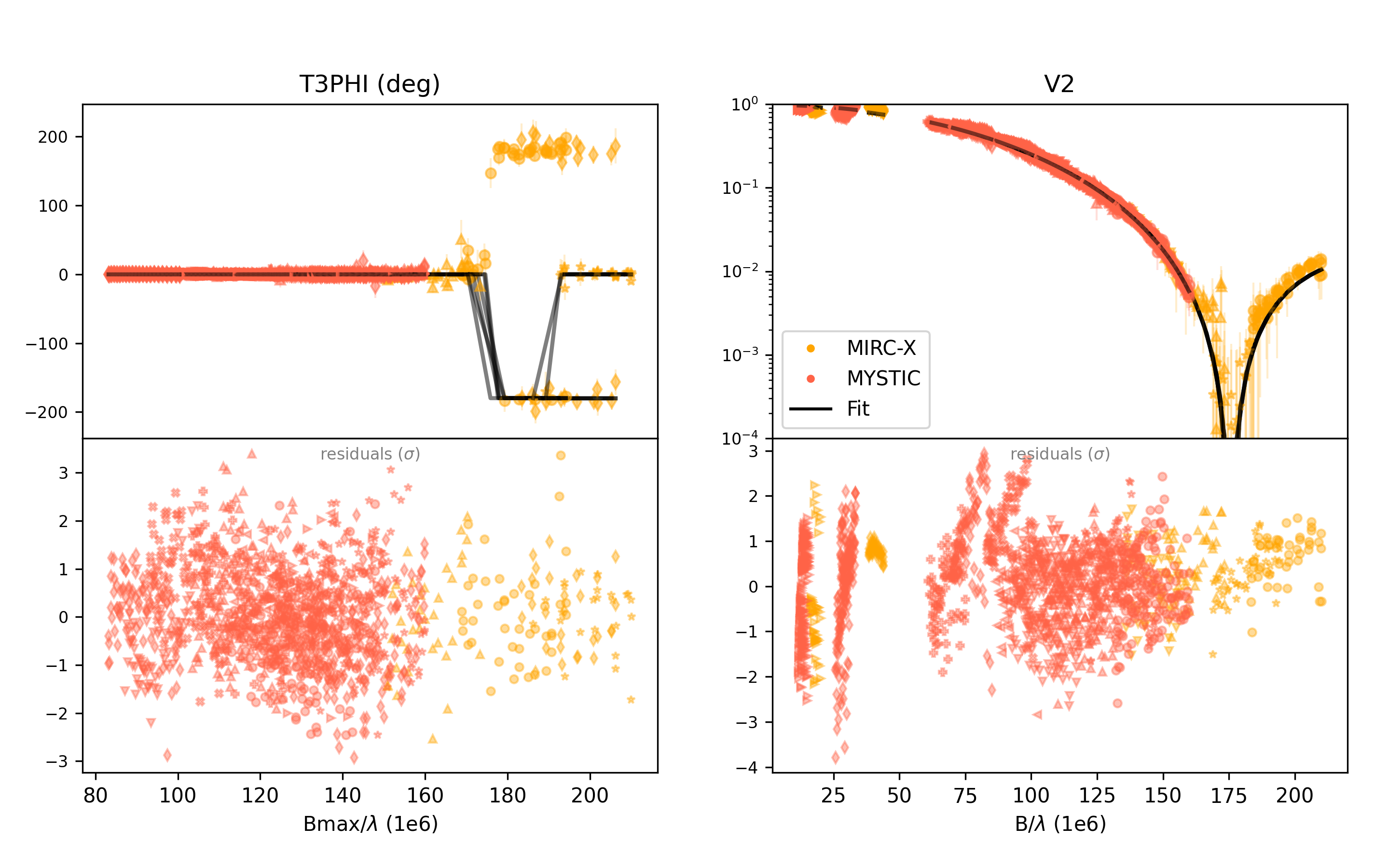}
    
    \begin{minipage}[t]{0.45\textwidth} \centering \textit{(c)} $\delta$ Cep: 2024-09-29 \end{minipage}
    \hfill
    \begin{minipage}[t]{0.45\textwidth} \centering \textit{(d)} $\delta$ Cep: 2024-09-30 \end{minipage}

    \includegraphics[width=0.45\textwidth]{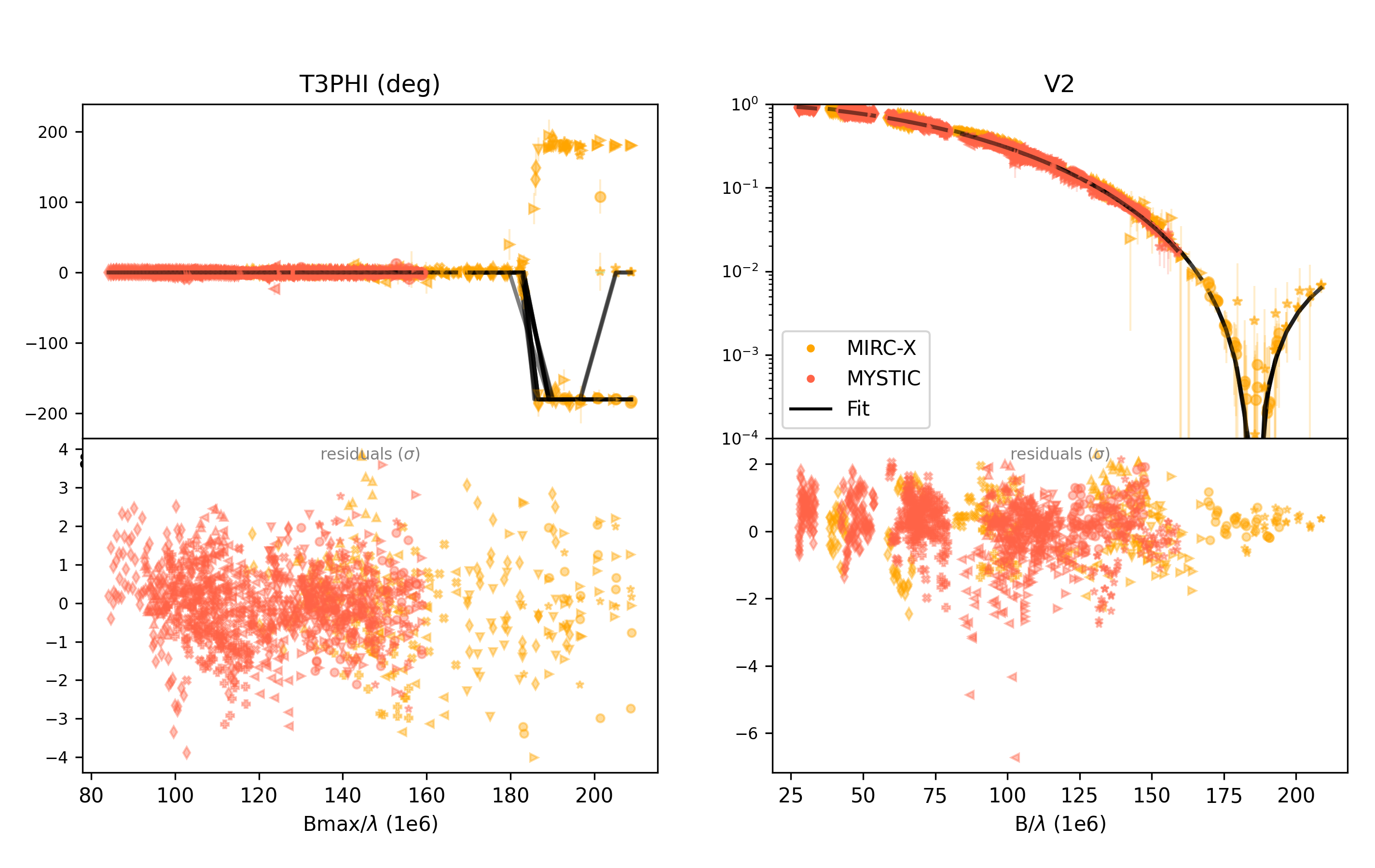}
    \hfill
    \includegraphics[width=0.45\textwidth]{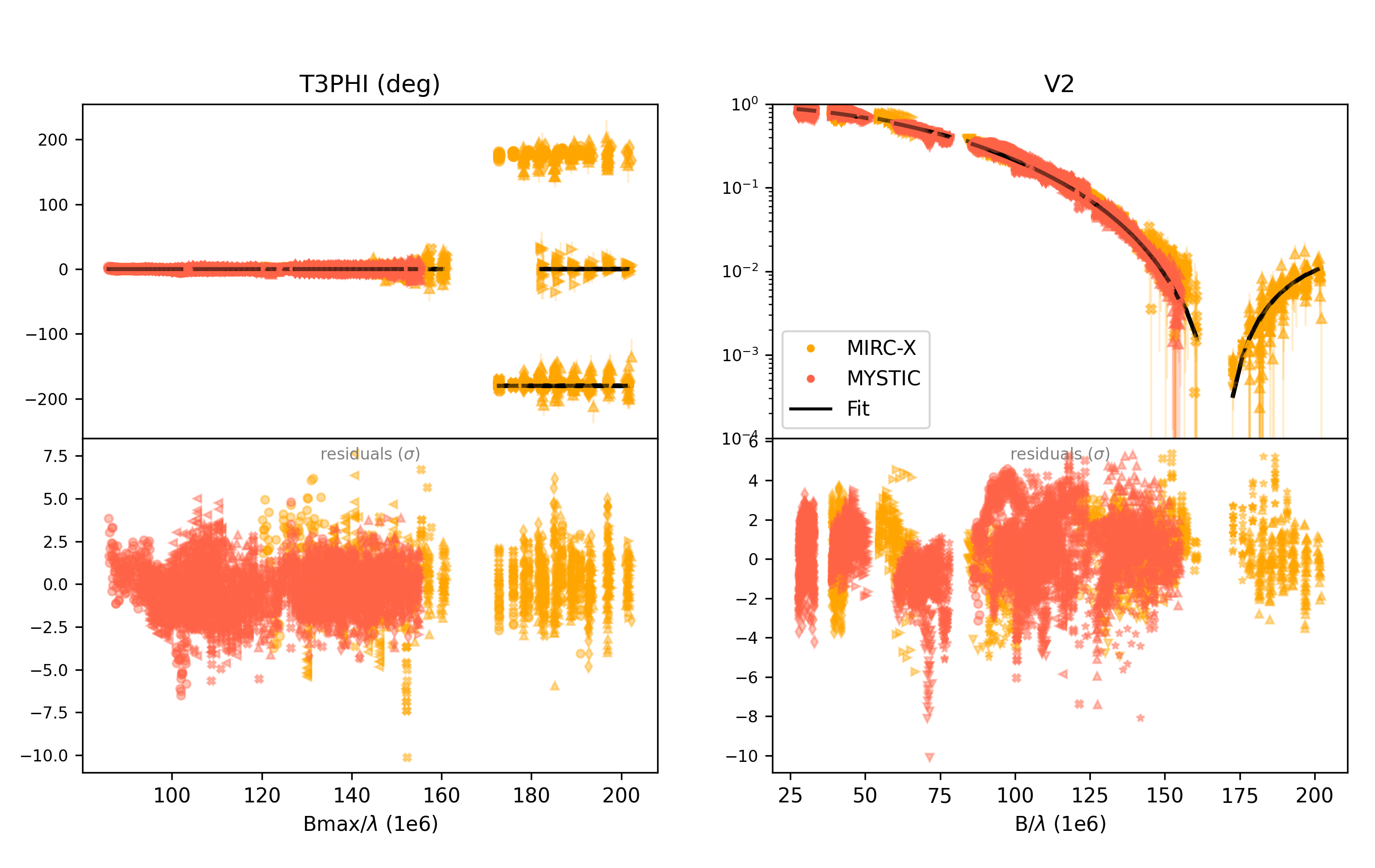}
    
    \begin{minipage}[t]{0.45\textwidth} \centering \textit{(e)} $\delta$ Cep: 2024-11-11 \end{minipage}
    \hfill
    \begin{minipage}[t]{0.45\textwidth} \centering \textit{(f)} $\delta$ Cep: 2024-11-14 \end{minipage}

    \includegraphics[width=0.45\textwidth]{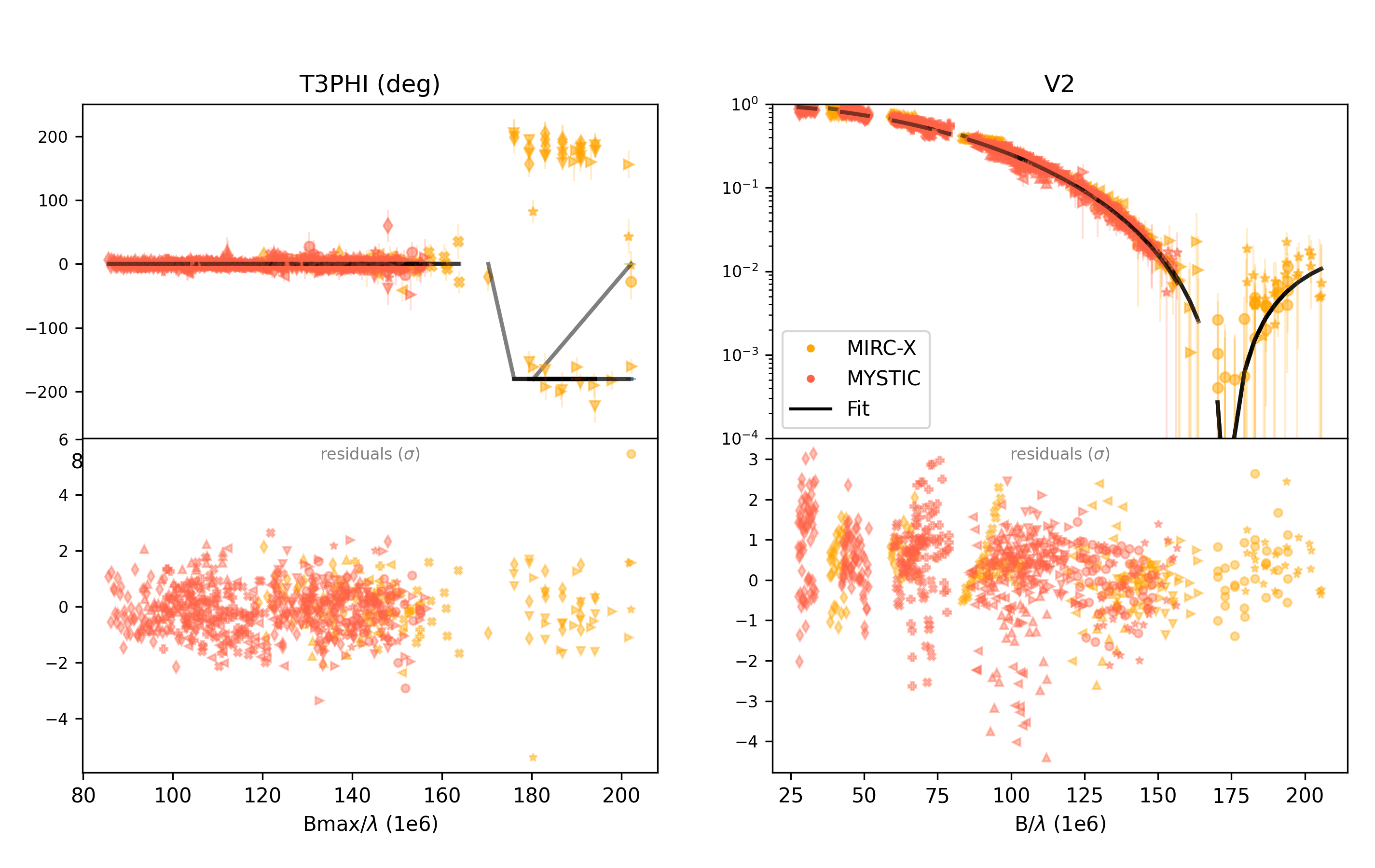}
    \hfill
    \includegraphics[width=0.45\textwidth]{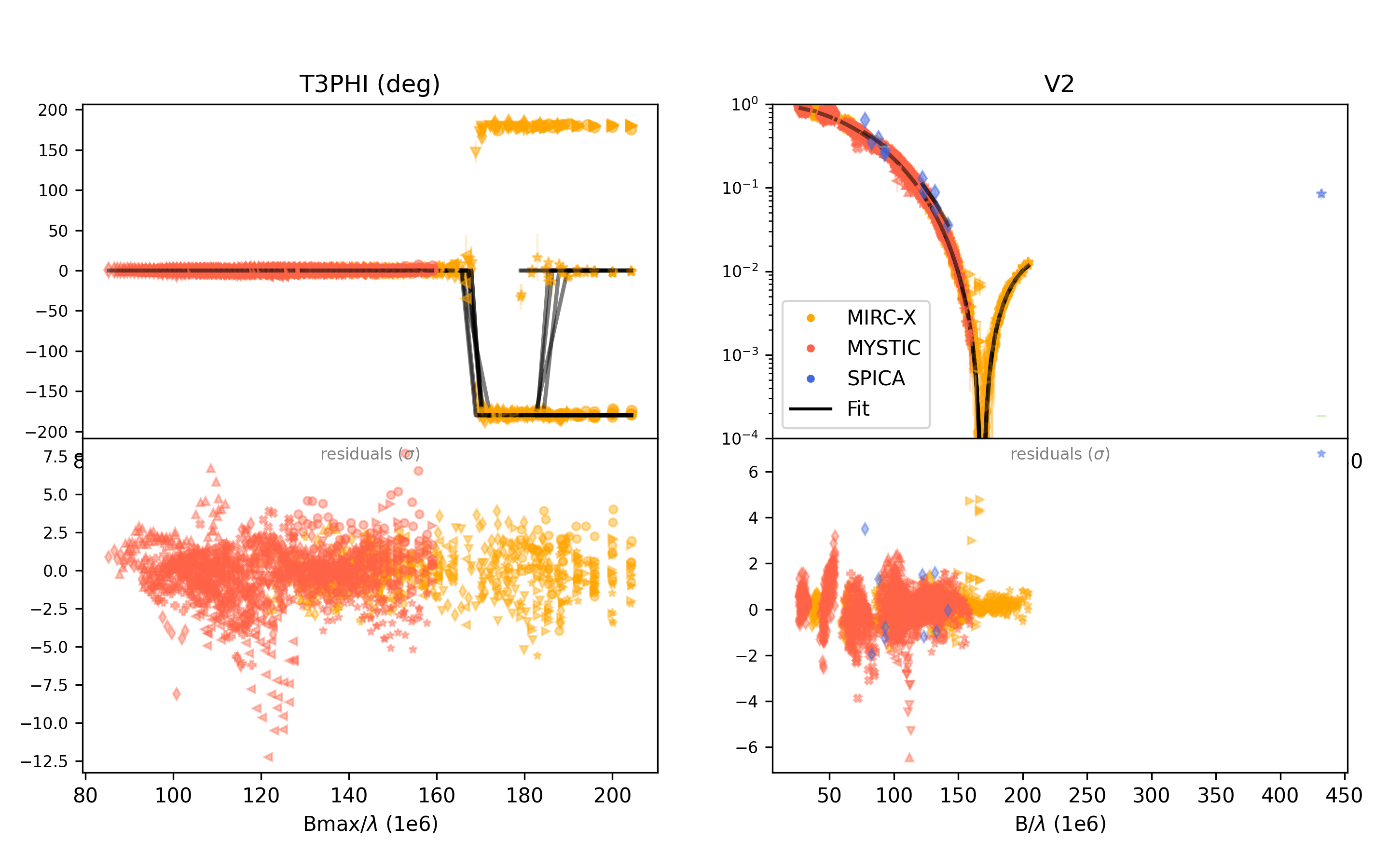}
    
    \begin{minipage}[t]{0.45\textwidth} \centering \textit{(g)} $\delta$ Cep: 2024-12-17 \end{minipage}
    \hfill
    \begin{minipage}[t]{0.45\textwidth} \centering \textit{(h)} $^{*}\delta$ Cep: 2024-08-19 \end{minipage}

    \captionsetup{type=figure}
    \caption{Results of the bootstrapping fits, night per night, for $\delta$ Cep, $\zeta$ Gem, and $\eta$ Aql. By default, we show the $\theta_{LD}^{\text{poly}}$ results (fixed LD coefficients). The asterisk (*) before the star’s name indicates that we are showing the results of the fits from Section \ref{sec:Measuring the limb-darkening of classical cepheids} (free LD coefficients).}
    \label{fig:all_V2_per_nights}
\end{center}

\begin{figure}[h!]
\centering
\scriptsize

\begin{subfigure}[t]{0.45\textwidth}
\includegraphics[width=\linewidth]{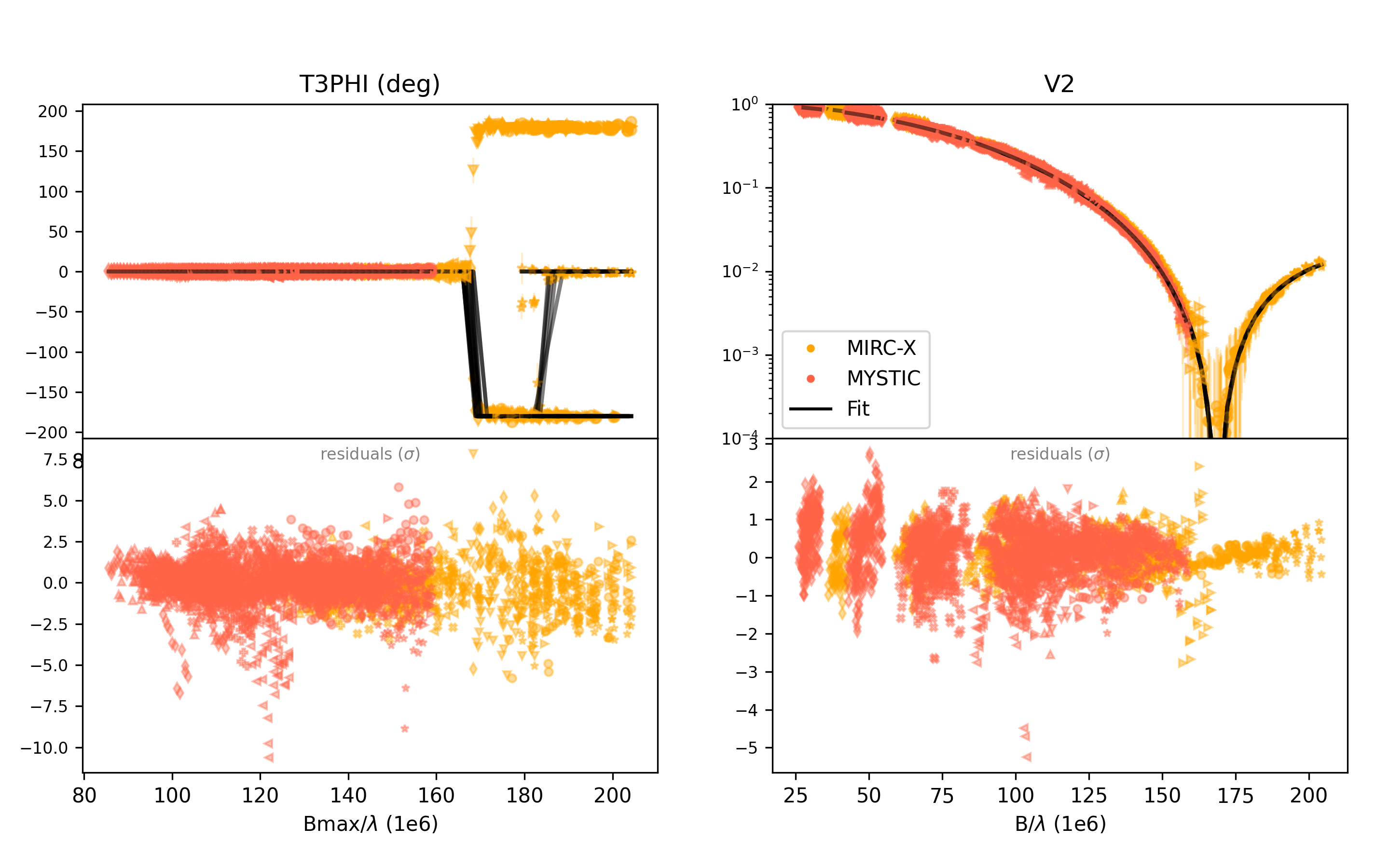}
\caption{$\delta$ Cep: 2025-08-20}
\end{subfigure}
\begin{subfigure}[t]{0.45\textwidth}
\includegraphics[width=\linewidth]{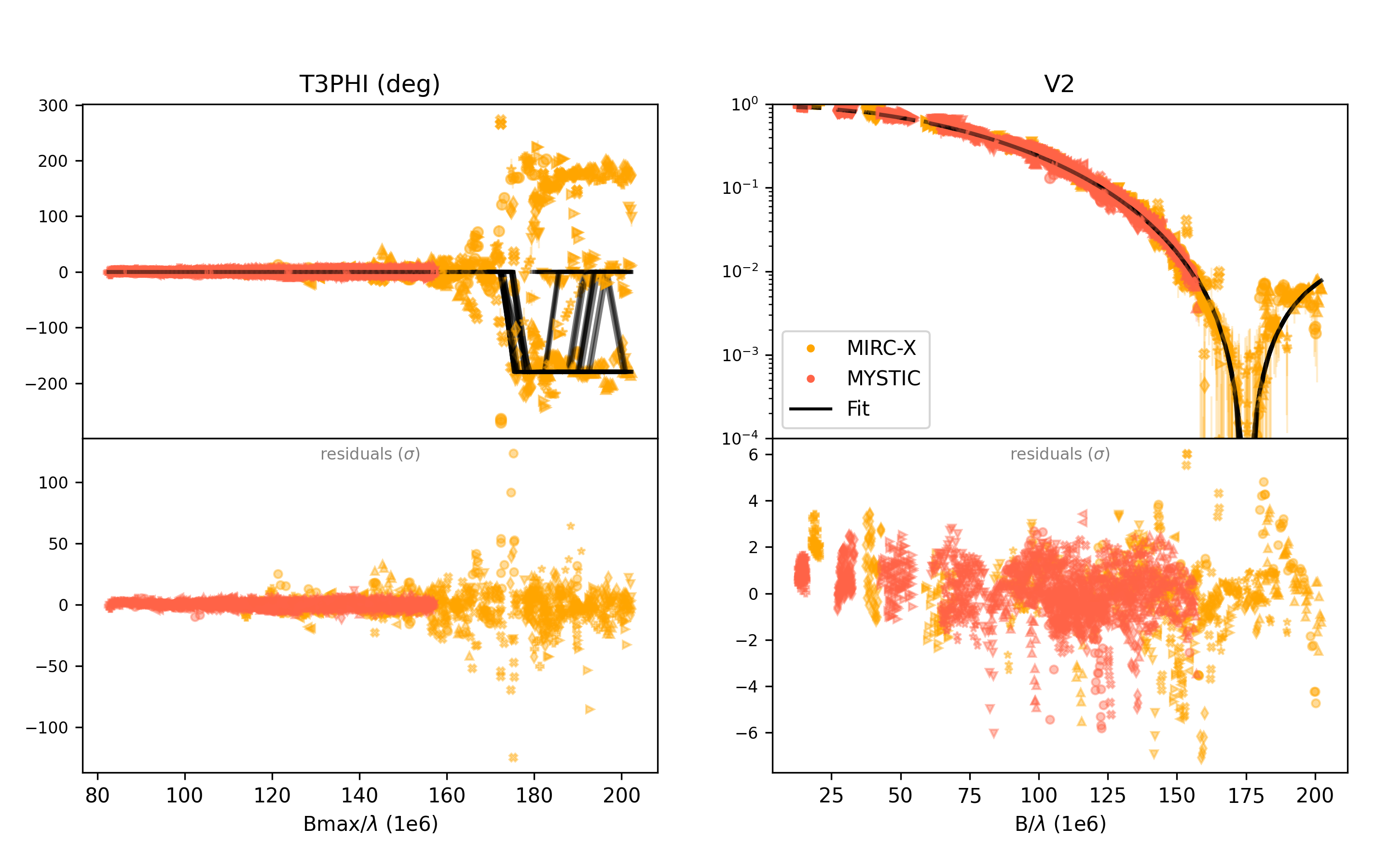}
\caption{$\delta$ Cep: 2025-08-21}
\end{subfigure}

\vspace{-2pt}

\begin{subfigure}[t]{0.45\textwidth}
\includegraphics[width=\linewidth]{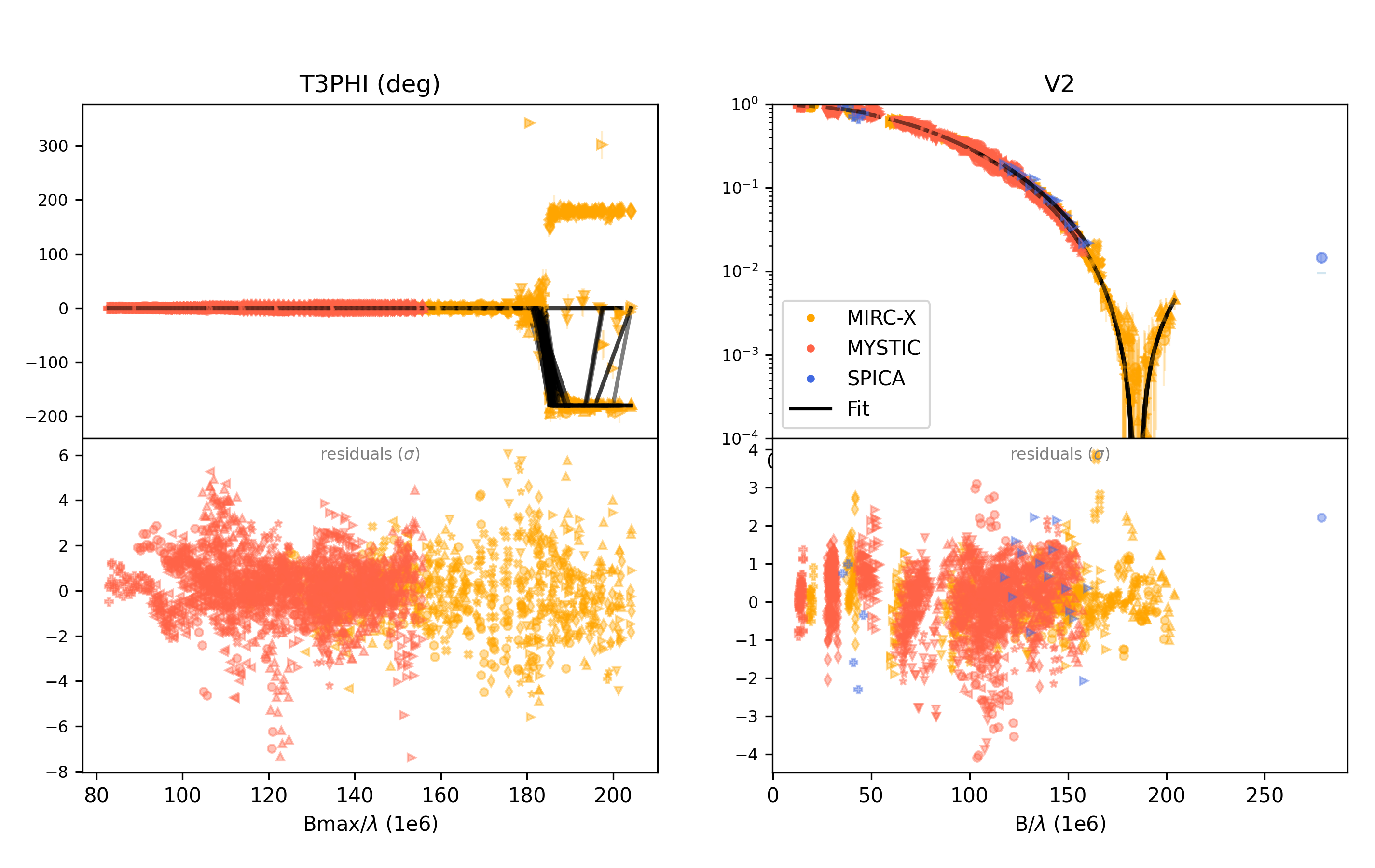}
\caption{$^{*}\delta$ Cep: 2025-08-22}
\end{subfigure}
\begin{subfigure}[t]{0.45\textwidth}
\includegraphics[width=\linewidth]{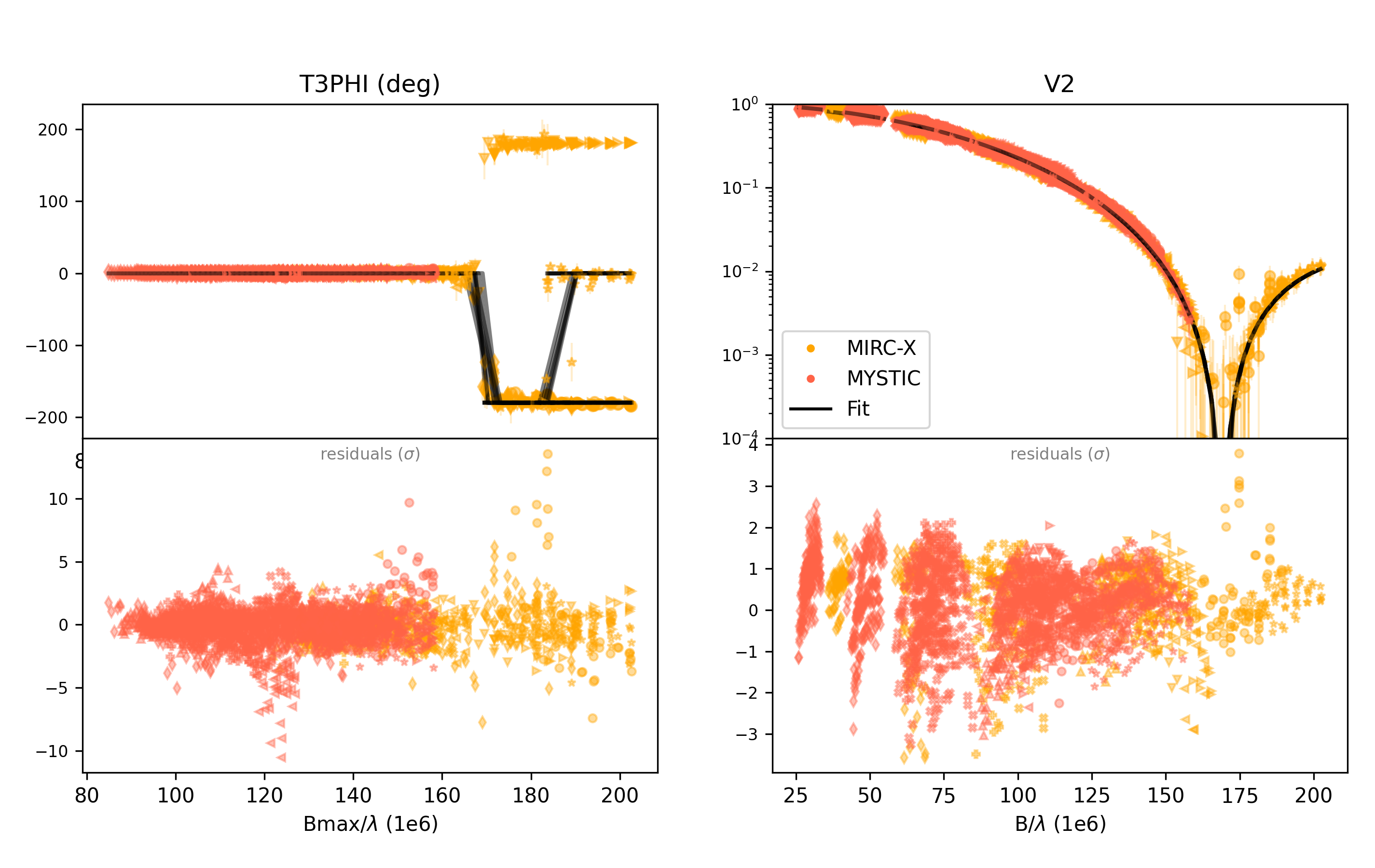}
\caption{$^{*}\delta$ Cep: 2025-10-13}
\end{subfigure}

\vspace{-2pt}

\begin{subfigure}[t]{0.45\textwidth}
\includegraphics[width=\linewidth]{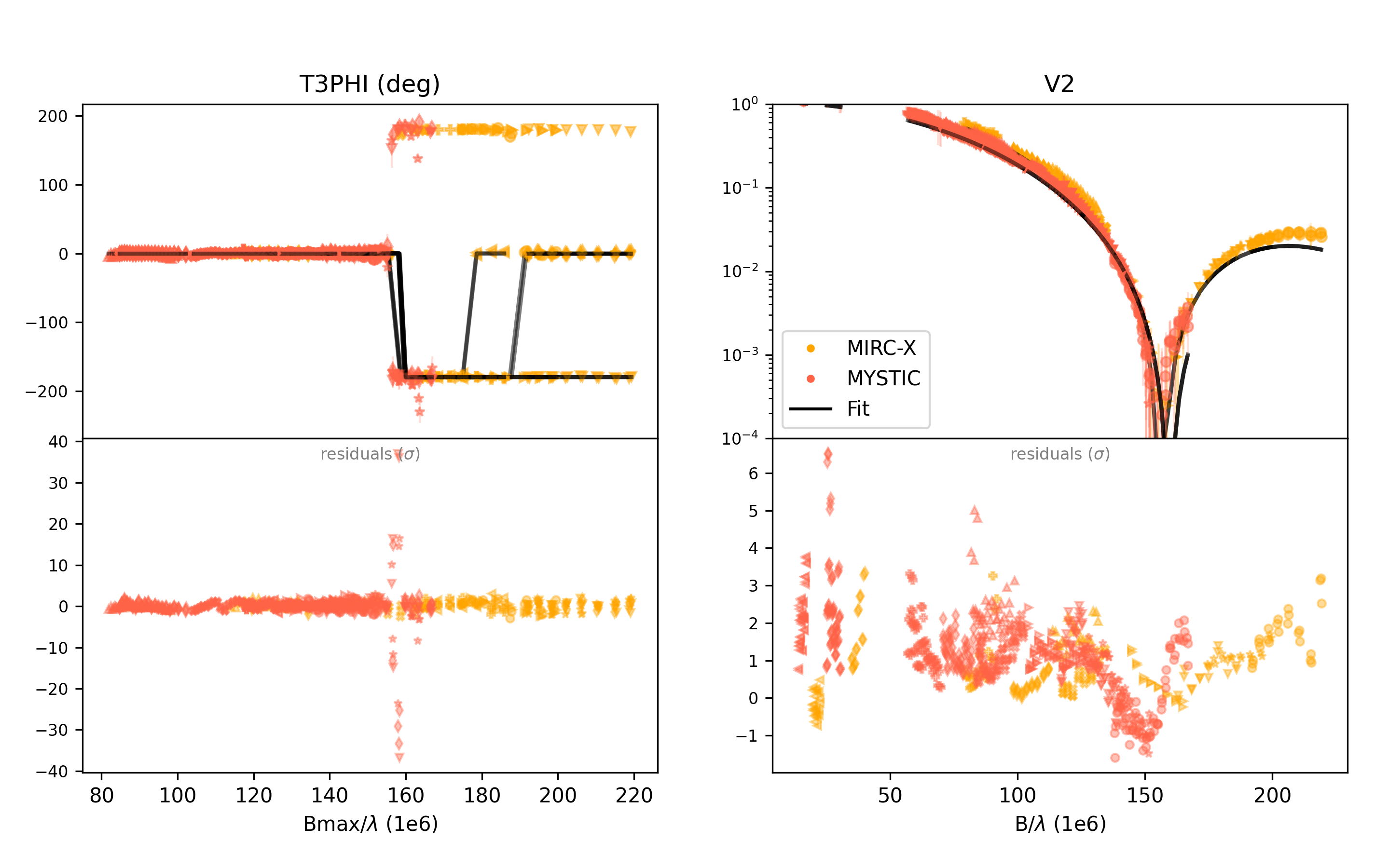}
\caption{$\zeta$ Gem: 2024-09-29}
\end{subfigure}
\begin{subfigure}[t]{0.45\textwidth}
\includegraphics[width=\linewidth]{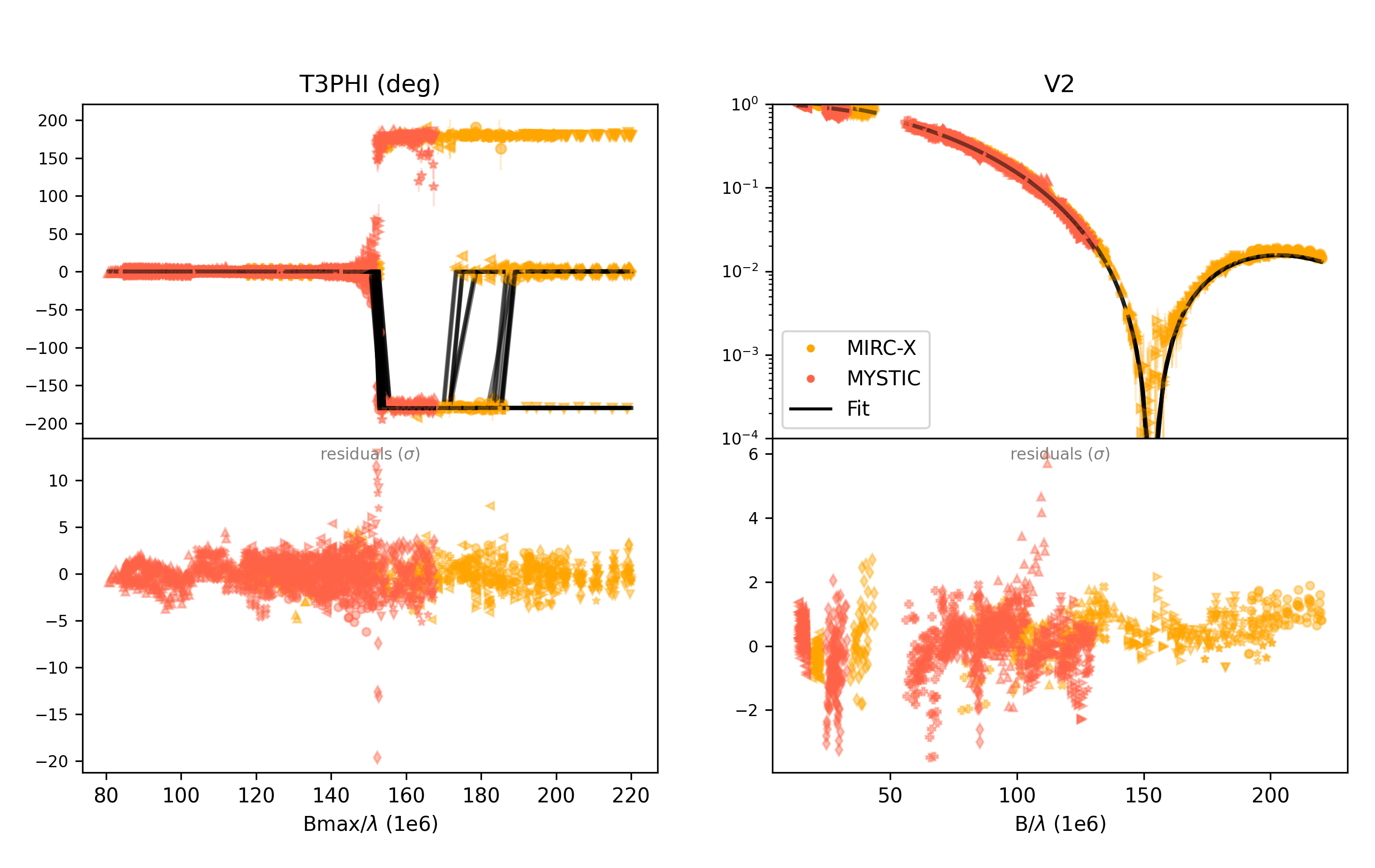}
\caption{$^{*}\zeta$ Gem: 2024-09-30}
\end{subfigure}

\vspace{-2pt}

\begin{subfigure}[t]{0.45\textwidth}
\includegraphics[width=\linewidth]{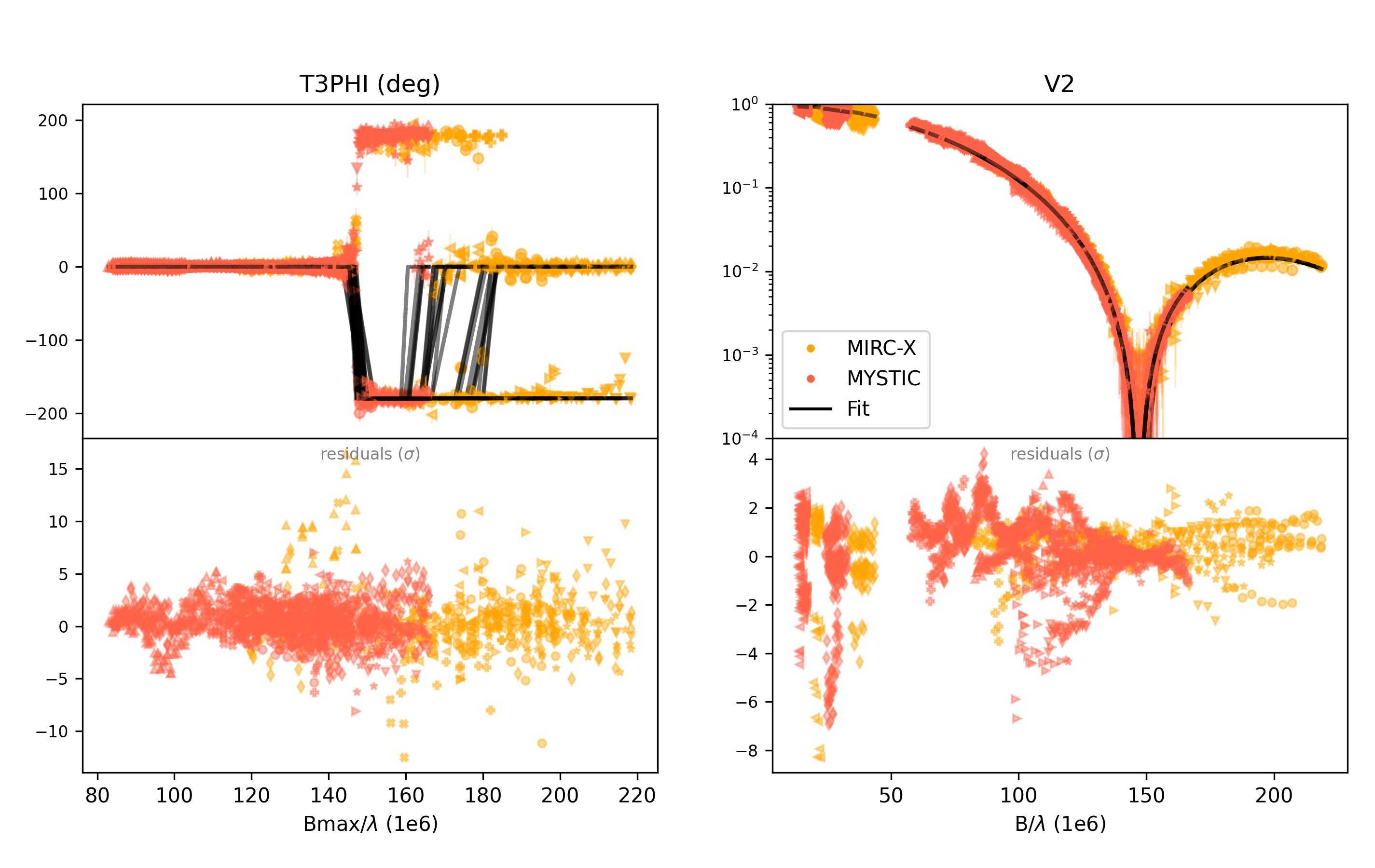}
\caption{$^{*}\zeta$ Gem: 2024-11-11}
\end{subfigure}
\begin{subfigure}[t]{0.45\textwidth}
\includegraphics[width=\linewidth]{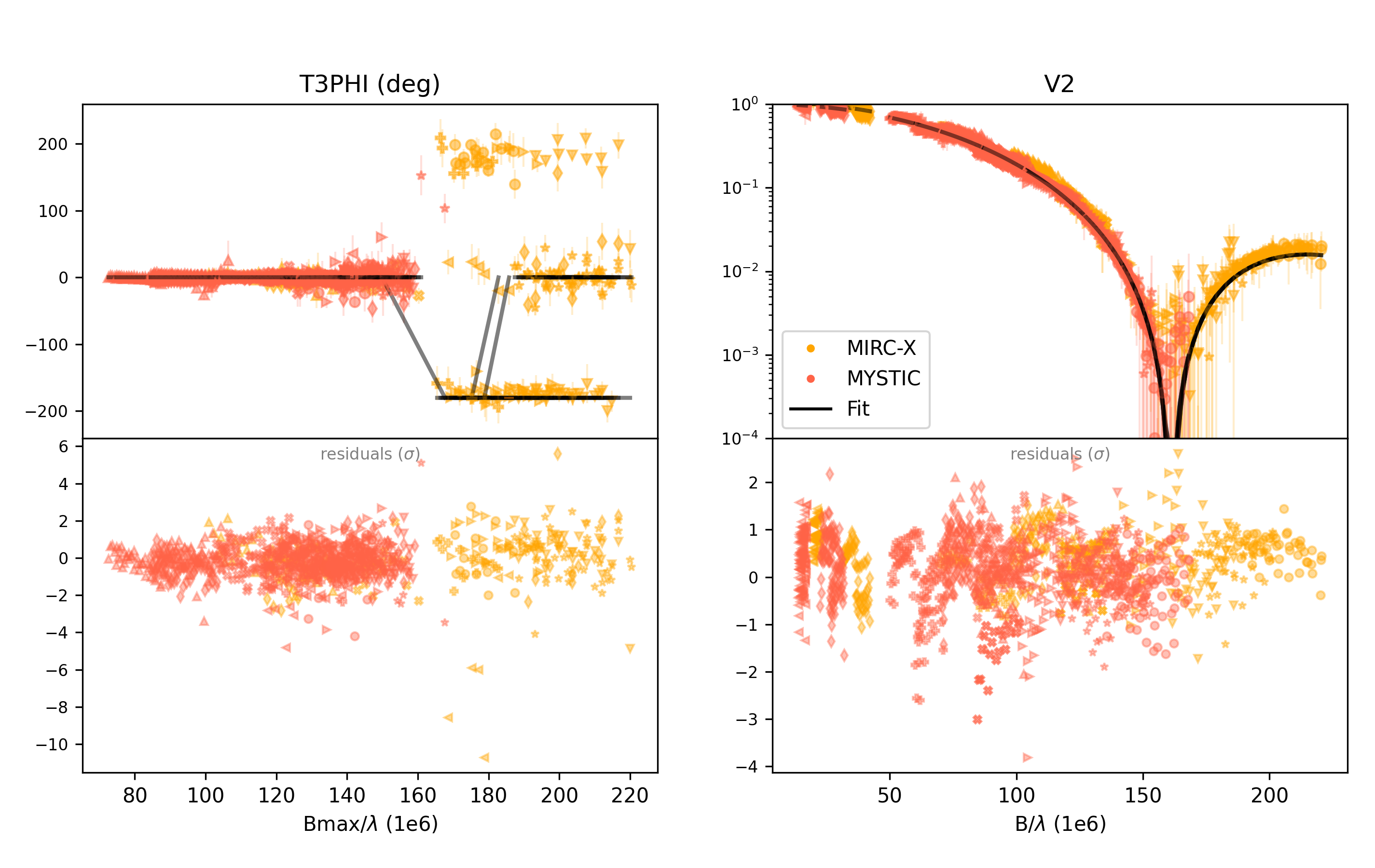}
\caption{$\zeta$ Gem: 2024-12-17}
\end{subfigure}
\caption{Figure \ref{fig:all_V2_per_nights}: Continued.}
\label{fig:all_V2_per_nights_2}
\end{figure}

\begin{figure}[h!]
\centering
\scriptsize

\begin{subfigure}[t]{0.45\textwidth}
\includegraphics[width=\linewidth]{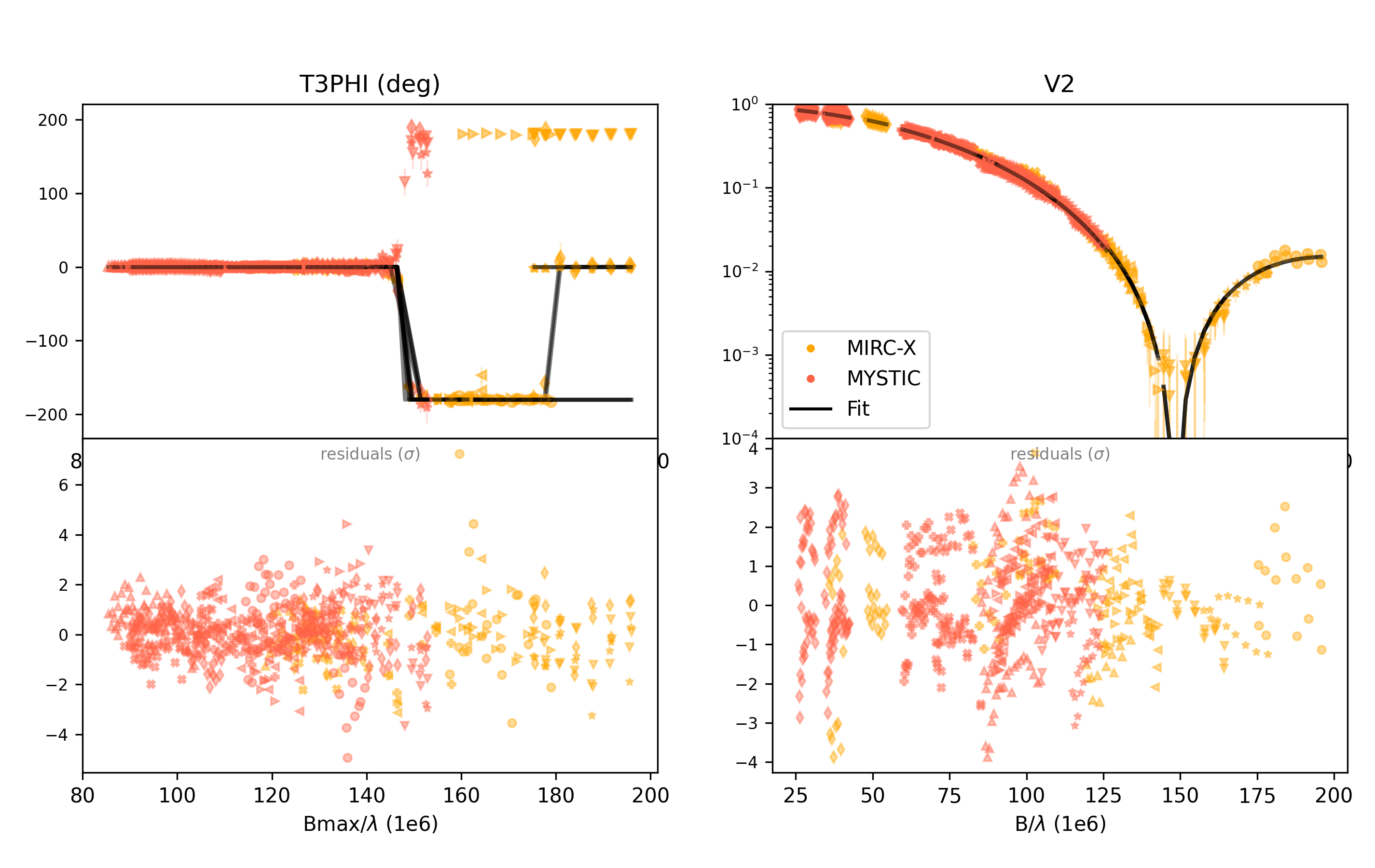}
\caption{$^{*}\zeta$ Gem: 2025-10-13}
\end{subfigure}
\begin{subfigure}[t]{0.45\textwidth}
\includegraphics[width=\linewidth]{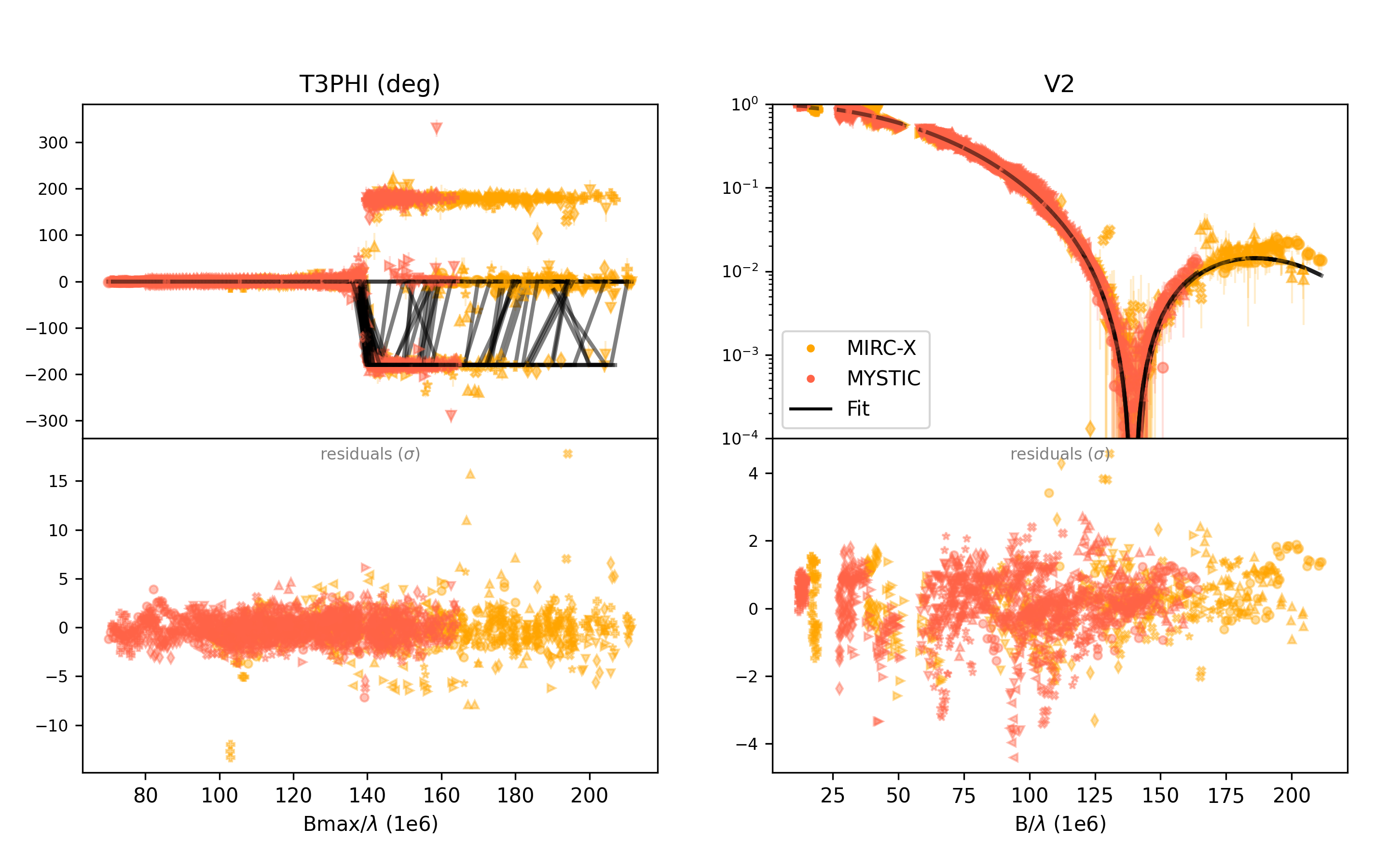}
\caption{$^{*}\eta$ Aql: 2025-05-16}
\end{subfigure}

\vspace{-2pt}

\begin{subfigure}[t]{0.45\textwidth}
\includegraphics[width=\linewidth]{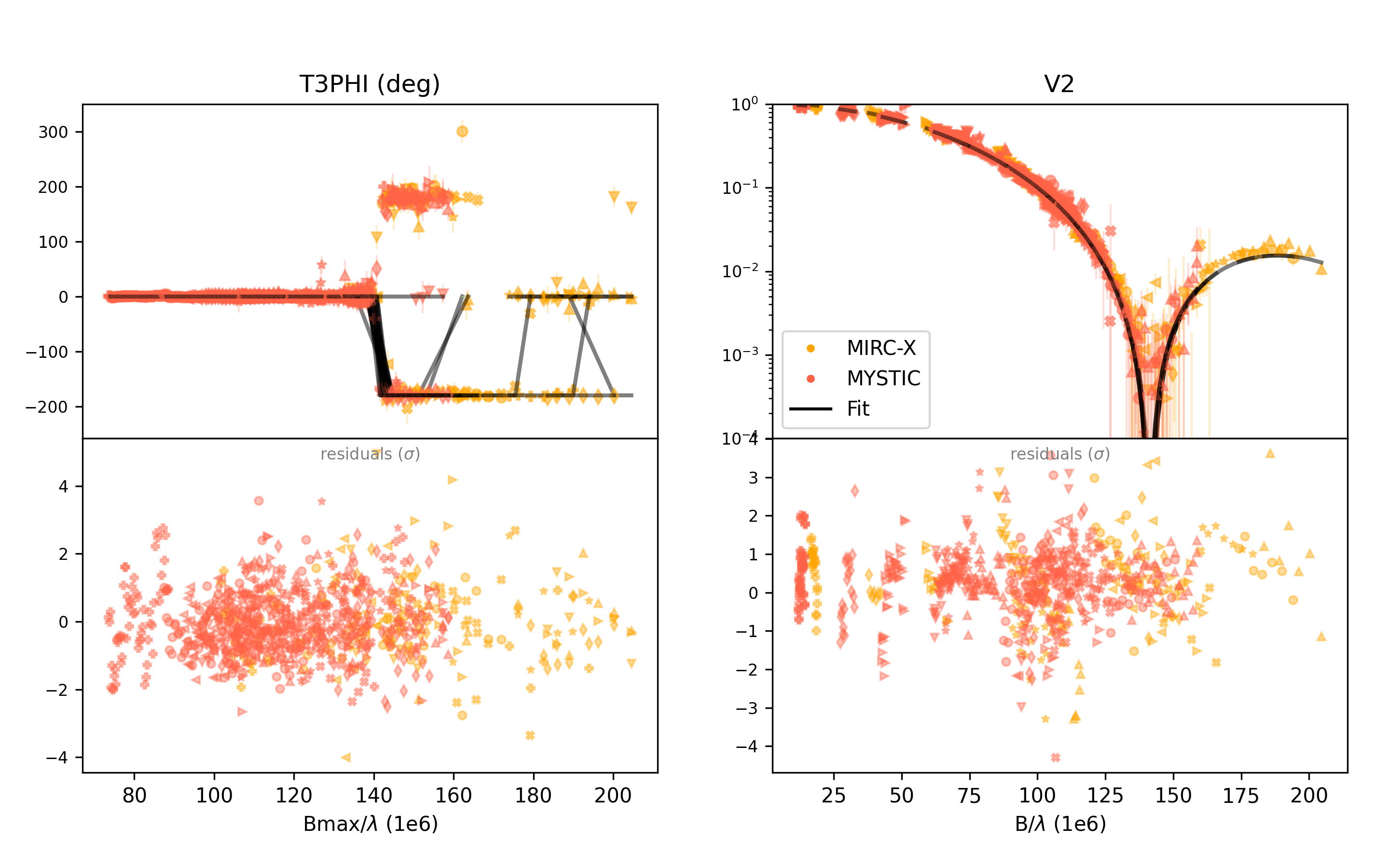}
\caption{$^{*}\eta$ Aql: 2025-05-17}
\end{subfigure}
\begin{subfigure}[t]{0.45\textwidth}
\includegraphics[width=\linewidth]{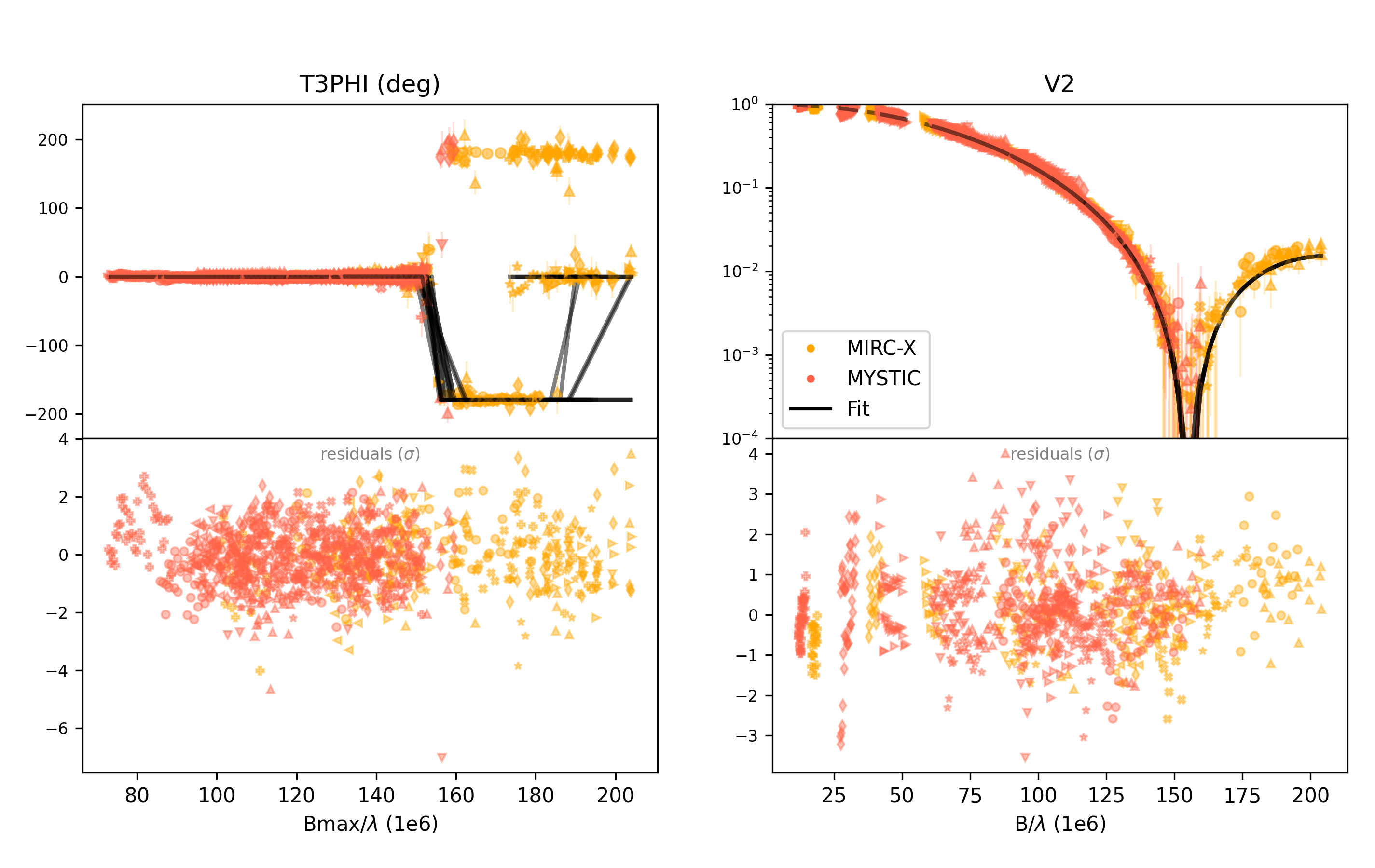}
\caption{$\eta$ Aql: 2025-06-17}
\end{subfigure}

\vspace{-2pt}

\begin{subfigure}[t]{0.45\textwidth}
\includegraphics[width=\linewidth]{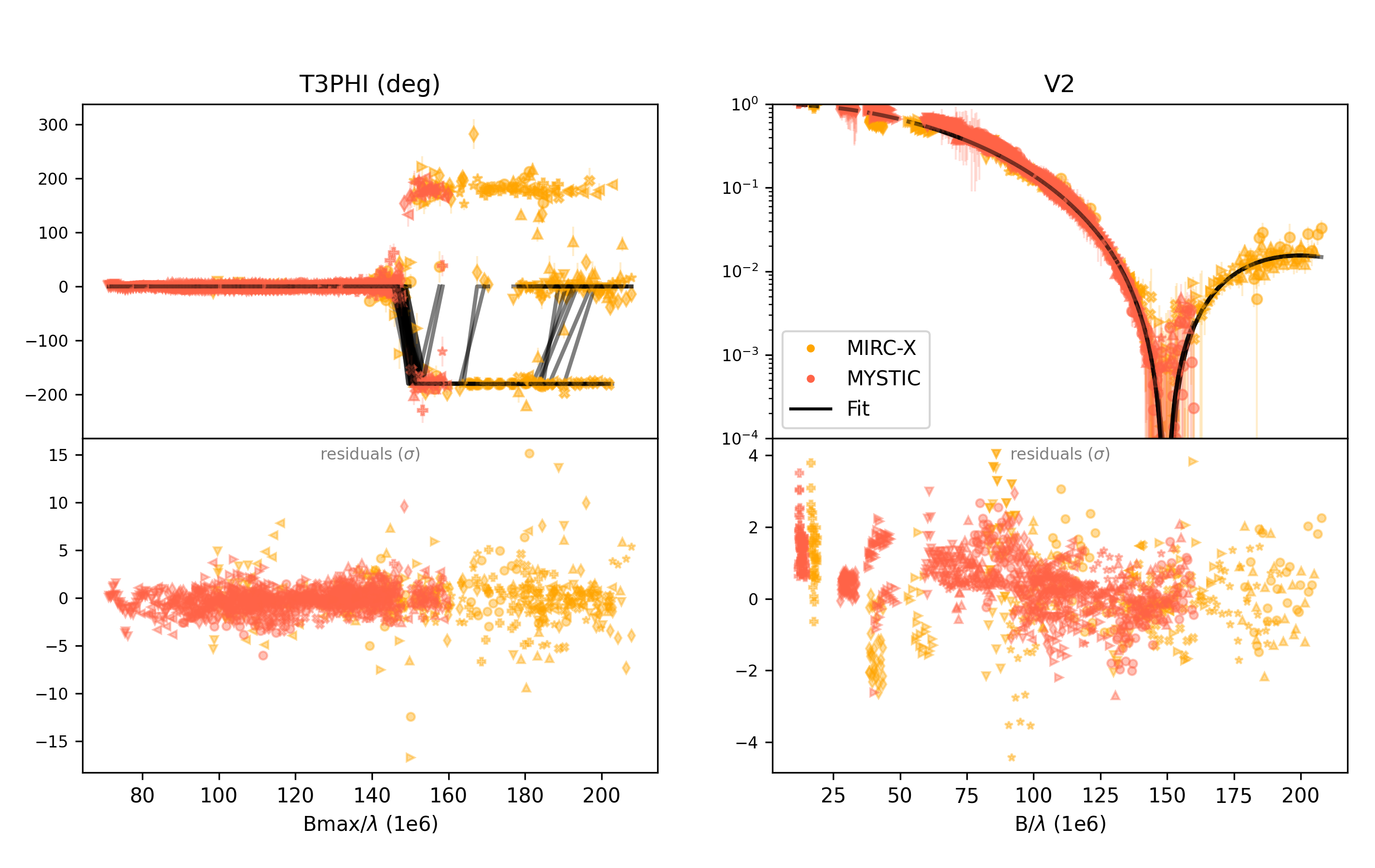}
\caption{$^{*}\eta$ Aql: 2025-06-18}
\end{subfigure}
\begin{subfigure}[t]{0.45\textwidth}
\includegraphics[width=\linewidth]{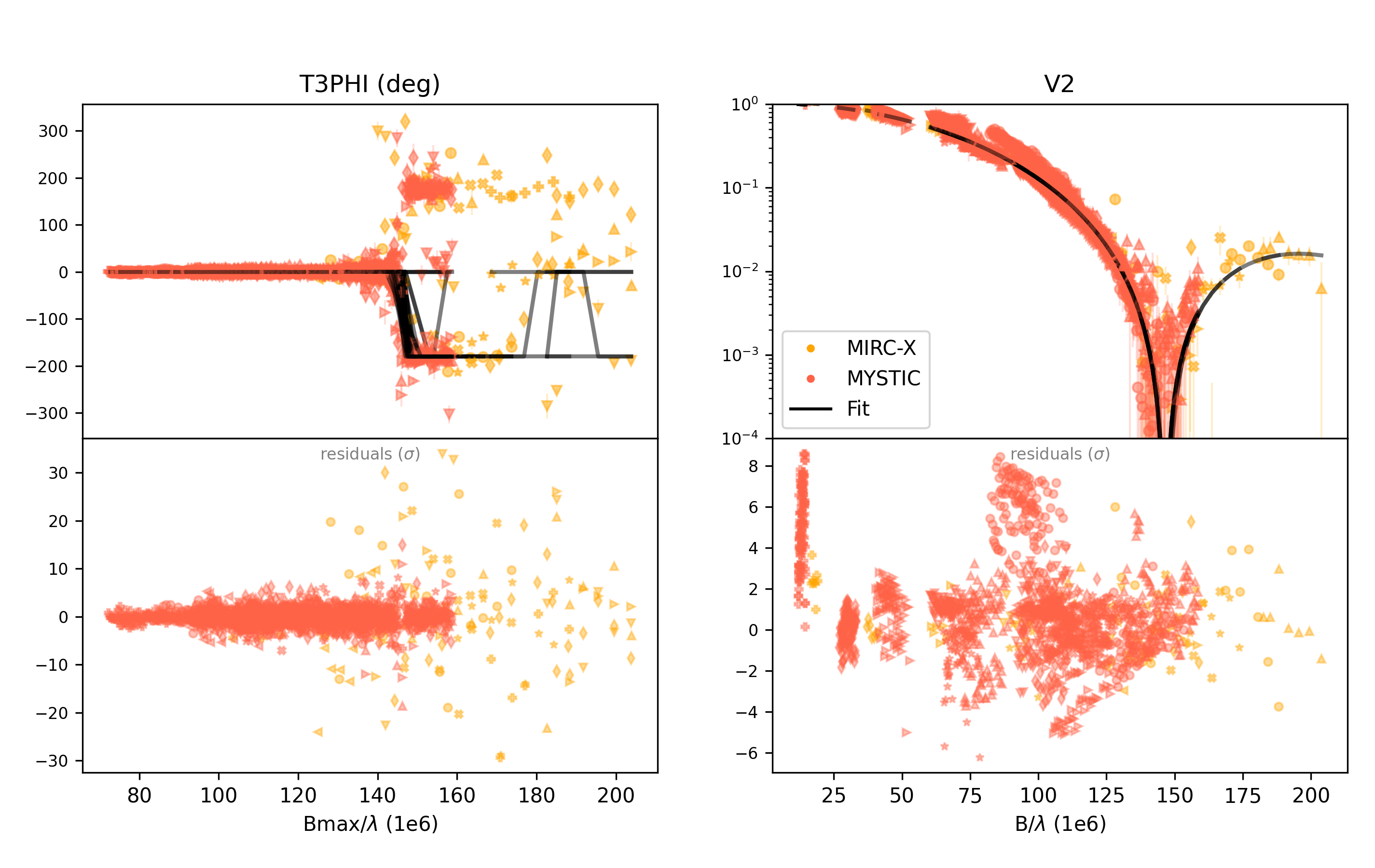}
\caption{$^{*}\eta$ Aql: 2025-07-10}
\end{subfigure}

\vspace{-2pt}

\begin{subfigure}[t]{0.45\textwidth}
\includegraphics[width=\linewidth]{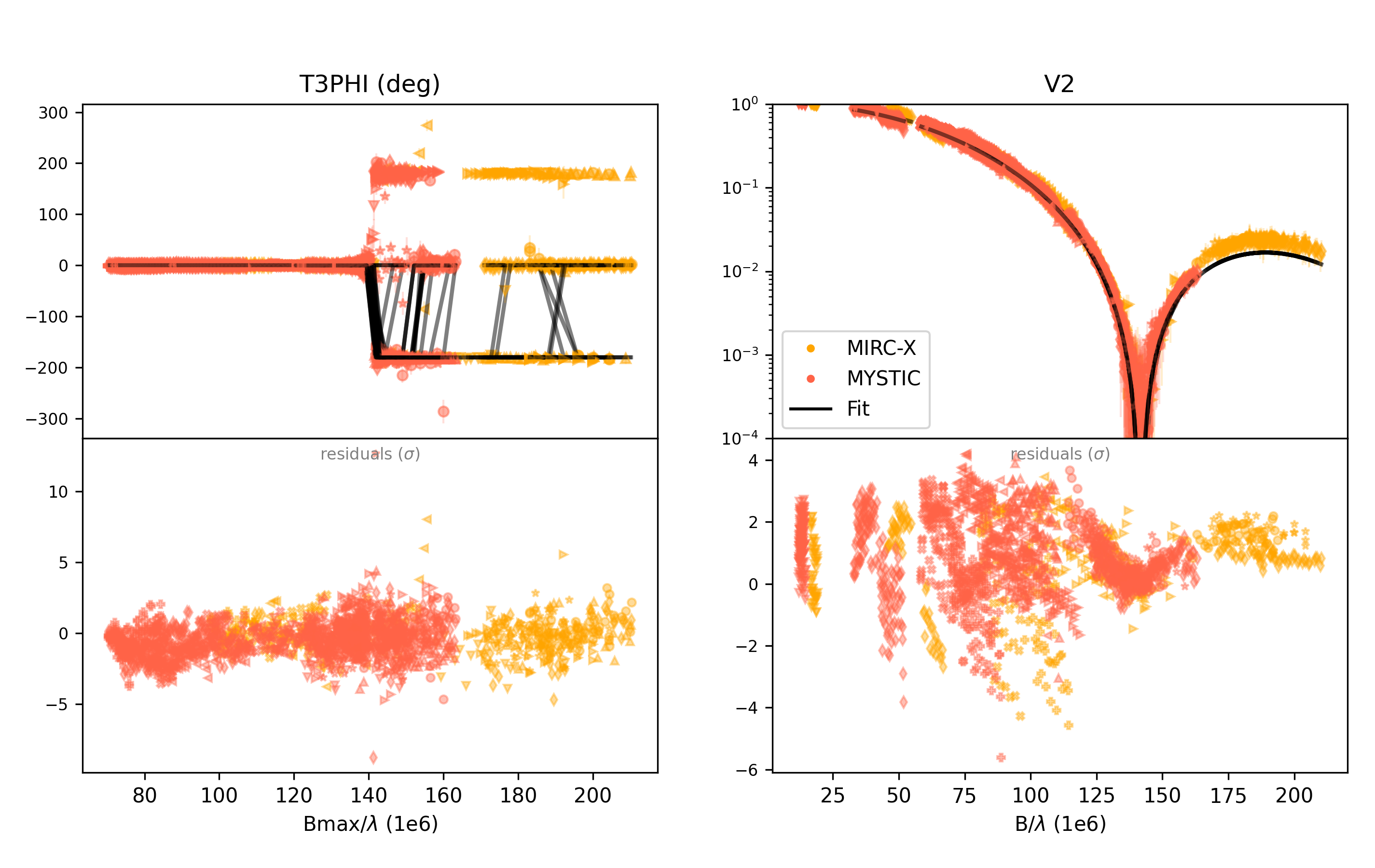}
\caption{$^{*}\eta$ Aql: 2025-07-11}
\end{subfigure}
\caption{Figure \ref{fig:all_V2_per_nights}: Continued.}
\label{fig:all_V2_per_nights_3}
\end{figure}


\begin{sidewaystable*}[ht]
\section{Limb-darkened angular diameters results}
\centering
\caption{Results for $\delta$ Cep. The columns showing a single result are common to the various instruments.}
\label{tab:results_delta_cep}
\begin{tabular}{ccccccccccccccccc}
\hline
Date & Phase & MJD & $T_{\text{eff}}$ & $\log g$ & Inst. & $c$ & $d$ & $\theta_{\text{UD}}$ & $\sigma_{\text{UD}}$ & $\chi^2_{\text{UD}}$ & $\theta_{\text{LD}}^{\text{root}}$ & $\sigma_{\text{root}}$ & $\chi^2_{\text{root}}$ & $\theta_{\text{LD}}^{\text{poly}}$ & $\sigma_{\theta_{\text{LD}}^{\text{poly}}}$ & $\chi^2_{\text{poly}}$ \\
\hline
\multirow{2}{*}{2024-08-26} & \multirow{2}{*}{0.730} & \multirow{2}{*}{60548.2669} & \multirow{2}{*}{5499} & \multirow{2}{*}{1.86} & MYSTIC & -0.148 & 0.587 & 1.491 & 0.003 & 1.20 & 1.524 & 0.003 & 1.21 & \multirow{2}{*}{1.517} & \multirow{2}{*}{0.002} & \multirow{2}{*}{1.11} \\
 & & & & & MIRCX & -0.178 & 0.688 & 1.469 & 0.001 & 1.27 & 1.510 & 0.001 & 1.16 & & & \\
\hline
\multirow{3}{*}{2024-08-27} & \multirow{3}{*}{0.911} & \multirow{3}{*}{60549.2823} & \multirow{3}{*}{6321} & \multirow{3}{*}{1.91} & MYSTIC & -0.155 & 0.527 & 1.406 & 0.002 & 1.09 & 1.431 & 0.002 & 1.15 & \multirow{3}{*}{1.431} & \multirow{3}{*}{0.002} & \multirow{3}{*}{1.13} \\
 & & & & & MIRCX & -0.173 & 0.602 & 1.399 & 0.001 & 1.01 & 1.431 & 0.001 & 1.03 & & & \\
 & & & & & SPICA & 0.028 & 0.693 & 1.376 & 0.015 & 24.24 & 1.444 & 0.016 & 14.60 & & & \\
\hline
\multirow{2}{*}{2024-09-29} & \multirow{2}{*}{0.051} & \multirow{2}{*}{60582.2137} & \multirow{2}{*}{6498} & \multirow{2}{*}{1.87} & MYSTIC & -0.129 & 0.482 & 1.385 & 0.005 & 3.10 & 1.408 & 0.005 & 3.04 & \multirow{2}{*}{1.404} & \multirow{2}{*}{0.004} & \multirow{2}{*}{2.59} \\
 & & & & & MIRCX & -0.136 & 0.544 & 1.366 & 0.004 & 1.95 & 1.396 & 0.004 & 1.77 & & & \\
\hline
\multirow{2}{*}{2024-09-30} & \multirow{2}{*}{0.247} & \multirow{2}{*}{60583.2823} & \multirow{2}{*}{5967} & \multirow{2}{*}{1.82} & MYSTIC & -0.149 & 0.538 & 1.443 & 0.002 & 0.86 & 1.470 & 0.002 & 0.96 & \multirow{2}{*}{1.470} & \multirow{2}{*}{0.002} & \multirow{2}{*}{0.89} \\
 & & & & & MIRCX & -0.171 & 0.623 & 1.436 & 0.003 & 0.74 & 1.471 & 0.003 & 0.92 & & & \\
\hline
\multirow{2}{*}{2024-11-11} & \multirow{2}{*}{0.055} & \multirow{2}{*}{60625.1864} & \multirow{2}{*}{6471} & \multirow{2}{*}{1.87} & MYSTIC & -0.129 & 0.482 & 1.368 & 0.001 & 0.85 & 1.391 & 0.002 & 0.82 & \multirow{2}{*}{1.389} & \multirow{2}{*}{0.001} & \multirow{2}{*}{0.94} \\
 & & & & & MIRCX & -0.136 & 0.544 & 1.357 & 0.003 & 1.36 & 1.388 & 0.003 & 1.37 & & & \\
\hline
\multirow{2}{*}{2024-11-14} & \multirow{2}{*}{0.623} & \multirow{2}{*}{60628.2318} & \multirow{2}{*}{5508} & \multirow{2}{*}{1.83} & MYSTIC & -0.148 & 0.587 & 1.500 & 0.001 & 2.48 & 1.534 & 0.001 & 2.48 & \multirow{2}{*}{1.537} & \multirow{2}{*}{0.001} & \multirow{2}{*}{2.74} \\
 & & & & & MIRCX & -0.178 & 0.688 & 1.495 & 0.001 & 3.14 & 1.541 & 0.001 & 3.23 & & & \\
\hline
\multirow{2}{*}{2024-12-17} & \multirow{2}{*}{0.749} & \multirow{2}{*}{60661.1415} & \multirow{2}{*}{5526} & \multirow{2}{*}{1.87} & MYSTIC & -0.148 & 0.587 & 1.459 & 0.004 & 0.98 & 1.490 & 0.004 & 1.08 & \multirow{2}{*}{1.492} & \multirow{2}{*}{0.003} & \multirow{2}{*}{0.93} \\
 & & & & & MIRCX & -0.178 & 0.688 & 1.457 & 0.002 & 0.88 & 1.497 & 0.002 & 0.83 & & & \\
\hline
\multirow{3}{*}{2025-08-19} & \multirow{3}{*}{0.450} & \multirow{3}{*}{60906.2960} & \multirow{3}{*}{5663} & \multirow{3}{*}{1.81} & MYSTIC & -0.147 & 0.560 & 1.502 & 0.002 & 1.75 & 1.533 & 0.002 & 2.38 & \multirow{3}{*}{1.534} & \multirow{3}{*}{0.001} & \multirow{3}{*}{2.00} \\
 & & & & & MIRCX & -0.174 & 0.654 & 1.495 & 0.001 & 1.34 & 1.535 & 0.001 & 1.09 & & & \\
 & & & & & SPICA & 0.113 & 0.635 & 1.437 & 0.042 & 3.82 & 1.525 & 0.047 & 3.25 & & & \\
\hline
\multirow{2}{*}{2025-08-20} & \multirow{2}{*}{0.636} & \multirow{2}{*}{60907.2945} & \multirow{2}{*}{5507} & \multirow{2}{*}{1.83} & MYSTIC & -0.148 & 0.587 & 1.503 & 0.001 & 1.23 & 1.536 & 0.001 & 1.09 & \multirow{2}{*}{1.536} & \multirow{2}{*}{0.001} & \multirow{2}{*}{1.26} \\
 & & & & & MIRCX & -0.178 & 0.688 & 1.494 & 0.001 & 1.33 & 1.537 & 0.001 & 1.26 & & & \\
\hline
\multirow{2}{*}{2025-08-21} & \multirow{2}{*}{0.825} & \multirow{2}{*}{60908.3262} & \multirow{2}{*}{5722} & \multirow{2}{*}{1.89} & MYSTIC & -0.147 & 0.560 & 1.455 & 0.002 & 1.29 & 1.485 & 0.002 & 1.21 & \multirow{2}{*}{1.476} & \multirow{2}{*}{0.001} & \multirow{2}{*}{32.97} \\
 & & & & & MIRCX & -0.174 & 0.654 & 1.452 & 0.001 & 3.02 & 1.491 & 0.001 & 2.82 & & & \\
\hline
\multirow{3}{*}{2025-08-22} & \multirow{3}{*}{0.010} & \multirow{3}{*}{60909.3275} & \multirow{3}{*}{6595} & \multirow{3}{*}{1.88} & MYSTIC & -0.129 & 0.482 & 1.371 & 0.001 & 1.45 & 1.394 & 0.001 & 1.50 & \multirow{3}{*}{1.394} & \multirow{3}{*}{0.001} & \multirow{3}{*}{1.59} \\
 & & & & & MIRCX & -0.136 & 0.544 & 1.363 & 0.001 & 1.77 & 1.393 & 0.001 & 1.90 & & & \\
 & & & & & SPICA & 0.112 & 0.580 & 1.341 & 0.012 & 0.86 & 1.411 & 0.014 & 1.26 & & & \\
\hline
\multirow{2}{*}{2025-10-13} & \multirow{2}{*}{0.672} & \multirow{2}{*}{60961.1439} & \multirow{2}{*}{5500} & \multirow{2}{*}{1.84} & MYSTIC & -0.148 & 0.587 & 1.497 & 0.001 & 1.43 & 1.530 & 0.001 & 1.25 & \multirow{2}{*}{1.531} & \multirow{2}{*}{0.001} & \multirow{2}{*}{1.44} \\
 & & & & & MIRCX & -0.178 & 0.688 & 1.489 & 0.002 & 2.26 & 1.533 & 0.002 & 2.11 & & & \\
\hline
\end{tabular}
\tablefoot{$T_{\text{eff}}$ and $\log g$ are derived from SPIPS model values. $c$ and $d$ are the LD coefficients of the square-root law, and are fixed parameters. For each measurement, we have given the diameter in mas, along with the associated uncertainty and the value of $\chi^{2}$, all derived from the bootstrap method. The MIRC-X values of $\theta_{UD}$ and $\theta_{LD}^{\text{root}}$ for the night of 21 August 2025 were calculated without taking the T3PHI data into account, due to their very poor quality.}
\end{sidewaystable*}

\begin{sidewaystable*}[ht]
\centering
\caption{Same as Table \ref{tab:results_delta_cep}, but for $\zeta$ Gem.}
\label{tab:results_zeta_gem}
\begin{tabular}{ccccccccccccccccc}
\hline
Date & Phase & MJD & $T_{\text{eff}}$ & $\log g$ & Inst. & $c$ & $d$ & $\theta_{\text{UD}}$ & $\sigma_{\text{UD}}$ & $\chi^2_{\text{UD}}$ & $\theta_{\text{LD}}^{\text{root}}$ & $\sigma_{\text{root}}$ & $\chi^2_{\text{root}}$ & $\theta_{\text{LD}}^{\text{poly}}$ & $\sigma_{\theta_{\text{LD}}^{\text{poly}}}$ & $\chi^2_{\text{poly}}$ \\
\hline
\multirow{2}{*}{2024-09-29} & \multirow{2}{*}{0.272} & \multirow{2}{*}{60582.4805} & \multirow{2}{*}{5380} & \multirow{2}{*}{1.51} & MYSTIC & -0.149 & 0.582 & 1.593 & 0.002 & 3.35 & 1.700 & 0.003 & 60.24 & \multirow{2}{*}{1.700} & \multirow{2}{*}{0.002} & \multirow{2}{*}{64.18} \\
 & & & & & MIRCX & -0.176 & 0.680 & 1.587 & 0.004 & 38.98 & 1.700 & 0.005 & 32.96 & & & \\
\hline
\multirow{2}{*}{2024-09-30} & \multirow{2}{*}{0.373} & \multirow{2}{*}{60583.4740} & \multirow{2}{*}{5256} & \multirow{2}{*}{1.51} & MYSTIC & -0.157 & 0.621 & 1.655 & 0.001 & 1.58 & 1.702 & 0.001 & 1.80 & \multirow{2}{*}{1.702} & \multirow{2}{*}{0.001} & \multirow{2}{*}{1.59} \\
 & & & & & MIRCX & -0.188 & 0.728 & 1.637 & 0.001 & 1.33 & 1.690 & 0.001 & 1.36 & & & \\
\hline
\multirow{2}{*}{2024-11-11} & \multirow{2}{*}{0.507} & \multirow{2}{*}{60625.3699} & \multirow{2}{*}{5222} & \multirow{2}{*}{1.53} & MYSTIC & -0.157 & 0.621 & 1.699 & 0.001 & 2.37 & 1.739 & 0.001 & 2.39 & \multirow{2}{*}{1.758} & \multirow{2}{*}{0.003} & \multirow{2}{*}{3.71} \\
 & & & & & MIRCX & -0.188 & 0.728 & 1.682 & 0.002 & 5.58 & 1.736 & 0.003 & 4.02 & & & \\
\hline
\multirow{2}{*}{2024-12-17} & \multirow{2}{*}{0.035} & \multirow{2}{*}{60661.2344} & \multirow{2}{*}{5765} & \multirow{2}{*}{1.55} & MYSTIC & -0.148 & 0.555 & 1.568 & 0.001 & 0.56 & 1.601 & 0.002 & 0.71 & \multirow{2}{*}{1.600} & \multirow{2}{*}{0.001} & \multirow{2}{*}{0.82} \\
 & & & & & MIRCX & -0.173 & 0.647 & 1.553 & 0.002 & 1.33 & 1.596 & 0.003 & 1.11 & & & \\
\hline
\multirow{2}{*}{2025-10-13} & \multirow{2}{*}{0.616} & \multirow{2}{*}{60961.4589} & \multirow{2}{*}{5411} & \multirow{2}{*}{1.57} & MYSTIC & -0.149 & 0.582 & 1.717 & 0.004 & 1.35 & 1.755 & 0.004 & 1.25 & \multirow{2}{*}{1.752} & \multirow{2}{*}{0.003} & \multirow{2}{*}{1.46} \\
 & & & & & MIRCX & -0.176 & 0.680 & 1.697 & 0.004 & 1.58 & 1.746 & 0.004 & 1.80 & & & \\
\hline
\end{tabular}

\vspace{2cm}

\centering
\caption{Same as Table \ref{tab:results_delta_cep}, but for $\eta$ Aql.}
\label{tab:results_eta_aql}
\begin{tabular}{ccccccccccccccccc}
\hline
Date & Phase & MJD & $T_{\text{eff}}$ & $\log g$ & Inst. & $c$ & $d$ & $\theta_{\text{UD}}$ & $\sigma_{\text{UD}}$ & $\chi^2_{\text{UD}}$ & $\theta_{\text{LD}}^{\text{root}}$ & $\sigma_{\text{root}}$ & $\chi^2_{\text{root}}$ & $\theta_{\text{LD}}^{\text{poly}}$ & $\sigma_{\theta_{\text{LD}}^{\text{poly}}}$ & $\chi^2_{\text{poly}}$ \\
\hline
\multirow{2}{*}{2025-05-16} & \multirow{2}{*}{0.575} & \multirow{2}{*}{60811.4064} & \multirow{2}{*}{5426} & \multirow{2}{*}{1.68} & MYSTIC & -0.149 & 0.582 & 1.813 & 0.006 & 1.48 & 1.853 & 0.006 & 1.56 & \multirow{2}{*}{1.851} & \multirow{2}{*}{0.002} & \multirow{2}{*}{1.91} \\
 & & & & & MIRCX & -0.176 & 0.680 & 1.795 & 0.002 & 3.17 & 1.849 & 0.002 & 2.79 & & & \\
\hline
\multirow{2}{*}{2025-05-17} & \multirow{2}{*}{0.719} & \multirow{2}{*}{60812.4712} & \multirow{2}{*}{5412} & \multirow{2}{*}{1.72} & MYSTIC & -0.149 & 0.582 & 1.804 & 0.005 & 1.00 & 1.839 & 0.005 & 0.88 & \multirow{2}{*}{1.829} & \multirow{2}{*}{0.002} & \multirow{2}{*}{0.95} \\
 & & & & & MIRCX & -0.176 & 0.680 & 1.782 & 0.006 & 1.33 & 1.833 & 0.006 & 1.40 & & & \\
\hline
\multirow{2}{*}{2025-06-17} & \multirow{2}{*}{0.026} & \multirow{2}{*}{60843.3809} & \multirow{2}{*}{6313} & \multirow{2}{*}{1.74} & MYSTIC & -0.119 & 0.485 & 1.631 & 0.004 & 1.26 & 1.660 & 0.004 & 1.19 & \multirow{2}{*}{1.661} & \multirow{2}{*}{0.002} & \multirow{2}{*}{1.09} \\
 & & & & & MIRCX & -0.128 & 0.559 & 1.623 & 0.003 & 1.10 & 1.663 & 0.003 & 1.24 & & & \\
\hline
\multirow{2}{*}{2025-06-18} & \multirow{2}{*}{0.162} & \multirow{2}{*}{60844.3478} & \multirow{2}{*}{5958} & \multirow{2}{*}{1.70} & MYSTIC & -0.154 & 0.541 & 1.687 & 0.003 & 1.51 & 1.719 & 0.003 & 1.54 & \multirow{2}{*}{1.721} & \multirow{2}{*}{0.002} & \multirow{2}{*}{2.13} \\
 & & & & & MIRCX & -0.173 & 0.620 & 1.680 & 0.003 & 4.27 & 1.723 & 0.003 & 4.90 & & & \\
\hline
\multirow{2}{*}{2025-07-10} & \multirow{2}{*}{0.225} & \multirow{2}{*}{60866.3412} & \multirow{2}{*}{5873} & \multirow{2}{*}{1.69} & MYSTIC & -0.148 & 0.555 & 1.724 & 0.005 & 5.34 & 1.759 & 0.006 & 5.64 & \multirow{2}{*}{1.759} & \multirow{2}{*}{0.004} & \multirow{2}{*}{10.29} \\
 & & & & & MIRCX & -0.173 & 0.647 & 1.716 & 0.005 & 83.88 & 1.759 & 0.005 & 55.79 & & & \\
\hline
\multirow{2}{*}{2025-07-11} & \multirow{2}{*}{0.358} & \multirow{2}{*}{60867.2584} & \multirow{2}{*}{5769} & \multirow{2}{*}{1.68} & MYSTIC & -0.148 & 0.555 & 1.782 & 0.004 & 2.30 & 1.818 & 0.004 & 2.85 & \multirow{2}{*}{1.818} & \multirow{2}{*}{0.002} & \multirow{2}{*}{2.29} \\
 & & & & & MIRCX & -0.173 & 0.647 & 1.771 & 0.004 & 2.21 & 1.817 & 0.004 & 1.96 & & & \\
\hline
\end{tabular}
\end{sidewaystable*}

\clearpage

\twocolumn
\section{Example of bootstraping results using \texttt{pmoired}}
\begin{figure}[h!]
    \centering

    \begin{subfigure}{0.5\textwidth}
        \centering
        \includegraphics[width=\linewidth]{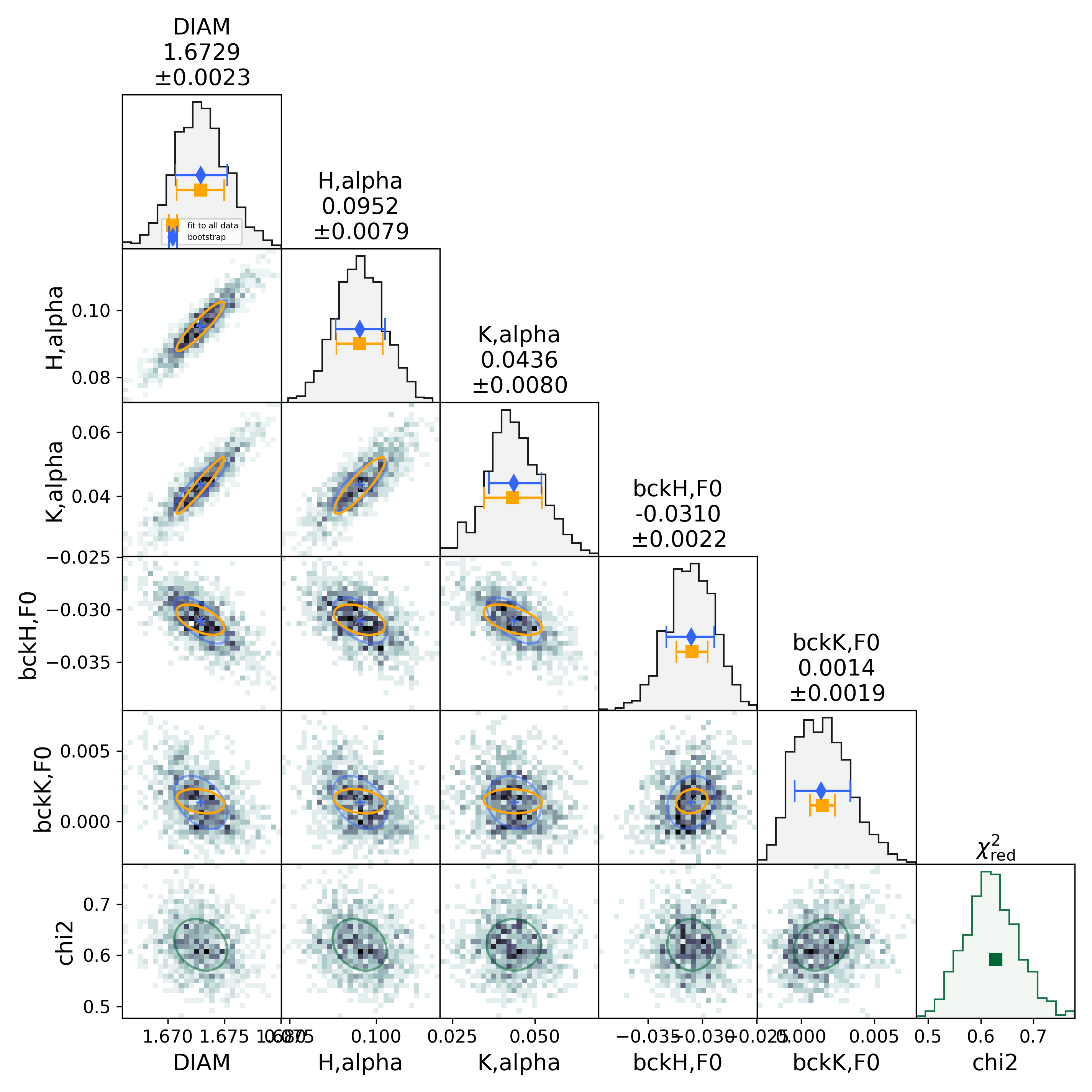}
        \label{fig:c1}
    \end{subfigure}
    \begin{subfigure}{0.25\textwidth}
        \centering
        \includegraphics[width=\linewidth]{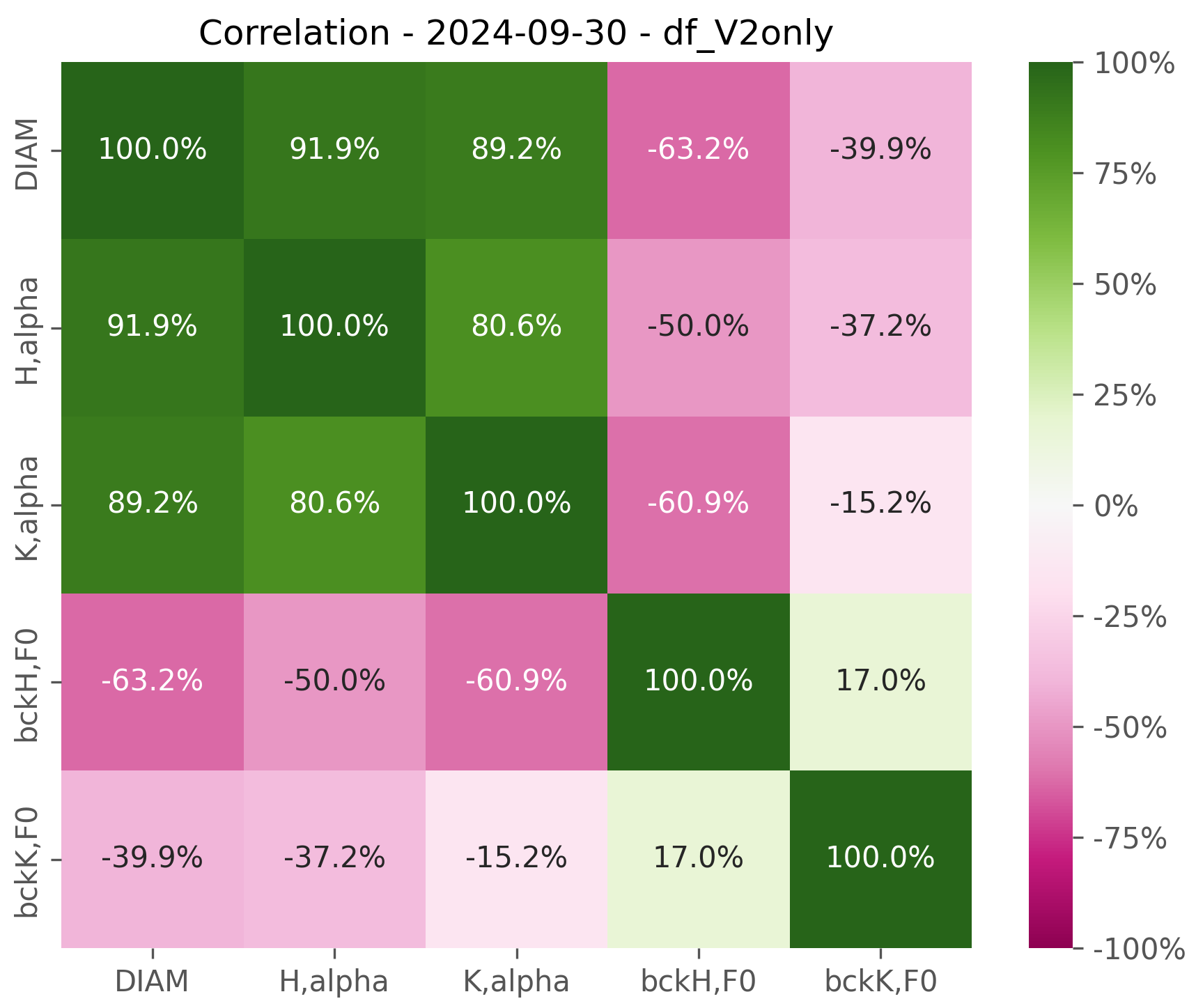}
        \label{fig:c2}
    \end{subfigure}

    \caption{Bootstrap fit for the night of 30 September 2024 for $\zeta$ Gem without taking the closure phase into account for the fit. The figure at the top show the results and the goodness of fit. The one at the bottom show the corresponding correlation matrices.}
    \label{fig:zeta_gem_20241111}
\end{figure}

\begin{figure}[h!]
    \centering

    \begin{subfigure}{0.5\textwidth}
        \centering
        \includegraphics[width=\linewidth]{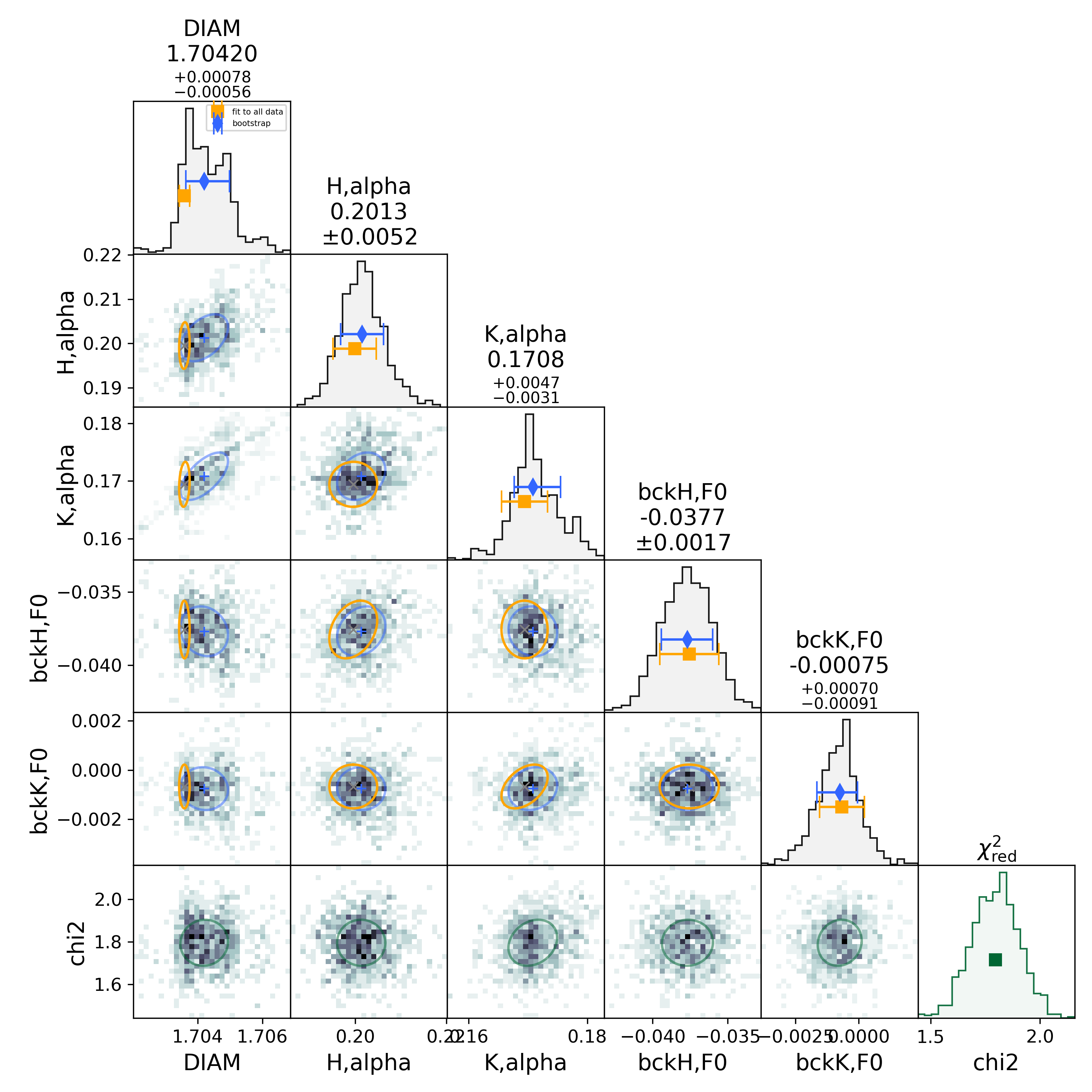}
    \end{subfigure}
    \begin{subfigure}{0.25\textwidth}
        \centering
        \includegraphics[width=\linewidth]{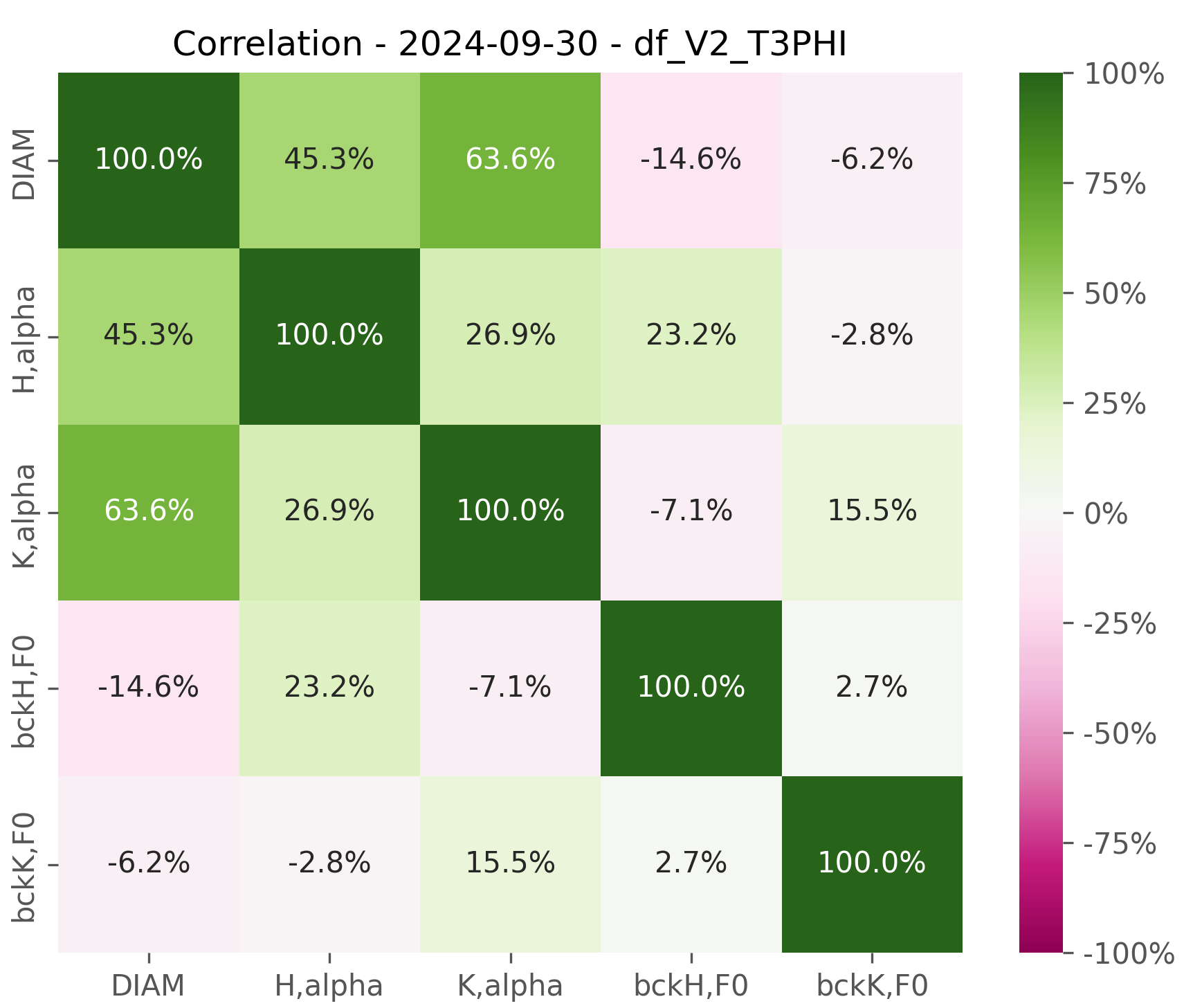}
    \end{subfigure}

    \caption{Same as Fig.~\ref{fig:zeta_gem_20241111} but taking into account the CP for the fit.}
    \label{fig:zeta_gem_20241111_wCP}
\end{figure}

\begin{figure}[h!]
    \centering
    \includegraphics[width=0.95\linewidth]{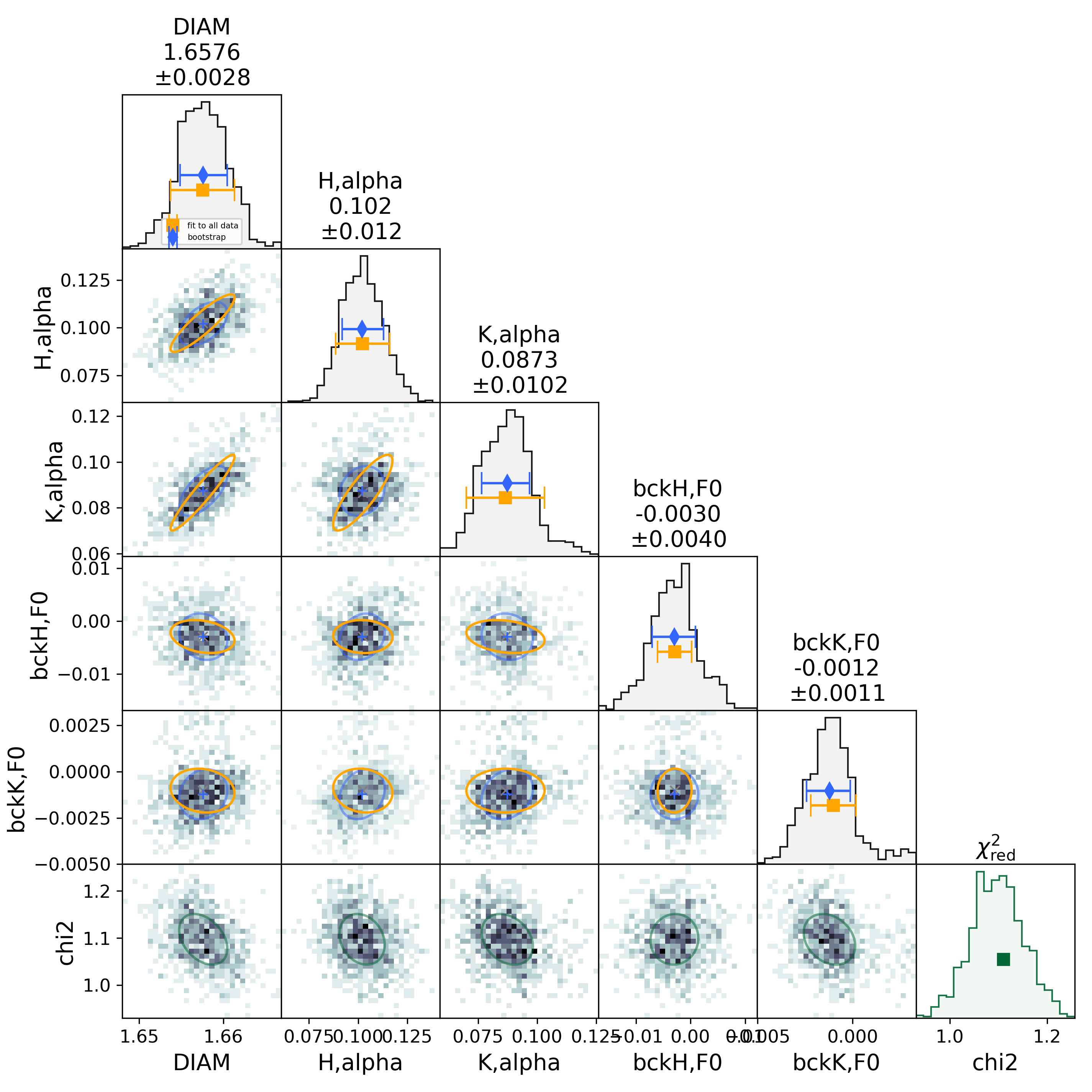}
    \caption{Bootstrap fit for the night of 17 June 2025 for $\eta$ Aql, with bad quality CP.}
    \label{fig:eta_aql_20250617}
\end{figure}

\clearpage
\section{Diameter comparison}
\begin{figure}[h!]
    \centering
    \includegraphics[width=0.8\linewidth]{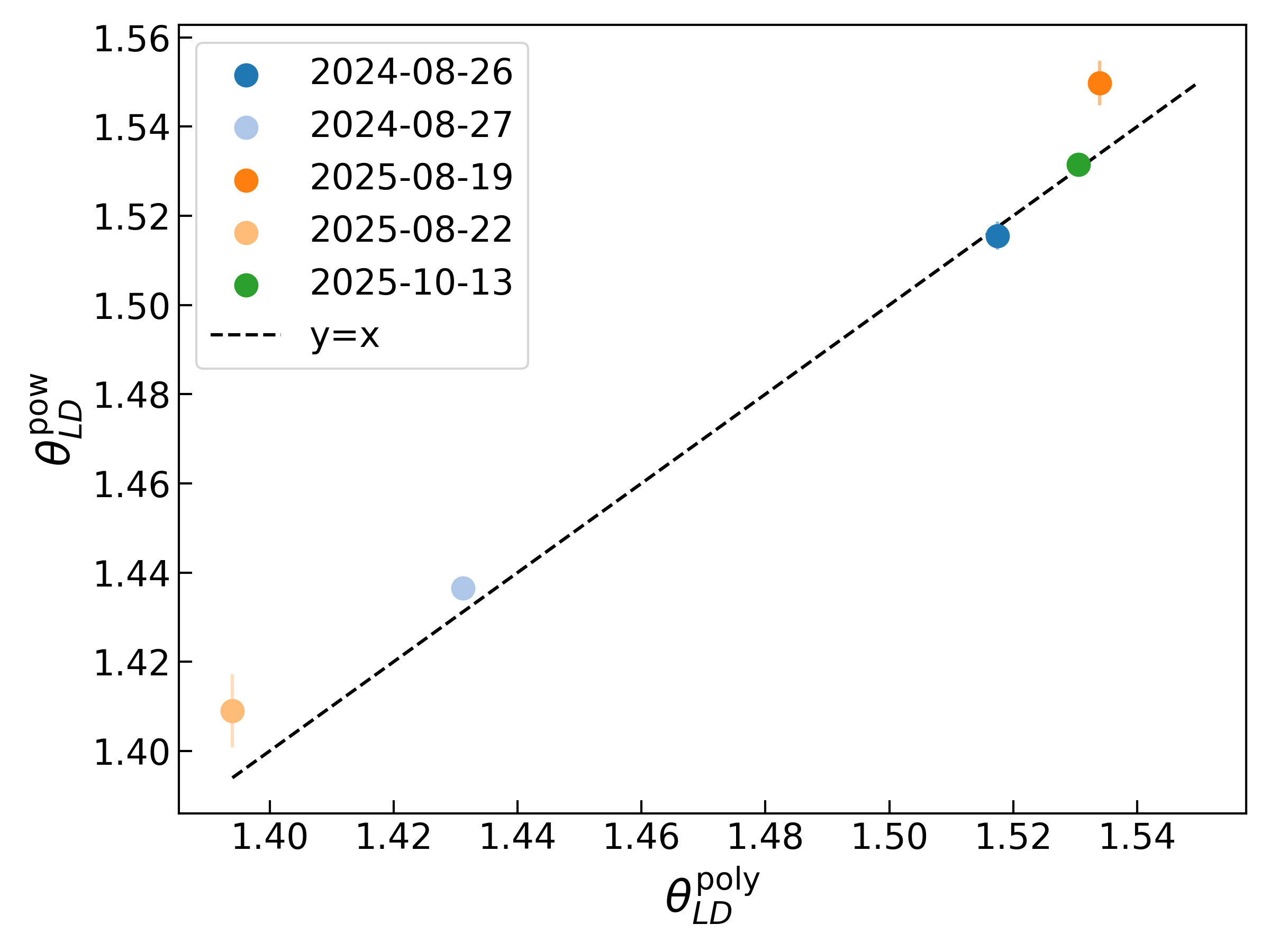}
    \includegraphics[width=0.8\linewidth]{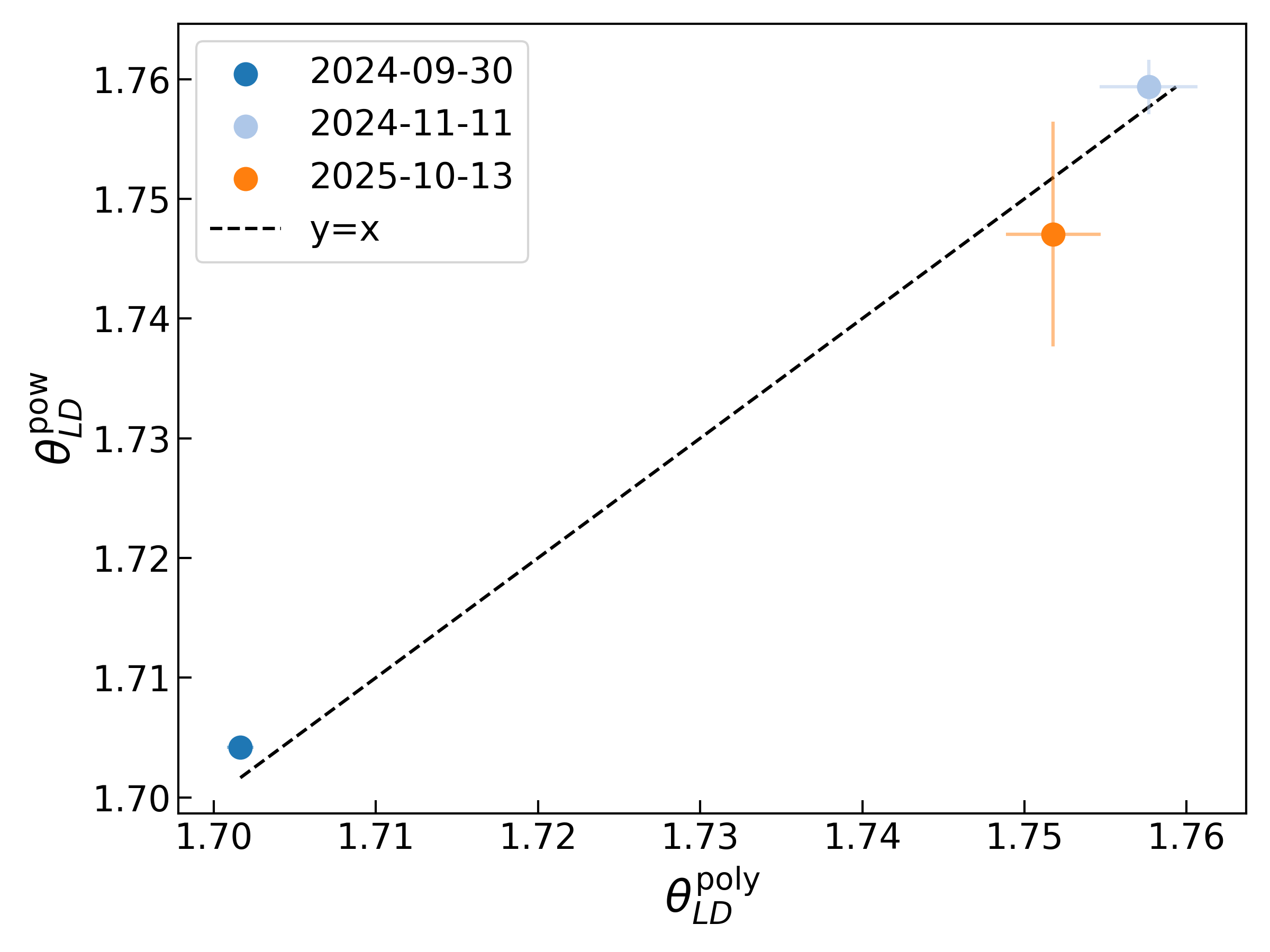}
    \includegraphics[width=0.8\linewidth]{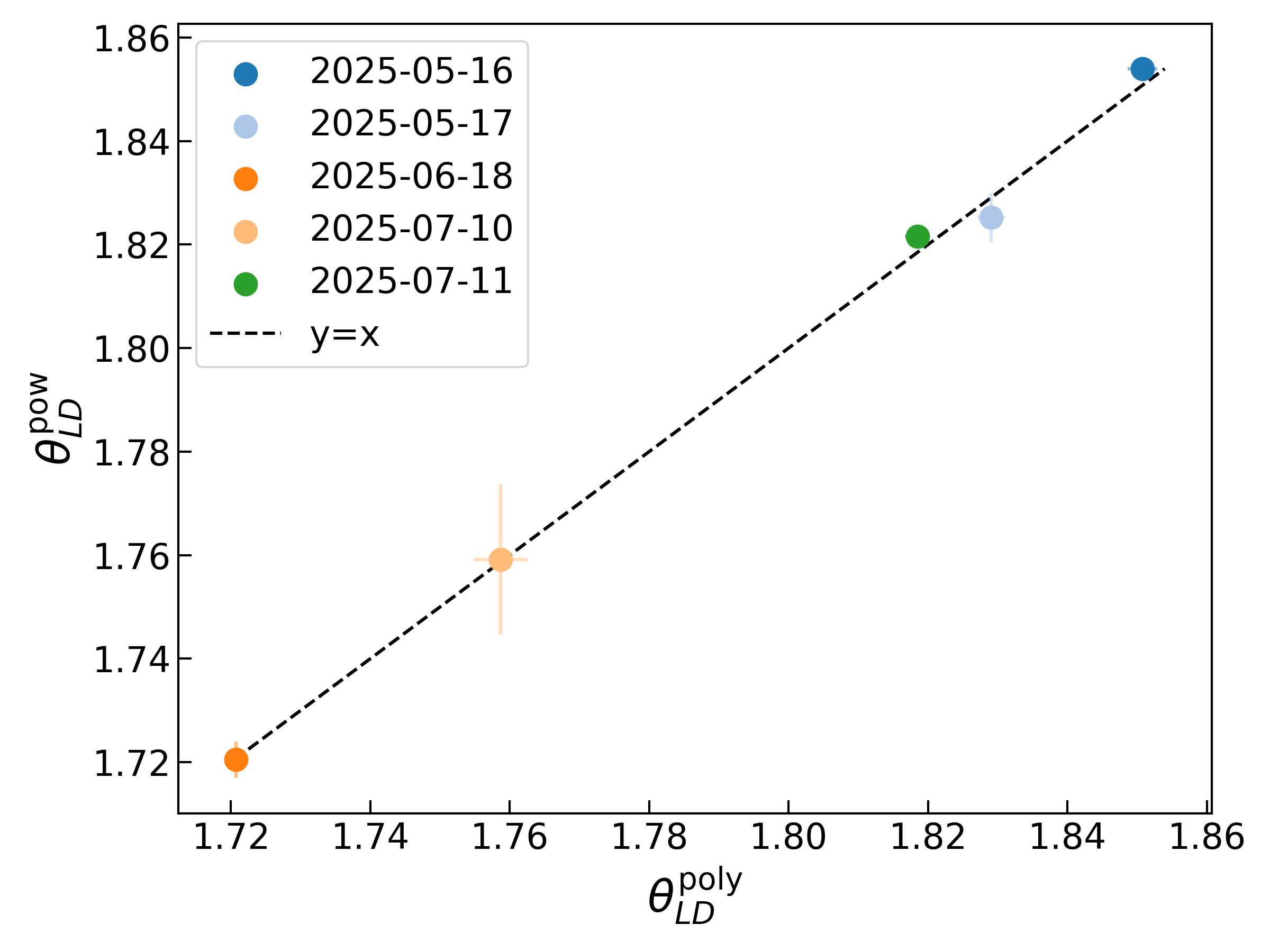}
    \caption{LD diameter obtained in Sect.~\ref{sec:Measuring the limb-darkening of classical cepheids} (LD coefficients fitted) compared to LD diameter obtained in Sect.~\ref{sec:Measurement of the limb-darkened diameter of Cepheids} (LD coefficients fixed). Top to bottom: $\delta$ Cep, $\zeta$ Gem and $\eta$ Aql.}
    \label{fig:diam_comparison}
\end{figure}

\section{Limb-darkening and geometric projection factor measurements for $\zeta$~Gem and $\eta$~Aql}

\begin{figure}[h!]
\centering
        \includegraphics[width=\linewidth]{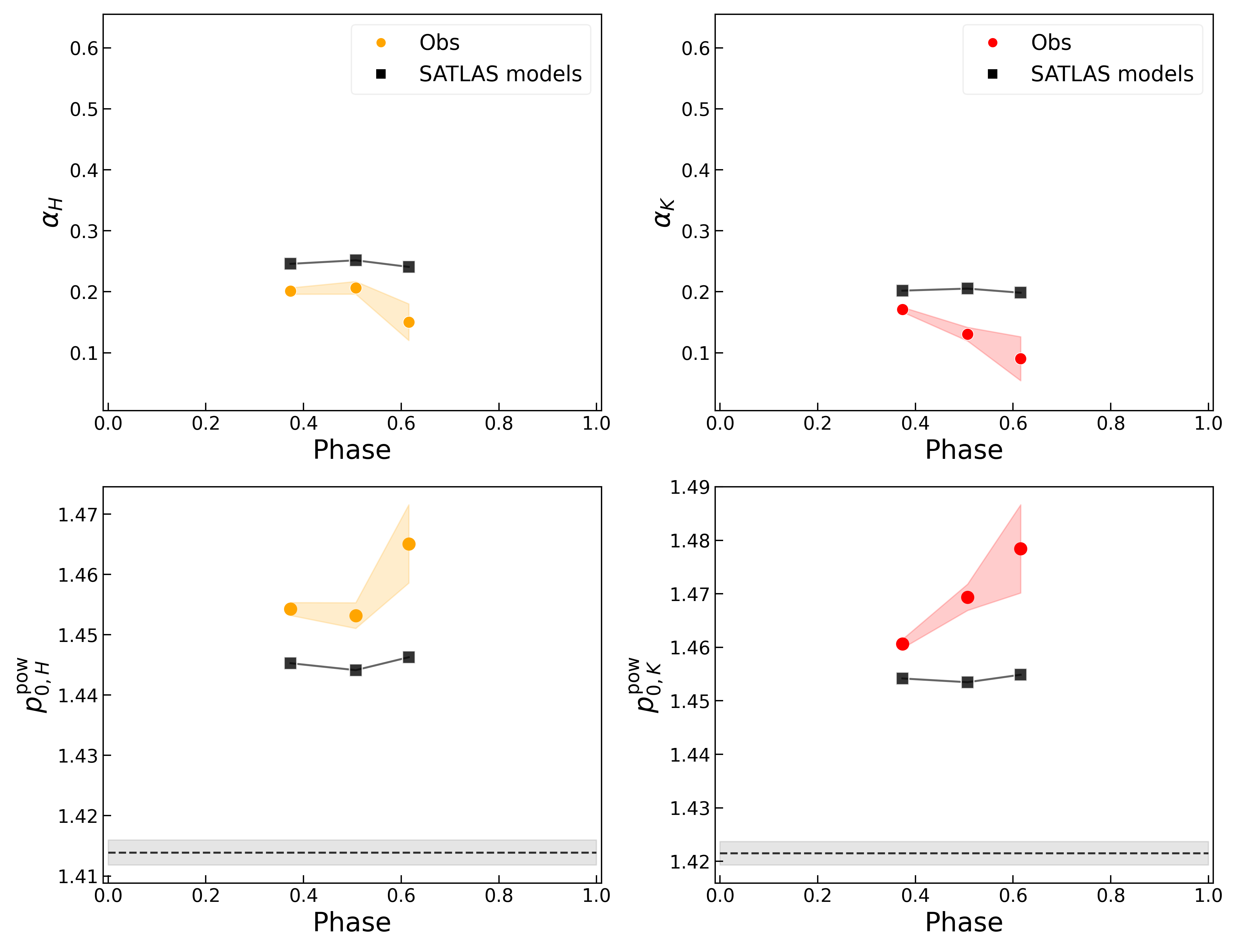}
        \caption{Same figure as Fig.~\ref{fig:LDs_p0_results_deltaCep} but for $\zeta$ Gem.}
        \label{fig:LDs_p0_results_zetaGem}
\end{figure}

\begin{figure}[h!]
\centering
        \includegraphics[width=\linewidth]{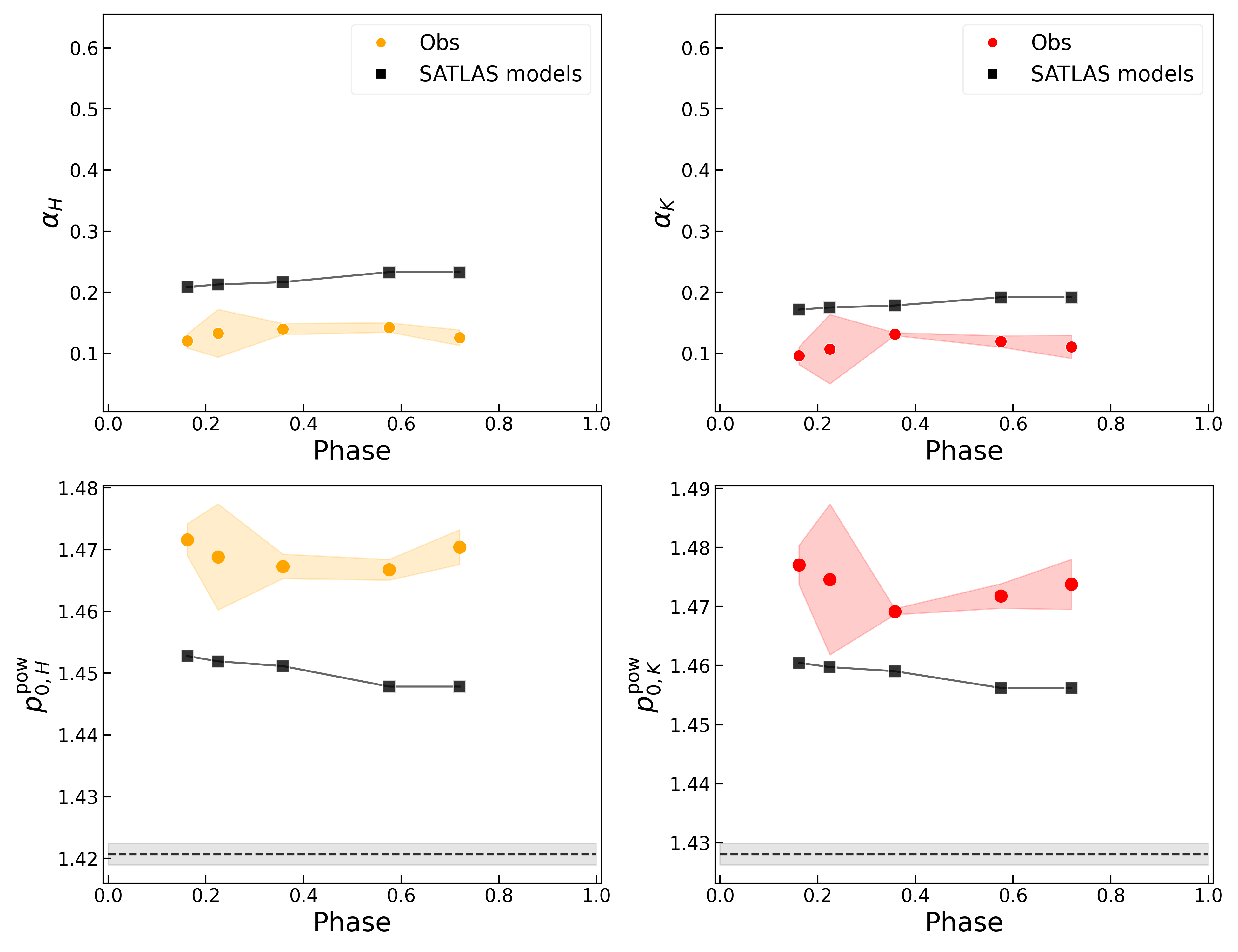}
        \caption{Same figure as Fig.~\ref{fig:LDs_p0_results_deltaCep} but for $\eta$ Aql.}
        \label{fig:LDs_p0_results_etaAql}
\end{figure}

\end{appendix}
\end{document}